\documentclass[sigconf,nonacm]{acmart}  

\usepackage{amsthm, amsfonts}
\usepackage{stmaryrd}

\usepackage{microtype}
\usepackage{multicol}
\usepackage{comment}
\usepackage{diagbox}
\usepackage{booktabs}
\usepackage{float}
\usepackage{hyperref}
\usepackage{paralist}
\usepackage{graphicx}
\usepackage{xcolor}
\usepackage[noend]{algorithm2e}
\usepackage{etoolbox}
\gappto{\UrlSpecials}{\do\_{\penalty\UrlBreakPenalty\mathchar`_}}

\definecolor{green}{RGB}{0,128,0}
\colorlet{orange}{orange!85!black}

\usepackage{balance}
\usepackage{dsfont}
\usepackage{mathtools}
\usepackage{paralist}
\usepackage{tikz}
\usepackage[appendix=append]{apxproof}

\usepackage{subcaption}
\usepackage{pifont}
\usepackage{enumitem}
\usepackage[normalem]{ulem}
\usepackage{xparse}
\usepackage{pgffor}

\def\websitesSinglePage{{ce}, {cl}, {ed}, {il}, {in}, {ju}, {nc}, {ok}, {wh}, {wo}}

\makeatletter
\newcommand{\getfullname}[1]{
  \csname fullname@#1\endcsname
}

\expandafter\def\csname fullname@1\endcsname{US Census (\emph{ce})}
\expandafter\def\csname fullname@2\endcsname{French Local Communities (\emph{cl}) }
\expandafter\def\csname fullname@3\endcsname{French Ministry of Education (\emph{ed})}
\expandafter\def\csname fullname@4\endcsname{UN International Labor Organization (\emph{il})}
\expandafter\def\csname fullname@5\endcsname{French Ministry of Interior (\emph{in})}
\expandafter\def\csname fullname@6\endcsname{French Ministry of Justice (\emph{ju})}
\expandafter\def\csname fullname@7\endcsname{US National Center for Education Statistics (\emph{nc})}
\expandafter\def\csname fullname@8\endcsname{Open Knowledge
Foundation (\emph{ok})}
\expandafter\def\csname fullname@9\endcsname{UN
World Health Organization (\emph{wh})}
\expandafter\def\csname fullname@10\endcsname{World Bank (\emph{wo})}
\makeatother

\newcommand{\0}{\phantom{0}}

\SetKwInput{KwInput}{Input}
\SetKwInput{KwOutput}{Output}

\newcommand{\cmark}{\ding{51}}
\newcommand{\xmark}{\ding{55}}

\RestyleAlgo{ruled}
\SetInd{0.5em}{0.3em}
\SetKwComment{Comment}{}{}

\newtheoremrep{theorem}{Theorem}
\newtheoremrep{proposition}[theorem]{Proposition}
\newtheorem{definition}[theorem]{Definition}
\newtheorem{problem}[theorem]{Problem}

\renewcommand\leq\leqslant
\renewcommand\geq\geqslant
\renewcommand\epsilon\varepsilon
\renewcommand\phi\varphi

\begin{document}

\newcommand{\extVersion}{false}

\newcommand{\printIfExtVersion}[2]
{
	\ifthenelse{\equal{\extVersion}{true}}{#1}{}
	\ifthenelse{\equal{\extVersion}{false}}{#2}{}
}

\title{Efficient~Crawling~for~Scalable~Web~Data~Acquisition (Extended~Version)}

\author{Antoine Gauquier}
\email{antoine.gauquier@ens.psl.eu}
\orcid{0009-0005-9573-6364}
\affiliation{
	\institution{DI ENS, ENS, CNRS, PSL, Inria}
	\city{Paris}
	\country{France}
}

\author{Ioana Manolescu}
\email{ioana.manolescu@inria.fr}
\orcid{0000-0002-0425-2462}
\affiliation{
  \institution{\makebox[0pt]{Inria \& Institut Polytechnique de Paris}}
	\city{Palaiseau}
	\country{France}
}

\author{Pierre Senellart}
\email{pierre@senellart.com}
\orcid{0000-0002-7909-5369}
\affiliation{
	\institution{DI ENS, ENS, CNRS, PSL, Inria}
	\city{Paris}
	\country{France}
}

\begin{abstract}
  Journalistic fact-checking, as well as social or economic research,
  require analyzing \emph{high-quality statistics datasets} (\emph{SDs}, in short).
  However, retrieving  SD corpora at scale may be hard, inefficient, or impossible,
  depending on how they are published online.
	To improve open statistics data accessibility, we present a \emph{focused Web crawling
  algorithm} that retrieves  as many \emph{targets}, i.e., resources of certain types,
  as possible, from a given website, in an efficient and scalable way, by crawling (much)
  less than the full website.  We show that optimally solving this problem is intractable,
  and propose an approach based on reinforcement learning, namely using sleeping bandits.
  We propose \textsf{\small SB-CLASSIFIER}, a crawler that  efficiently learns \emph{which
  hyperlinks lead to pages that link to many targets}, based on the paths leading to the
  links in their enclosing webpages. Our experiments on websites with millions of webpages
  show that our crawler is \emph{highly efficient}, delivering high fractions of a site's
  targets while crawling only a small part.
\end{abstract}


\maketitle

\section{Introduction}\label{section:introduction}

Openly accessible data has long been shared through the Web; in
particular, many HTML pages contain \emph{entity-centric tables}, where
each line is about an entity, e.g., an album, and describes its attributes such as artist or year. Entity-centric tables have been leveraged to extract large-scale, open knowledge bases \cite{DBLP:conf/wsdm/LockardSDH20,DBLP:conf/semweb/AuerBKLCI07,DBLP:conf/esws/TanonWS20,DBLP:journals/ftdb/WeikumDRS21}, and for question answering \cite{DBLP:journals/pacmmod/WangF23,herzig-etal-2021-open,Christmann2023CompMixAB,Hulsebos2024ItTL}.
They are also frequent in \emph{data lakes}, \cite{fan_semantics-aware_2023,deng_lakebench_2024,christensen_fantastic_2025}.
In this work, we focus on \emph{openly accessible} Web data, independently of their licensing status. Among them, \emph{statistics datasets} (\textbf{SDs}, in short) form a distinct and highly valuable category.
They are compiled by domain specialists typically working for national
or international governing bodies, such as the
\href{https://www.un.org/}{United Nations}, the
\href{https://www.imf.org/}{IMF}, but also by companies, NGOs, etc.
Some organizations, e.g.,
\href{https://ec.europa.eu/eurostat/}{Eurostat}, publish hundreds of
thousands of SDs, updated yearly; others, e.g., an NGO focused on equitable justice, or a pollution watchdog in a certain area, may only publish a few dozen highly specialized datasets.

SDs~significantly differ from entity-oriented Web tables.
SD~content is
mostly numeric, e.g.,  birth and death counts by age in a
country over decades, or chip production per country and type of chip.
From a data management perspective, SDs are multidimensional
	aggregates, sometimes data cubes.
Also, SDs are mostly  published as standalone files, encoded in a
variety of formats, e.g., CSV, TSV, spreadsheets, XML dialects such as
SDMX~\cite{sdmx}, JSON, etc. Open SDs are also sometimes embedded in PDF
documents.

\emph{Answering queries or questions over tables}, and
  in our context, over SDs, is as the very core of the data management
  science and industry. The wide availability of statistic datasets on the Web requires tools that make
such data reachable in order to realize its potential; as the paper shows, such tools are currently lacking.

SDs are of great public utility. Officials rely on them to
inform decisions and support debates; social scientists use them to
analyze policy impacts and detect trends; newsrooms use them to check the
accuracy of public statements, make comparisons across countries, etc.
Also, the \emph{statistical literacy} of the general public needs to be improved: for instance, US voters vastly misrepresent the size of various fractions of
the US population, e.g., how many are transgender (estimated: 21\%,
true: 0.6\%)~\cite{wrongstats}. Thus, \emph{building and using warehouses of
statistics data} is central. Research works seeking to automate the recognition and checking of claims leveraging on retrieved corpora of SDs include, e.g.,~\cite{DBLP:journals/pvldb/Karagiannis0PT20,DBLP:journals/debu/0002P21,DBLP:conf/webdb/CaoMT18,balalau2024star, claimbuster} in the Information Retrieval and Data Management, and \cite{vlachos-riedel-2015-identification} in NLP.

Finding all SDs published by an organization is desirable in
order to study them as a whole, compare among organizations, populate a data lake for fact-checking, etc. For example, a work historian may want to retrieve all SDs published by the \href{https://www.ilo.org}{ILO}; a journalist may need all \href{https://ec.europa.eu/eurostat/}{Eurostat} SDs on poverty.  Unfortunately, current methods for this are either lacking, or quite inefficient.

Some websites publish an API to retrieve all their SDs, e.g.,
\href{https://ec.europa.eu/eurostat/}{Eurostat},
yet writing
custom code for each such API is cumbersome and brittle when APIs change.
Worse, many sites hosting numerous Open SDs, e.g., the ILO, do not
provide APIs to access all their data.
\emph{Search engines} (SEs, in short) and associated data
portals turn out to provide access to only a tiny fraction of
existing SDs (see Sec.~\ref{subsubsection:search_engines}); also, how these are selected is opaque to users.
A  \textbf{na\"ive, exhaustive website crawl is extremely inefficient} for large websites:
\begin{inparaenum}[(i)]
	\item acquiring a full
	website takes space and time;
	\item {\em crawling ethics} requires to wait
	(typically 1~second) between two successive HTTP requests; for
	a site of 1~million pages, such waits, alone, take 11
	days.
\end{inparaenum}
Thus, we are interested in a focused crawl, seeking to acquire within a given website as many \textbf{targets} as possible, while
\textbf{minimizing the resources consumed}, e.g., HTTP requests or
volume of transferred data.
Motivated by our SD retrieval problem, we define \emph{targets} as \emph{data files (CSV, spreadsheet, etc.)}. This can generalize to \emph{any} definition of target. Also, we aim for an efficient method, feasible on a single machine, as opposed to SE-scale parallel infrastructure.

Focused crawlers have been extensively
studied~\cite{chakrabarti1999focused,diligenti2000focused,gouriten2014scalable,meusel2014focused,faheem2015adaptive,han2018focused,kontogiannis2022tree}.
However, existing approaches often rely on \emph{assumptions that may not
hold in our setting}, e.g., that all pages on a
topic tend to be close within a website~\cite{10.1145/345508.345597}.
Early work includes generic focused crawling
approaches~\cite{chakrabarti1999focused,diligenti2000focused} (that we
will illustrate with a \textsf{\small FOCUSED} baseline); other study
large-scale website selection~\cite{meusel2014focused} (as opposed to SDs
within a given website), topic-driven page retrieval methods such as
\textsf{\small TRES}~\cite{kontogiannis2022tree}
or~\cite{chakrabarti1999focused,gouriten2014scalable,han2018focused}, and
non-topical objectives such as textual
diversity~\cite{faheem2015adaptive} (adapted to our setting
as a \textsf{\small TP-OFF} baseline).

Motivated by the large numbers of open, valuable SDs accessible
via navigation, we address the challenging problem of
\textbf{retrieving SDs in a website  as efficiently as possible, 
without relying on prior knowledge of the website's structure or content.}

\setlength{\fboxsep}{2pt}
\paragraph*{Contributions} Our contributions are as follows. \begin{inparaenum}[(1)]
	\noindent\item We formalize
	our graph crawling problem and show that optimally solving it is
	intractable (Sec.~\ref{sec:problem-model}).
	\noindent\item We propose a novel approach \textsf{\small SB-CLASSIFIER} based on
		reinforcement learning (RL) in Sec.~\ref{section:solution}, relying on two crucial  hypotheses: \begin{inparaenum}[(i)]
		\item 
		\emph{links found on similar \fbox{tag paths} within HTML web pages} (a tag path~\cite{miao2009extracting, furche2012turn} goes from the page root, to a hyperlink tag such as \texttt{\small $<$a$>$}) \emph{lead to similar content},
		\item \emph{we can learn, from the tag path structure, which
			tag paths are likely to lead to links towards targets}.
	\end{inparaenum}
	\noindent\item We demonstrate the effectiveness and efficiency of our approach
	through extensive experiments on diverse
		websites, totalling 22.2 millions pages (Sec.~\ref{sec:results}). \textsf{\small SB-CLASSIFIER} outperforms all baselines.
	On some websites we experimented on, in particular very large ones,
	\textbf{our crawler retrieves 90\% of the targets  accessing only 20\% of the
		webpages} (Sec.~\ref{subsection:dataset_characteristics}).
\end{inparaenum}

We discuss related work in
Sec.~\ref{section:related_work}. This paper is an extended version of the conference paper~\cite{gauquier2026efficient2}. A Github repository~\cite{github_repository} provides code to reproduce the experiments. 

\section{Problem Statement and Modeling}
\label{sec:problem-model}

We formalize our SD acquisition as a \emph{graph crawling problem}. We show that a relaxed version thereof is NP-hard, even if the website is
known before the crawl (it is not!) in
Sec.~\ref{section:problem_statement}. Then we detail the mapping of our problem into the graph crawling framework, including a crucial choice of labels for the graph edges, in Sec.~\ref{section:graph_to_website}.

\subsection{Graph Crawling Problem}
\begin{toappendix}
	\begin{figure}[h]
		\centering
		\begin{tikzpicture}[level distance=1cm,
				level 1/.style={sibling distance=5cm},
				level 2/.style={sibling distance=2.5cm}]

			\node[circle] (root) {$r$};
			\node[circle] (s1) at (-2,-1.3) {$s_1$};
			\node[circle] (s2) at (-0.67,-1.3) {$s_2$};
			\node at (0.67,-1.3) {$\dots$};
			\node[circle] (sn) at (2,-1.3) {$s_n$};

			\draw[->] (root) -- (s1);
			\draw[->] (root) -- (s2);
			\draw[->] (root) -- (sn);

			\node[circle] (a1) at (-3.33,-2.6) {$u_1$};
			\node[circle] (a2) at (-2,-2.6) {$u_2$};
			\node[circle] (a3) at (-0.67,-2.6) {$u_3$};
			\node at (0.67,-2.5) {$\dots$};
			\node[circle] (am1) at (2,-2.6) {$u_{m-1}$};
			\node[circle] (am) at (3.33,-2.6) {$u_m$};

			\draw[->] (s1) -- (a1);
			\draw[->] (s1) -- (a2);
			\draw[->] (s1) -- (am1);

			\draw[->] (s2) -- (a3);
			\draw[->] (s2) -- (am);

			\draw[->] (sn) -- (a2);
			\draw[-] (sn) -- ++(0,-1) coordinate (extended) -- (am1);
			\draw[->] (sn) -- (extended);
			\draw[->] (sn) -- (am);

		\end{tikzpicture}
		\caption{Graphical summarization of the graph $G_{\textrm{sc}}$}
		\label{fig:graph_set_cover}
	\end{figure}

	\subsection{Graph Crawling Problem}
\end{toappendix}
\label{section:problem_statement}

We model a website
as a rooted,
node-weighted, edge-labeled directed graph. Each node represents a webpage, and each edge is a hypertext link leading from one to another.
We fix a countable set~$\mathcal{L}$ of labels, e.g., finite sequences of
character strings.

\begin{definition}
	A \emph{website graph} is a tuple $G = (V, E, r, \omega, \lambda)$, with: $V$ a finite set of \emph{nodes}
	(representing webpages);
	$E\subseteq V^2$ a set of \emph{edges} (representing hyperlinks);
	$r \in V$ the \emph{root} of the graph (the input webpage);
	$\omega: V\to\mathbb{R}^+$ a \emph{cost} function assigning a positive
	weight to every node (the cost of	retrieving that page);
	$\lambda: E\to\mathcal{L}$ a \emph{labeling} function, assigning a label to	each link found in a page.
\end{definition}

On such a graph, we define a \emph{crawl}
and its \emph{cost} as follows:
\begin{definition}
	A \emph{crawl} of a website graph $G$
	is an $r$-rooted subtree $T=(V',E')$ of $(V,E)$.
	Its total cost, $\omega(T)$, is defined as $\sum_{\substack{u\in V'}} \omega(u)$.
\end{definition}

The graph crawling problem is formalized as follows:

\begin{problem}
\label{problem:graph-crawling}
Given a website graph $G=(V,E,R,\omega,\lambda)$ and a subset~$V^*$
of \emph{targets} in~$V$
, the graph crawling problem is to find a crawl $T=(V',E')$ of~$G$ with $V^*\subseteq V'$, of minimal total cost.
\end{problem}

\begin{figure}
	\centering
	\includegraphics[width=\linewidth]{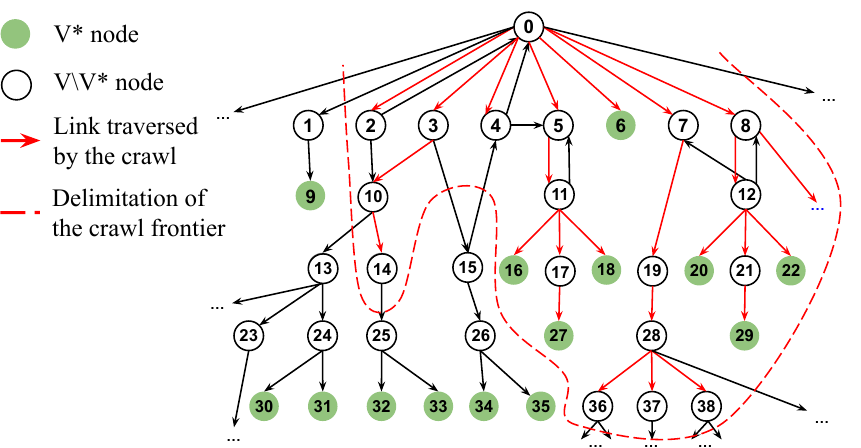}
	\caption{Sample website, crawl, and frontier}
	\label{fig:example_graph_crawling}
\end{figure}

We define the \emph{frontier} of a crawl $\left\{u\in V\setminus V'\mid
\left(v,u\right)\in E, v\in V'\right\}$: nodes that have not
been crawled but are pointed to from nodes that have been. Figure~\ref{fig:example_graph_crawling} shows a website graph as well as a possible crawl and its frontier. Even assuming the graph is fully known in advance, the graph crawling problem is intractable:

\begin{propositionrep}
	\label{prop:np-complete}
	Given a website graph $G=(V,E,r,\omega,\lambda)$, a subset
	$V^*\subseteq V$ and some $B\in\mathbb{R}^+$, determining
	whether there exists a crawl~$T=(V',E')$ of $G$ such that $V^*\subseteq V'$
	and $\omega(T)\leq B$ is
	NP-complete; hardness holds even when $\omega$ is a constant function.
  This holds even when nodes in $V^*$ do not have any out 
    links in~$G$.
\end{propositionrep}

This is the decision variant of the graph crawling problem. Hardness
is shown by reduction from the set cover problem~\cite{johnson1979computers2}; the proof can be found in the extended version in~\cite{github_repository}.

\begin{proof}
	To show NP-completeness, we must show that the problem belongs to NP and is NP-hard.

	Let us start with the upper bound. Given a graph
	$G=(V,E,r,\omega,\lambda)$, we guess a subgraph $T=(V',E')$ of~$G$
	(which is a polynomial-sized guess). In polynomial time, we check
	whether $T$ is a $r$-rooted tree (i.e., whether it is connected,
	includes~$r$, $r$ has indegree 0 and other nodes indegree 1), we
	check that $V'$ contains all nodes of $V^*$, and we check that
	$\omega(T)\leq B$. We accept if and only if these conditions are
	all satisfied. This yields a nondeterministic polynomial-time
	algorithm, meaning the problem is in NP.

	We now move to the lower bound.
	Our
	crawling problem can be seen as a directed variant of the well-known
	NP-complete \emph{Steiner Tree} \cite{johnson1979computers}
	problem. NP-hardness of the directed Steiner tree problem is mentioned in
	the literature (see, e.g., \cite{watel2016greedy}), but as it is not
	formally shown there, we prefer for completeness of the presentation
	reducing from the set cover problem, a classic NP-hard
	problem~\cite{johnson1979computers}. We denote $\mathcal{U} =
		\left\{u_1, \dots, u_m\right\}$ a set of $m$ elements called the
	universe. We also define a collection $\mathcal{S} = \left\{s_1,
		\dots, s_n\right\}$ of $n$ non-empty subsets, each of them containing some elements of $\mathcal{U}$, such that:
	\[
		\bigcup_{s \in \mathcal{S}} s = \mathcal{U}.
	\]
	In its decision version, the set cover consists in given such a
	universe and collection, given a natural integer~$B$, determining
	whether there exists a \emph{cover} $\mathcal{C} \subseteq
		\left\{s_1, \dots, s_n\right\}$ such that $\left\lvert\mathcal{C}\right\rvert \leq B$ and:
	\[
		\bigcup_{s \in \mathcal{C}} s = \mathcal{U}.
	\]
	\def\sc{\mathrm{sc}}

	We now propose a polynomial-time many-one reduction of the set cover problem to an instance of
	the graph crawling problem. We create a website graph
	$G_\sc = \left(V_\sc, E_\sc, r, \omega,
		\lambda\right)$ as follows. We set $V_\sc$ to be
	$\{\,u_1,\dots,u_m,r,s_1,\dots,s_n\,\}$, including representations for
	every element of the universe~$\mathcal{U}$, every set of the
	collection~$\mathcal{S}$, as well as a distinct root~$r$ (by abuse of
	notation, we do not distinguish between elements of $\mathcal{U}$,
	$\mathcal{S}$ and the way they are represented in~$V_\sc$).
	We define $E_\sc$ as $\left\{\left(r, s_i\right) \, |
		\, i \in \left\{1, \dots, n\right\}\right\} \cup \left\{\left(s_i,
		u\right) \, | \, u \in s_i, i \in \left\{1, \dots, n\right\}\right\}$.
	In other words, in $G_\sc$ from the origin
	(root) $r$, we model each element of $\mathcal{S}$ as a vertex, that
	can be reached following a dedicated (directed) edge. Finally, for
	each new vertex $s_i \in \mathcal{S}$, we have as many outgoing edges
	as there are elements of $\mathcal{U}$ in $s_i$. Finally, we set
	$\omega$ to be the constant function that assigns cost~$1$ to every
	vertex, and $\lambda$ to be some constant function.
	The result is a graph in the form of a tree of depth 2, depicted in
	Figure~\ref{fig:graph_set_cover}. We fix $V^*$ to be
  $\mathcal{U}$. Observe that nodes in $\mathcal{U}$ do not have
  any outgoing links. We state that there exists
	$\mathcal{C} \subseteq \left\{s_1, \dots, s_n\right\}$ such that
	$\left\lvert \mathcal{C} \right\rvert \leq B$ and $\bigcup_{i \in
			\mathcal{C}} s_i = \mathcal{U}$ if and only if there exists a crawl
	$T_\sc$ of $G_\sc$ containing all elements of $V^*$ and of total cost
	$\omega(T_\sc) \leq \left\lvert \mathcal{U} \right\rvert + B + 1$.

	Let us explain why this reduction is
	polynomial-time. In set cover, the universe $\mathcal{U}$ can be
	described by the number $m$ of its elements, with representation size
	$\Theta(\log m)$. Each set of $\mathcal{S}$ needs to list every element
	within this set, so a set $s_i$ has representation size $\Theta(\log
		m\times |s_i|)$. Finally, $B$ has representation size $\Theta(\log B)$.
	This yields a total input size of $\Theta((\log
		m)\left(\sum_{i=1}^n|s_i|+1\right)+\log B)$. Note that $\sum_{i=1}^n
		|s_i|\geq \max(m,n)$ so this is $\Omega(\max(m,n)\log m+\log B)$.
	But then, the construction depicted in
	Figure~\ref{fig:graph_set_cover} can clearly be done in time
	polynomial in $m$ and $n$ (namely, in $O(m\times n)$ in the
	worst case where every set of the collection contains every element).
	The reduction is therefore polynomial-time.

	We now proceed to show equivalence between the initial problem known
	to be NP-hard (set cover) and the graph crawling instance presented
	above. First, suppose that
	there exists $\mathcal{C} \subseteq \left\{s_1, \dots,
		s_n\right\}$ such that $\left\lvert \mathcal{C} \right\rvert \leq B$
	and $\bigcup_{i \in \mathcal{C}} s_i = \mathcal{U}$.
	Then consider the crawl $T_\sc$ of $G_\sc$ formed by including
	$r$, every element of
	$\mathcal{C}$ using the edge from $r$ to that element, and every
	edge from an element of $\mathcal{C}$ to an element of $\mathcal{U}$.
	Since $\mathcal{C}$ is a cover, this includes all elements of
	$\mathcal{U}$. The total cost of this crawl
	$\omega(T_\sc)=1+|\mathcal{C}|+|\mathcal{U}|\leq |\mathcal{U}|+B+1$.

	Now suppose that there exists a crawl $T_\sc$ of $G_{\textrm{sc}}$ of
	total cost $\omega(T_\sc) \leq \left\lvert \mathcal{U} \right\rvert
		+ B + 1$. Note that by definition of $\omega$, the cost is just the
	number of nodes in $T_\sc$, and this crawl necessarily includes the
	root~$r$ as well as all vertices of $\mathcal{U}$. The remaining
	$\leq B$ vertices are therefore vertices of~$\mathcal{S}$. We pose
	$\mathcal{C}$ to be those. Then $|\mathcal{C}|\leq B$ and since
	$T_\sc$ is a crawl, for every $u\in\mathcal{U}$, there exists at
	least one $s\in\mathcal{C}$ such that the edge $(s,u)$ is in $T_\sc$,
	meaning that $u\in s$. We indeed have
	$\mathcal{U}=\bigcup_{s\in\mathcal{C}} s$.
\end{proof}
Optimal methods being out of
reach (websites may have millions of
pages),
we need heuristic methods with a low cost in practice.

\subsection{Data Acquisition as Graph Crawling}\label{section:graph_to_website}
\begin{toappendix}
	\subsection{Data Acquisition as Graph Crawling}
\end{toappendix}

We complete our modeling of Web data acquisition as an instance of the
graph crawling problem by detailing the graph~$G$, the target subset
$V^*$, and the edge labeling function~$\lambda$. The choice of $\lambda$ will turn out to be crucial for the performance of our approach.

We identify pages by their URL and fix $r$, the website root, as the start of the crawl.
Next, we need to identify which pages belong to the website graph. Lacking a commonly agreed definition of a website's boundary~\cite{senellart2005identifying,alshukri2010website}, we use a pragmatic approach. A~URL is considered to be part of the same website as the root $r$ if \emph{its hostname} (the part of the URL which is after the scheme but before the path, with a potential ``www.'' prefix excluded) \emph{is a subdomain of the hostname of~$r$}.
$V$ contains all such URLs. For instance, if $r$ is \url{https://www.A.B.com/index.php}, the URLs \url{https://www.A.B.com/folder/content.php} and \url{https://www.C.A.B.com/page.html} are part of $V$, but \url{https://www.B.com/page.php} and \url{https://edbticdt2026.github.io/?contents=EDBT_CFP.html} are not.\footnote{The reason for the special handling of a ``www.'' prefix is that many (but not all\dots) websites use it as a prefix for the domain name of the
	Web server.} An edge $(u,v) \in E$ exists if $u$ links to $v$~via HTML
tags like \texttt{<a>}, \texttt{<area>},\texttt{<iframe>}.

Since our goal is to find SDs,
target pages are those whose \emph{Multipurpose Internet Mail
	Extensions} (MIME) type is in a
\emph{user-defined} list (e.g., 
{\small\verb|text/csv|}, {\small\verb|application/pdf|},
{\small\verb|application/vnd.ms-excel|}).
Non-target types include {\small\verb|text/html|},
{\small\verb|video/*|}, {\small\verb|audio/*|}, {\small\verb|image/*|}, etc. 
The MIME type can be obtained via HTTP HEAD requests; to generalize to other content than SDs, 
any other target MIME type set can be used. The full list of MIME types
we use 
is in~\cite{github_repository}.

\begin{figure}
	\centering
	\includegraphics[width=0.65\linewidth]{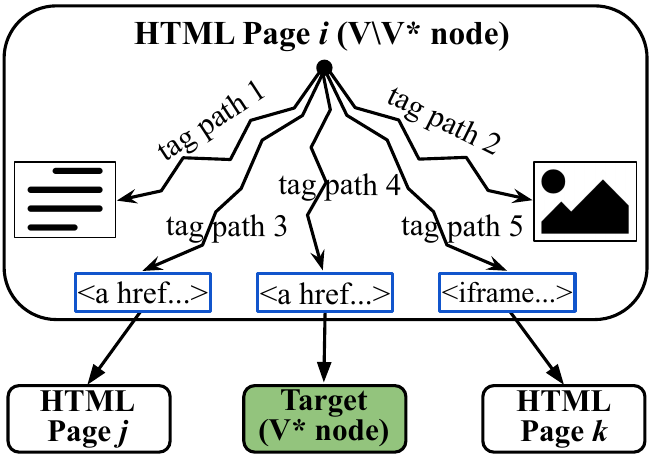}
	\caption{Tag paths in an HTML page}
	\label{fig:tag_paths_example}
\end{figure}

To each edge $e$, the edge labeling function $\lambda$ associates as label $\lambda(e)$ the \textbf{tag path}
derived from the HTML page's DOM structure~\cite{DOM}, the standard tree-based representation of an HTML page.
The tag path includes the \textbf{full path of HTML tags from the root 
to the hyperlink tag} (\texttt{a}, \texttt{area}, \texttt{iframe}, etc.), along with class and id attributes
describing additional structural and styling information. For example, a
label might be ``\texttt{html} \texttt{body} \texttt{div\#main}
\texttt{ul.datasets} \texttt{li} \texttt{a}'', where `\texttt{\#}' prefixes
the HTML ID, and `\texttt{.}' indicates a class. This labeling makes it very likely that \emph{links labeled with similar tag paths lead to similar content types}, even across different webpages of the same website. Figure~\ref{fig:tag_paths_example} zooms within an HTML page. It shows five tag paths (the tags are omitted for readability), leading to: a text paragraph, an image and 3 hyperlinks to other pages (1 target, and~2~HTML pages). 
$\lambda$ captures exactly the tag paths leading to each hyperlink.
We consider two cost functions $\omega$: \begin{inparaenum}[(i)] \item counting HTTP requests, $\omega(u)$$=$$1$ for all $u$$\in$$V$; \item measuring the exchanged data volume, especially data volume the crawler receives with $\omega(u)$ as the page size, which is, in theory, unbounded. \end{inparaenum}
We also account for the cost $c$ of \emph{determining if a vertex is
	in~$V^*$}, typically via an HTTP HEAD request. If $\omega$ counts
requests, then $c(u)$$=$$1$; if $\omega$ measures volume, $c(u)$ is based on
HEAD response size (much smaller than $\omega(u)$). Sec.~\ref{subsection:url_classifier} discusses a MIME type prediction method, resulting in a small, amortized $c$ cost, which we can view as constant.

\begin{toappendix}
	Here is the full list of the 38 MIME types used to identify targets in 
	our implementation:

	\bigskip

	{\ttfamily\begin{tabular}{l}
			application/csv                                                           \\
			application/json                                                          \\
			application/msword                                                        \\
			application/octet-stream                                                  \\
			application/pdf                                                           \\
			application/rdf+xml                                                       \\
			application/rss+xml                                                       \\
			application/vnd.ms-excel                                                  \\
			application/vnd.ms-excel.sheet.macroenabled.12                            \\
			application/vnd.oasis.opendocument.presentation                           \\
			application/vnd.oasis.opendocument.spreadsheet                            \\
			application/vnd.oasis.opendocument.text                                   \\
			application/vnd.openxmlformats-officedocument.presentationml.presentation \\
			application/vnd.openxmlformats-officedocument.spreadsheetml.sheet         \\
			application/vnd.openxmlformats-officedocument.wordprocessingml.document   \\
			application/vnd.openxmlformats-officedocument.wordprocessingml.template   \\
			application/vnd.rar                                                       \\
			application/x-7z-compressed                                               \\
			application/x-csv                                                         \\
			application/x-gtar                                                        \\
			application/x-gzip                                                        \\
			application/xml                                                           \\
			application/x-pdf                                                         \\
			application/x-rar-compressed                                              \\
			application/x-tar                                                         \\
			application/x-yaml                                                        \\
			application/x-zip-compressed                                              \\
			application/yaml                                                          \\
			application/zip                                                           \\
			application/zip-compressed                                                \\
			text/comma-separated-values                                               \\
			text/csv                                                                  \\
			text/json                                                                 \\
			text/plain                                                                \\
			text/x-comma-separated-values                                             \\
			text/x-csv                                                                \\
			text/x-yaml                                                               \\
			text/yaml                                                                 \\
		\end{tabular}}
	\clearpage
\end{toappendix}

\section{Crawling based on Reinforcement Learning}\label{section:solution}

Our goal in efficient crawling is to retrieve as many targets as possible at minimal cost, without prior knowledge of the website's size or structure. We hypothesize that \emph{an edge's label} $\lambda$ (its tag path in the page containing the hyperlink) \emph{can be used to learn which edges leads to target-rich pages}. Obtaining and leveraging such insignhts require both \emph{exploration} (to find promising paths) and \emph{exploitation} (to retrieve targets), which we achieve via \emph{reinforcement learning} (RL)~\cite{sutton2018reinforcement}. 
Below, we present the environment (states and actions) of our crawling task (Sec.~\ref{subsection:states_actions}), and the learning algorithm (or \emph{ agent}) that evolves in it (Sec.~\ref{subsection:group-links}).
We describe the URL classifier we use to compute the agent's rewards in 
Sec.~\ref{subsection:url_classifier}, and the overall crawling algorithm in Sec.~\ref{subsection:crawling_algorithm}.

\subsection{Environment: States and Actions}\label{subsection:states_actions}

In RL, an \emph{agent} evolves in an \emph{environment}. It starts from an initial \emph{state}, where it can choose among a set of \emph{actions}. Each \emph{action} leads to a new \emph{state}, where other actions can be chosen. Each choice leads to a \emph{reward}. The  goal of the agent is to learn a \emph{policy} (a mapping from states to actions), that maximizes the total reward. This model is known as a \emph{Markov Decision Process} (MDP) \cite{howard1960dynamic}.
For our crawler, one might consider using ``the currently crawled webpage'' as \emph{state}. However, this is not suitable: the current webpage does not reflect previous choices, as it may be reachable via multiple paths. Also, an efficient crawler should not visit any webpage twice, whereas learning supposes multiple visits to a state to refine and then leverage information learned in that state.

To keep the model simple and lightweight (see Sec.~\ref{section:related_work} for discussion), we use a \textbf{single-state model}, where the agent is perpetually in exactly only one state. What matters then is only the set of \emph{actions} that the agent can follow at each crawl step. Intuitively, an action is to navigate along a link. We delve into action details below.

\subsection{Grouping Links into Actions}\label{subsection:group-links}

Following our hypothesis that \emph{links with similar labels (tag paths) have similar values}, i.e., likelihood of leading to targets, we model an action as \emph{a group
	of similar links}, with the semantics that taking the action amounts to
      crawling along one of these links.

Each link is labeled with its tag path, so we need a measure of similarity between tag paths.
To that effect, we represent each tag path by a \emph{numerical bag-of-words (BoW) vector~$p$}, over the $n$-grams vocabulary built from all tag paths encountered so far.
Note that $n$-grams preserve the order of HTML node names appearing in tag paths; in our context, this order is significant (as Sec.~\ref{sec:results} confirms).
Since the $n$-grams vocabulary is \emph{dynamically} built during the crawl, BoW vectors for tag paths encountered at different times have different lengths. To still compute distances between them, we first \emph{project each tag path into a fixed-dimensional vector} of size $D = 2^m$ for a chosen $m > 0$. We do this as follows:

\begin{asparaenum}
	\item We fix a \emph{hash function} $h: x\mapsto\left\lfloor\frac{(\Pi \times x) \mod 2^w}{2^{w-m}}\right\rfloor$ which maps
	any integer $x$ to a number between $0$ and $D-1$. Its parameters $\Pi$ (a large prime number) and $w$ are fixed, and chosen such that $w>m$. 
	\item  Calling~$h$ on each position between $0$ and $d-1$, where $d$ is
	the length of the BoW vector~$p$, maps it to a new position between $0$
  and $D-1$. This enables us to transform $p$ into a $D$-dimensional
  $p_D$ vector, where for every $0\leq i<d$, $p_D\left[h(i)\right]=p\left[i\right]$.
	\item Potential collisions of $h$ (i.e., $i_1\neq i_2$ with
	$h\left(i_1\right)=h\left(i_2\right)$) are resolved by setting $p_D\left[h(i_1)=h(i_2)\right]$
	to the \emph{mean} of all elements of $p$, at positions which collide with
	$i_1,i_2$. 
	Positions $0\le j<D$ not hit by any $i$ ($h(i)\ne j$ for all $0\le i<d$) are set to $p_D[j]=0$. This happens, e.g., at the beginning of the crawl, when the set of HTML element names seen so far is small (thus $d$ is small).
\end{asparaenum}

\begin{figure}[t!]
	\includegraphics[width=1\linewidth]{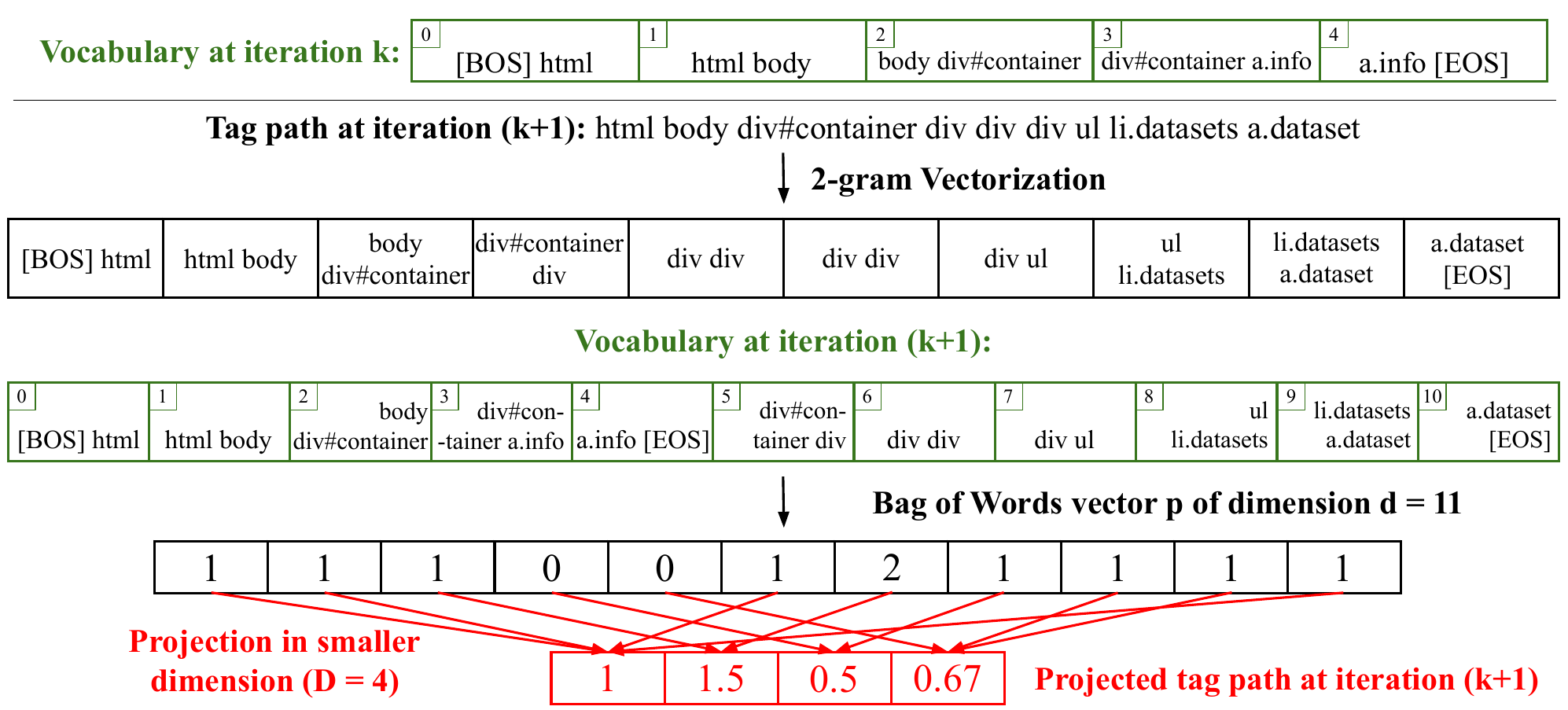}
	\caption{Mapping of a tag path into a fixed-size vector.\label{fig:sample-link-vectors}}
\end{figure}

Figure~\ref{fig:sample-link-vectors} illustrates tag path
projection for 
$D=2^m=4$, $w=11$ and $\Pi=766\,245\,317$. At iteration~$k$, the 2-gram vocabulary contains $d_k=5$ elements
(including special tokens \texttt{BOS} and \texttt{EOS} denoting
beginning and end of stream). At iteration~$k+1$, a new tag path is
vectorized into a 2-gram vector of dimension~$10$, and the
vocabulary is updated accordingly, totaling 
$d_{k+1}=11$ elements. 
From it, we compute the
$d_{k+1}$-dimensional BoW vector $p_{k+1}$.
The hash function~$h$ maps
each input dimension $i \in
\left\{0,\dots,d_{k+1}{-}1\right\}$ to
$\left\{0,\dots,D{-}1\right\}$, e.g.,
$h(2)=\left\lfloor \frac{(766\,245\,317 \times 2) \mod 2048}{512}
\right\rfloor=1$. 
Collisions can occur, e.g.,
$h(4)$$=$$h(8)$$=$$h(9)$$=$$3$, so ${p_D}_{k+1}[3]$$=$
$\frac{p_{k+1}[4] + p_{k+1}[8] + p_{k+1}[9]}{3}$$\approx$$0.67$. This yields a 4-dimensional ${p_D}_{k+1}$.

Our next task is to find, for each projected tag path $p_D$, the nearest action $a_c$  among all known actions~$A$.
Conceptually, an action is an evolving cluster of similar
tag paths with a centroid: the mean of all projected tag paths.
For efficiency, we only store the centroid, which updates as the action's set of tag paths evolves.

Algorithm \ref{alg:url_mapping} computes the nearest action of each $p_D$ if it is close enough; otherwise, it creates a new action.
Action centroids are stored in a \emph{Hierarchical Navigable Small Worlds} (HNSW)~\cite{malkov2018efficient} index~$\mathcal{I}$, chosen for its highly efficient updates of centroids as new tag paths join. While maintaining and querying this index might be costly in CPU time, it is negligible compared to  crawl time (waiting between two requests, or webpages download). The index also estimates the \emph{cosine similarity}
between $p_D$ and its closest centroid: if above a
threshold $\theta$, the tag path is added to the action; 
otherwise, a new action composed of only $p_D$ is created. As our experiments show (Sec.~\ref{sec:results}), the choice of $\theta$ significantly impacts  the agent performance. In extreme cases, $\theta=0$ groups all tag paths into a single action, preventing learning as the agent selects paths randomly, whereas $\theta=1$ creates a separate action for each path, so the agent only explores and never exploits. A useful threshold balances enough actions to separate interesting from uninteresting paths while still enabling learning and exploitation.

\begin{algorithm}[t!]
	\caption{Finding the action for a hyperlink $l$, $\lambda(l)$$=$$p$}\label{alg:url_mapping}
	\KwInput{Action set $A$, projected tag path~$p_D$}
	\KwOutput{Action $a$}

	$a_c\gets$ Approx.\ nearest neighbor (from index $\mathcal{I}$) of $p_D$\;
	$s_c \gets$ Cosine similarity between $a_c$ and $p_D$\;
	\eIf{$s_c \geq \theta$}{
		Update centroid of $a_c$ in $\mathcal{I}$ with respect to $p_D$\;
	}{
		Add a new entry $a_{\textrm{new}}$ to $\mathcal{I}$, with value $p_D$\;
		Add $a_{\textrm{new}}$ to $A$\;
		$a_c \gets a_{\textrm{new}}$\;
	}
	\Return $a_c$
\end{algorithm}

The family of reinforcement models for single-state environments is called \emph{Multi-Armed Bandits} (MABs)~\cite{sutton2018reinforcement}. The standard, deterministic, MAB algorithm optimizing the exploration--\\exploitation trade-off is \emph{Upper-Confidence Bound} (UCB)~\cite{auer2002finite}, which implements \emph{optimism under the face of uncertainty}: 
at each step, a score is computed for each action, and the highest-scoring action is selected.

The MAB score of an action $a$ at time $t+1$, denoted $ s_{\textrm{MAB}}^{t+1}(a)$, combines an exploitation term (the mean of rewards obtained so far for $a$), and an exploration term, reflecting how often $a$ has been selected relative to the total crawl history:
\(
	s_{\textrm{MAB}}^{t+1}(a) = R_a^t + \alpha \sqrt{\frac{\log(t)}{N_t(a)
  + \epsilon}}
\), 
where $R_a^t$ is the mean reward for action $a$ up to time $t$, $\alpha$~weights exploration vs. exploitation, $N_t(a)$~counts how many times $a$ has been selected, and $\epsilon > 0$ prevents division by zero when $N_t(a)=0$.

Since each URL is visited only once, an action may become empty after all its URLs are visited. The  \emph{Awake Upper-Estimated Reward} (AUER)~\cite{kleinberg2010regret} adapts UCB to cope with actions becoming unavailable,or \emph{sleeping}. Following the AUER model, our score becomes:
\(
	s_{\textrm{SB}}^{t+1}(a) = \mathds{1}_a(t) \times \left(R_a^t + \alpha \sqrt{\frac{\log(t)}{N_t(a) + \epsilon}}\right)
\)
where $\mathds{1}_a(t)=1$ if action $a$ has remaining unvisited links at time $t$, and $0$ otherwise. After selecting $a$, our crawler \emph{randomly chooses an unvisited link $l \in a$ to traverse next with equal probability}.
How should we reward the crawling along $l\in a$? As our goal is to find \emph{new targets}, the reward should reflect the number of \emph{not-yet-discovered links to targets}, contained in the HTML page $h$ reached via $l$: this ``novelty'' condition encourages the agent to focus on unseen target.
For instance, if $h$ has 12 links, 5 leading to targets, and 2 already retrieved, the reward is $3$. This simple choice performs well, as shown in Sec.~\ref{sec:results}.

Yet, a challenge remains: we must compute the reward of all links in $h$ before following any, since each traversal incurs a crawling cost. We therefore need a fast \emph{estimation}, applicable to \emph{any} link, even if some are never followed. For this, we introduce an \emph{online URL classifier} that estimates rewards by inspecting links in the newly acquired page $h$. We detail it in Sec.~\ref{subsection:url_classifier}.

Last, we discuss the parameter $\alpha$ in our learning agent.
UCB and AUER are optimal for 
$\alpha=2\sqrt{2}$~\cite{auer2002finite,kleinberg2010regret}, but only under standard conditions where the rewards are normally
distributed in $[0, 1]$. In our case, rewards (counts of new targets)
 are unbounded and typically do not follow a normal
distribution (see Sec.~\ref{sec:exp-sb-algo}), which is desirable: otherwise, it would mean that most HTML pages contain links to targets. If this held, we would have to visit the entire website, or at least a vast share, to retrieve most targets.
Setting up $\alpha$ to guarantee optimality in our case would require knowing the reward distribution, which is not possible.
We therefore keep $\alpha=2\sqrt{2}$, which gives good results in practice,  across very varied reward distributions (as shown in Sec.~\ref{subsection:dataset_characteristics}, in particular in Figure~\ref{fig:10_sites_plots}).

\subsection{Estimating Rewards with URL Classifier}\label{subsection:url_classifier}

Determining the reward of an action taken by our agent (crawl a link) requires assessing a page's MIME type from its URL.
Doing HTTP HEAD requests is costly and requires spacing for crawling ethics, while using the URL extension (e.g., \texttt{.html}, \texttt{.csv}) fails for extensionless links such as \url{https://www.justice.gouv.fr/en/node/9961} or others within ILO\footnote{E.g., \url{https://www.ilo.org/publications/elevating-potential-rural-youth-paths-decent-jobs-and-sustainable-futures-1}.}.
To be efficient and generic, we devise an
online URL classifier\footnote{Observe that here we classify
\emph{URLs}, whereas in the previous section we clustered \emph{edge labels}, which, crucially to our method, are \emph{tag paths} within HTML pages (Fig.~\ref{fig:tag_paths_example}).} to estimate rewards, as follows.

For our purposes, any URL belongs to one of three classes:  ``HTML'',
``Target'' or ``Neither''. HTML pages are added to the crawl frontier, while a target contributes to the crawler's reward. The ``Neither'' class includes non-HTML pages, URLs with non-target MIME types, or URLs lacking a MIME type from the server. In our experience, over 92\% of ``Neither'' cases correspond to HTTP \textbf{4xx} or \textbf{5xx} codes, designating errors on client or server.

We initially set our classifier to assign one of these three classes to each link found in a newly crawled page $h$.
However, it proved unable to distinguish URLs leading to 4xx or 5xx errors  from those that resolve fine, Intuitively, as these are often very similar to accessible ``HTML'' or ``Target'' URLs. However, \emph{different classification errors have different impacts on crawling}:
\begin{inparaenum}
	\item Misclassifying a ``Neither'' URL as ``HTML'' or ``Target'' only incurs the moderate cost of following it (later, respectively, immediately), as it will likely throw an error. To reduce such visits, we apply filters based on user-defined MIME types and URL extensions (see Sec.~\ref{subsection:crawling_algorithm}).
	\item Misclassifying ``HTML'' or ``Target'' as ``Neither''  \emph{completely withdraws the page from the crawl}: if it is a target, we miss it; if it is HTML, we miss all URLs that might be discovered by following it. This can make the crawler miss huge parts of the website.
\end{inparaenum}
To avoid the second kind of errors, our classifier is trained on just two classes (``HTML'' and ``Target''), despite knowing some URLs are neither.

Algorithm~\ref{alg:online_url_classifier} details the training and usage of our classifier $\mathcal{C}$. It is a logistic regression model~\cite{james2013introduction}, iteratively trained through epochs of \emph{Stochastic Gradient Descent} (SGD)~\cite{bottou2010large}, with batch size~$b$. It takes as input a vector of character-wise 2-grams of the URL. For instance, the URL \url{https://www.A.com/data/file.csv} is transformed into a list
$\left[\textrm{ht}, \textrm{tt}, \textrm{tp}, \dots, \textrm{.c}, \textrm{cs}, \textrm{sv}\right]$. We associate to each pair of usual ASCII characters\footnote{ASCII digits, upper and lower case letters and main special characters.} an ID, and encode each URL as a BoW over this vocabulary. The choice of model and features is discussed in Sec.~\ref{subsubsection:metaparameters_result}.

\begin{algorithm} [t!]
	\caption{Online URL classifier procedure}\label{alg:online_url_classifier}
	\KwInput{$U$, a URL}
	\KwOutput{$l_U\in \{$``HTML'', ``Target''$\}$}

	\uIf{$\left\lvert X \right\rvert \geq b$}{
		$X_{\textrm{mat}} \gets$ $2$-gram bag-of-words of each URL in $X$\;
		$\mathcal{C}$ is incrementally trained on batch $\left(X_{\textrm{mat}}, y\right)$\;
		$\left(X, y\right) \gets \left(\emptyset, \emptyset\right)$\;
		$\mathrm{initial\_training\_phase} \gets$ False\;
	}
	\eIf{$\mathrm{initial\_training\_phase}$}{
		$l_U \gets$ MIME type class from HTTP HEAD on $U$\;
		Add $\left(U, l_U\right)$ to $\left(X, y\right)$\;
		\Return $l_U$
	}{
		$U_{\textrm{vec}} \gets$ $2$-gram bag-of-words vector of $U$\;
		\Return $\mathcal{C}\left(U_{\textrm{vec}}\right)$
	}
\end{algorithm}

As Algorithm~\ref{alg:online_url_classifier} shows, in the first training epoch, we label $b$ URLs via HTTP HEAD requests, with $X$ the URL set and $y$ their labels. Based on this, we compute rewards and assign URLs to the frontier~$\mathcal{F}$ or targets~$V^*$.
After this epoch, we set $\textrm{initial\_training\_phase}$ to false, and use the classifier to infer URL classes without additional HTTP HEAD requests.
The classifier is further improved through \emph{online training}: during execution, each HTTP GET issued by the crawler contributes an annotated (URL, class) pair, added to $X$ and $y$ for incremental training when the batch size $b$ is reached. In this way, after the first epoch, we get labels at no extra cost.

Since we want to both \emph{use} the classifier as soon as possible and \emph{train} it often to improve its accuracy, we must set $b$ relatively small.
This online method adapts to potential changes in the form of the URLs, e.g.,  when the crawl discovers new parts of the website where URLs are formatted differently. An off-line method would not be able to adapt, even if trained on many examples initially.

\subsection{Crawling Algorithm}\label{subsection:crawling_algorithm}

We now present the overall crawling procedure (Algorithm~\ref{alg:crawling_algorithm}). Initially, the tree~$T$ of crawled pages, the set of
actions~$A$, and the frontier $\mathcal{F}$ are empty. The budget~$\beta$
spent by the crawler and the crawling step $t$ are set to~$0$. The crawler starts by visiting the original link~$r$, continuing until all links are visited (the website is completely
crawled) or the maximum budget $B$ is reached. At each step, if $A$ is non-empty,
the learning agent chooses an action $a_c$ following the Sleeping-Bandit (SB)
procedure (Sec.~\ref{subsection:group-links}). and chooses a link from $\mathcal{F}$ associated with $a_c$ uniformly at random. Then, it crawls the page behind this link
with Algorithm~\ref{alg:crawl_next_page}, and if its MIME type matches a predefined blocklist during download, its retrieval is
immediately interrupted; this is typically used to avoid 
multimedia content.

\begin{algorithm}[t!]
	\caption{Efficient target retrieval}\label{alg:crawling_algorithm}
	\KwInput{$r$ the initial page to crawl}
	$A, \mathcal{F}, T \gets \emptyset$; $\beta, t \gets 0$; $u \gets r$; Add $r$ to $\mathcal{F}$\;
	\While{$\left\lvert\mathcal{F}\right\rvert > 0$ and $\beta \leq B$}{
		\eIf{$\left\lvert A\right\rvert > 0$}{
			$a_{\textrm{c}} \gets \arg\max_{a \in A} \mathds{1}_a(t)
				\left(R_{\textrm{mean}}(a) + \alpha \sqrt{\frac{\log(t)}{N_t(a) +
						\epsilon}}\right)$\;
			$u \gets$ Select a link from $\mathcal{F}$ mapped with action
			$a_{\textrm{c}}$ uniformly at random, and $N_t(a_c)\gets N_t(a_c)+1$\;
		}{
			$u \gets$ Select a link from $\mathcal{F}$ uniformly at random\;
		}
		crawl\_next\_page$\left(u\right)$ (Algorithm
		\ref{alg:crawl_next_page})\;
	}
\end{algorithm}

\begin{algorithm}[t!]
	\caption{Crawl next page}\label{alg:crawl_next_page}
	\KwInput{$u$ the URL to be crawled}
	Add URL $u$ to $T$, and $t \gets t+1$\;
	$\mathrm{http\_response}, \mathrm{mime\_type}, \mathrm{body} \gets$ Result of the request on URL $u$ with HTTP GET, and update $\beta$ accordingly\;
	\textbf{if} request interrupted (banned MIME type) \textbf{then return}\;
	\textbf{if} $\mathrm{http\_response}$  \emph{is 4xx or 5xx (error)} \textbf{then} \textbf{return}\;
	\ElseIf{$\mathrm{http\_response}$ is 2xx (success)}{
		\If{$\mathrm{mime\_type} \ne \emptyset$}{
			\If{``$\mathrm{HTML}$'' $\subset \mathrm{mime\_type}$}{
				Add $(u$, ``HTML''$)$ to $(X, y)$\;
				$U_{\mathrm{new}}$ $\gets$ All hyperlinks in $\mathrm{body}$
				pointing to the same website as $r$\;
			}
			\ElseIf{$\mathrm{mime\_type} \in L$}{
				Add $(u$, ``Target''$)$ to $(X, y)$\;
				Add target (the content of $\mathrm{body}$) to $V^*$\;
			}
			\Return\;
		}
	}
	\ElseIf{$\mathrm{http\_response}$ is 3xx (redirection)} {
		$u \gets$ Location from HTTP GET result on $u$\;
		\textbf{if} $u \notin T \cup \mathcal{F}$ \textbf{then} crawl\_next\_page$(u)$ and \textbf{return}\;
	}
	reward $\gets 0$\;
	\For{$u_{\mathrm{new}} \in U_{\mathrm{new}}$ \textnormal{such that} $u_{\mathrm{new}} \notin T \cup \mathcal{F}$}{
		\textbf{if} has\_blocklisted\_extension$\left(l_{u_{\mathrm{new}}}\right)$ \textbf{then continue}\;
		$l_{u_{\mathrm{new}}} \gets$ class\_of\_url$\left(u_{\mathrm{new}}\right)$ (Algorithm~\ref{alg:online_url_classifier})\;
		Update $\beta$ if HTTP HEAD request was made\;
		\If{$l_{u_{\mathrm{new}}} =$ ``HTML''}{
			$a_c \gets $ map\_link\_to\_action$\left(A, u_{\mathrm{new}}\right)$ (Algorithm~\ref{alg:url_mapping})\;
			Add $u_{\mathrm{new}}$ to $\mathcal{F}$\;
		}\ElseIf{$l_{u_{\mathrm{new}}} =$ ``Target''}{
			crawl\_next\_page$\left(u_{\mathrm{new}}\right)$, and reward $\gets \mathrm{reward} + 1$\;
		}
	}
	\textbf{if} $u \ne r$ \textbf{then} $R_{\mathrm{mean}}\left(a_c\right) \gets R_{\mathrm{mean}}\left(a_c\right) + \frac{\mathrm{reward} - R_{\mathrm{mean}}\left(a_c\right)}{N_t\left(a_c\right)}$\;
\end{algorithm}

Algorithm~\ref{alg:crawl_next_page} starts by retrieving 
the HTTP GET response: status, MIME type from the header, and body (content) of the webpage. The request cost is added to the budget~$\beta$. The 
status may fall into one of three categories:
\begin{inparaenum}
	\item Errors (on the client or server side) marked by \textbf{4xx} or \textbf{5xx} statuses, do not yield new links or targets.
	\item Redirections, indicated by \textbf{3xx} statuses, include an extra ``Location'' header with the redirection URL: if uncrawled, the crawler follows and processes it.
	\item A \textbf{2xx} status means the server successfully responds with either an HTML page, a target (whose MIME type is in the list $L$ of Sec.~\ref{section:graph_to_website}), or neither. In the first two cases, $X$ and $y$ are updated. For HTML pages, new hyperlinks $U_{\textrm{new}}$ are extracted from the page, keeping only in-website links.
\end{inparaenum}
Each new link not in $\mathcal{F}$ or $T$ is classified using Algorithm~\ref{alg:online_url_classifier}, except when its extension matches a predefined blocklist established before the crawl.

If it points to an HTML page, it is mapped to an action using Algorithm~\ref{alg:url_mapping}. The set $A$ is updated by either \emph{changing an action's centroid} or \emph{adding a new action}, and the chosen action $a_c$ is mapped to the link, which is then added to frontier~$\mathcal{F}$. If the link leads to a target, we immediately retrieve it by recursively calling Algorithm~\ref{alg:crawl_next_page}, and the reward is updated. Finally, if the page crawled by the current call of Algorithm~\ref{alg:crawl_next_page} is not the starting page, the mean reward for $a_c$ is updated to reflect the last computed reward.

\section{Experimental Results}\label{sec:results}

We now present our experimental results: websites used (Sec.~\ref{subsection:dataset_characteristics}), poor search engine performance for this task (Sec.~\ref{subsubsection:search_engines}), baselines (Sec.~\ref{subsection:baselines}), crawling setup (Sec.~\ref{sec:exp-crawl-modalities}),
comparison of our crawler to baselines (Sec.~\ref{subsubsection:main_result}), and impact of hyper-parameters (Sec.~\ref{subsubsection:metaparameters_result}). Sec.~\ref{sec:exp-sb-algo} examines our Sleeping Bandit (SB) algorithm's effectiveness in grouping links into coherent and reward-based groups, and Sec.~\ref{sec:early_stopping} presents an early stopping mechanism.

\subsection{Websites}\label{subsection:dataset_characteristics}

\begin{table*}
	\caption{Main characteristics of websites (see detailed crawling methodology in Sec.~\ref{sec:exp-crawl-modalities})}

  \begin{tabular}{cl@{}ccc@{\hspace{.75em}}c@{\hspace{.75em}}ccccc}
		\toprule
		          & \bfseries Starting URL                                & \bfseries Mlg.       & \bfseries F. C.  & \bfseries \#Available (k)                             &
    \bfseries \#Target (k) & \bfseries HTML to T. (\%) & \bfseries
    Target Size (MB) & \bfseries Target Depth                                                            \\
		\midrule
		\emph{ab} & \tiny\url{https://www.abs.gov.au/}                    & \xmark               & \xmark                        & 952.26                       & 263.26               & \08.86 & \04.50 ($\pm$\056.04)  & \0\08.94 ($\pm$\02.56) \\
		\emph{as} & \tiny\url{https://www.assemblee-nationale.fr/}        & \xmark               & \xmark                        & 949.42                       & 155.94               & \04.34 & \00.54 ($\pm$\0\06.38) & \0\05.84 ($\pm$\01.07) \\
		\emph{be} & \tiny\url{https://www.bea.gov/}                       & \xmark               & \cmark                        & \031.23                      & \015.84              & 32.19  & \02.03 ($\pm$\0\06.99) & \0\05.73 ($\pm$\03.21) \\
		\emph{ce} & \tiny\url{https://www.census.gov/}                    & \xmark               & \xmark                        & 988.37                       & 257.68               & \03.47 & \01.51 ($\pm$\015.77)  & \0\04.23 ($\pm$\00.48) \\
		\emph{cl} & \tiny\url{https://www.collectivites-locales.gouv.fr/} & \xmark               & \cmark                        & \0\05.54                     & \0\03.70             & \05.40 & \01.15 ($\pm$\0\04.91) & \0\02.80 ($\pm$\00.82) \\
		\emph{cn} & \tiny\url{https://www.cnis.fr/}                       & \xmark               & \cmark                        & \012.80                      & \0\07.49             & 13.87  & \00.43 ($\pm$\0\01.74) & \0\04.26 ($\pm$\01.59) \\
		\emph{ed} & \tiny\url{https://www.education.gouv.fr/}             & \xmark               & \cmark                        & 102.71                       & \010.47              & \03.95 & \01.00 ($\pm$\0\03.07) & \011.89 ($\pm$13.22)   \\
		\emph{il} & \tiny\url{https://www.ilo.org/}                       & \cmark               & \xmark                        & 990.71                       & \081.01              & \02.53 & 13.40 ($\pm$110.01)    & \0\04.26 ($\pm$\01.28) \\
		\emph{in} & \tiny\url{https://www.interieur.gouv.fr/}             & \xmark               & \cmark                        & 922.46                       & \022.98              & \01.54 & \01.12 ($\pm$\0\03.06) & \066.94 ($\pm$39.43)   \\
		\emph{is} & \tiny\url{https://www.insee.fr/}                      & \cmark               & \cmark                        & 285.55                       & 168.88               & 41.34  & \03.13 ($\pm$\021.43)  & \0\05.20 ($\pm$\01.81) \\
		\emph{jp} & \tiny\url{https://www.soumu.go.jp/}                   & \cmark               & \xmark                        & 993.87                       & 328.83               & \06.30 & \00.80 ($\pm$\0\04.49) & \0\05.18 ($\pm$\01.29) \\
		\emph{ju} & \tiny\url{https://www.justice.gouv.fr/}               & \xmark               & \cmark                        & \056.61                      & \014.85              & \04.85 & \00.48 ($\pm$\0\01.34) & \086.91 ($\pm$86.30)   \\
		\emph{nc} & \tiny\url{https://nces.ed.gov/}                       & \xmark               & \cmark                        & 309.97                       & \084.94              & 18.87  & \01.10 ($\pm$\011.56)  & \0\03.63 ($\pm$\01.66) \\
		\emph{oe} & \tiny\url{https://www.oecd.org/}                      & \cmark               & \cmark                        & 222.58                       & \045.04              & 15.61  & \02.31 ($\pm$\023.37)  & \0\06.28 ($\pm$\05.65) \\
		\emph{ok} & \tiny\url{https://okfn.org/}                          & \cmark               & \cmark                        & 423.12                       & \012.95              & \00.74 & \00.04 ($\pm$\0\00.24) & \0\02.64 ($\pm$\02.89) \\
		\emph{qa} & \tiny\url{https://www.psa.gov.qa/}                    & \cmark               & \cmark                        & \0\04.36                     & \0\02.45             & \04.15 & \02.97 ($\pm$\019.28)  & \0\03.03 ($\pm$\00.61) \\
		\emph{wh} & \tiny\url{https://www.who.int/}                       & \cmark               & \xmark                        & 351.86                       & \055.59              & 14.19  & \01.26 ($\pm$\011.14)  & \0\04.43 ($\pm$\00.62) \\
		\emph{wo} & \tiny\url{https://www.worldbank.org/}                 & \cmark               & \xmark                        & 223.67                       & \023.10              & \02.38 & \02.80 ($\pm$\027.16)  & \0\04.52 ($\pm$\00.69) \\
		\bottomrule
	\end{tabular}
	\label{tab:websites_characteristics}
\end{table*}

\def\websites{{as},{cl},{cn},{ed},{il},{in},{jp},{ju},{qa}}

Table~\ref{tab:websites_characteristics} lists the 18 websites used in our experiments.
As our primary application is a collaboration with French journalists~\cite{DBLP:conf/cikm/BalalauEGMMDGKP22} from \href{https://www.radiofrance.fr/}{RadioFrance},
six websites are French, public, and governmental: the \href{https://www.assemblee-nationale.fr/}{French National Assembly} (\emph{as}), \href{https://www.collectivites-locales.gouv.fr/}{French Local Communities} (\emph{cl}), \href{https://www.cnis.fr/}{French Council for Statistical Information} (\emph{cn}), and the ministries of \href{https://www.interieur.gouv.fr/}{Interior} (\emph{in}), \href{https://www.education.gouv.fr/}{Education} (\emph{ed}), and \href{https://www.justice.gouv.fr/}{Justice} (\emph{ju}). Each contains material on government branches, including SDs (our crawler's targets).
We also include eight multilingual websites: the UN's \href{https://www.ilo.org/}{International Labor Organization} (\emph{il}), \href{https://www.insee.fr/}{French Official Statistical Institute} (\emph{is}), \href{https://www.soumu.go.jp/}{Japan's Ministry of Interior} (\emph{jp}), \href{https://www.oecd.org/}{OECD} (\emph{oe}), \href{https://okfn.org/}{Open Knowledge Foundation} (\emph{ok}), \href{https://www.psa.gov.qa/}{Qatar's Official Statistical Service} (\emph{qa}), the UN \href{https://www.who.int/}{World Health Organization} (\emph{wh}), and the \href{https://www.worldbank.org/}{World Bank} (\emph{wo}).
Finally, we add national websites in English: the \href{https://www.abs.gov.au/}{Australian Bureau of Statistics} (\emph{ab}), \href{https://nces.ed.gov/}{US National Center for Education Statistics} (\emph{nc}), the \href{https://www.census.gov/}{US Census} (\emph{ce}), and the \href{https://www.bea.gov/}{US Bureau of Economic Analysis} (\emph{be}). Websites  in at least two languages have a \cmark{} in the ``\textbf{Mlg.}'' column.
The websites range from a few thousand to over 1M pages, excluding those resulting in 4xx or 5xx HTTP errors (see ``\textbf{\#Available (k)}'' column). A \xmark{} in the ``\textbf{F. C.}'' (``Fully Crawled'') column indicates incomplete crawls (Sec.~\ref{sec:exp-crawl-modalities}).

The following metrics assess the \emph{difficulty} and \emph{interest} of a focused crawl. For partially crawled websites, metrics are computed on the BFS-visited subset (Sec.~\ref{sec:exp-crawl-modalities}). ``\textbf{\#Target~(k)}'' is the total number of identified targets, and the ratio $\frac{\textrm{\#Target (k)}}{\textrm{\#Available (k)}}$ measures \emph{target density}; extremes are 66.78\% (\emph{cl}) and 2.49\% (\emph{in}). ``\textbf{HTML to T. (\%)}'' is the percentage of HTML pages linking to at least one target, reflecting the \emph{density of target-pointing pages} (e.g., \emph{cl} has the highest density, but all targets are concentrated in 5.40\% of HTML pages). ``\textbf{Target Size (MB)}'' gives the mean and standard deviation (STD) of target file sizes, and ``\textbf{Target Depth}'' the mean and STD of depths (shortest link navigation from the root). \emph{Mean depths vary greatly, from 3 to nearly 90}: highest values arise on portals requiring page-to-page navigation, e.g. \emph{ju}.
Table~\ref{tab:websites_characteristics} shows our websites are \emph{extremely diverse}: small or huge, varying in depth, target number and  locations, and over 20 languages. This diversity demands \emph{adaptable crawling strategies} to efficiently retrieve targets. A manual analysis of typical target locations on \emph{ju}, \emph{il}, \emph{wh}, and \emph{nc} is provided in the extended version in~\cite{github_repository}.

\begin{toappendix}

\subsection{Websites}

To further emphasize the diversity of crawling efforts, we manually analyzed the types of pages where targets are typically found. In \emph{ju}, targets appear under various resource lists and data portals disseminated across the website. Some are huge (e.g., \url{https://www.justice.gouv.fr/documentation/bulletin-officiel}), while others are smaller (e.g., \url{https://www.justice.gouv.fr/documentation/ressources/conventions-judiciaires-dinteret-public}). Some require multi-step navigation to reach deeper pages that contain one or a few target links. In \emph{il}, most targets are found in small, dispersed data catalogs: some gathered in ``ILOSTAT'' (\url{https://ilostat.ilo.org/}), others accessible only through further exploration. These targets are generally downloadable directly from the catalogs. In \emph{wh}, the majority of targets require multi-step navigation through a hierarchy of subjects and sub-subjects, with only a subset leading to targets: efficient crawling thus requires quickly identifying interesting parts. In \emph{nc}, all previously mentioned typologies are present: data portals (e.g., \url{https://nces.ed.gov/national-center-education-statistics-nces/products}), exhaustive lists (e.g., \url{https://nces.ed.gov/surveys/spp/results.asp}), and deep pages requiring multi-step navigation.
\end{toappendix}

\subsection{Search Engines and Dataset Search}
\label{subsubsection:search_engines}

We attempted to retrieve SDs using search engines (SEs), in particular by testing three popular Google services:
\href{https://www.google.com/}{classical search} (GS), \href{https://datasetsearch.research.google.com/}{Google Datasets Search} (GDS), and the \href{https://www.google.com/publicdata/directory}{Google Public Data Explorer} (GPDE) providing data from international
organizations.

GS allows filtering by website (\textsf{\small site:}) and file type (\textsf{\small filetype:}), via its Web interface and \href{https://programmablesearchengine.google.com/}{API}. However, SEs, including Google, do not fully index websites and GS caps results at 1k, often truncating them. 
For instance, querying \emph{ju} for PDFs yields only 302 results, although more than 9k exist; similarly, only 240 ODS files are returned (out of 910), and on \emph{in} 38 XLS (out of 1\,546). Even worse: TSV is not recognized at all despite 11\,097 files on \emph{ju}. On \emph{il}, GS lists 641 results instead of over 49k, and no ZIP archives instead of at least 2\,239.
GDS trims results even further, e.g., 109 tabular files on
\emph{ju} (vs. 1\,188); 93 datasets on \emph{il} (vs.\ $\geq$170k found by our crawler); and 312 on \emph{ce} (vs.\ 800k), etc. 
GPDE is built for human readers, indexing from a closed, fixed set of providers, such as \href{https://ec.europa.eu/eurostat/}{Eurostat}, \emph{wo}, \href{https://www.statcan.gc.ca/}{Statistics Canada}, and \emph{oe}, but excludes key sites like UN’s \emph{wh}, \emph{il}, and US agencies (\emph{be}, \emph{nc}, \emph{ce}).	
GPDE only allows manual exploration, plotting datasets guided by users' filters but offering neither downloads nor links to originals: e.g., it displays 150 OECD statistics across 37 countries and 100 years, but only references a 272-page PDF (\href{https://www.oecd-ilibrary.org/}{OECD iLibrary}).

Overall, \textbf{SEs lack transparency and control over the retrieved SDs}, a critical limitation for our task. In contrast, our crawler retrieves all data files within a specified budget, with explainable decisions and explicit choice of target MIME types.
\subsection{Baselines}
\label{subsection:baselines}

We compare our best-performing \textsf{\small SB-CLASSIFIER} with an adapted version in which the URL classifier is replaced by an (unrealistic) perfect oracle~(\textsf{\small SB-ORACLE}), as well as with several baseline methods. First, we consider four simple crawlers.
\begin{inparaenum}[(i)]
\item The \textsf{\small RANDOM} crawler selects the next hyperlink to visit
	uniformly at random from the crawl frontier.
	\item The
    \emph{Breadth-First Search} (\textsf{\small BFS}) exhaustive crawler maintains the
	frontier as a \emph{First-In-First-Out queue} data structure. It crawls
	all pages reachable by a path of length $\ell$ before any page reachable
	only by longer paths ($\ell' > \ell$).
	\item \emph{Depth-First
    Search} (\textsf{\small{DFS}}) maintains the frontier in a \emph{Last-In-First-Out stack}.
	It is rarely used for exhaustive crawling since it may fall
        into
	\href{https://en.wikipedia.org/wiki/Spider_trap}{robot traps}.
	\item The unrealistic \emph{Omniscient Crawler} (\textsf{\small OMNISCIENT}) knows all target URLs ($V^*$) from the start and crawls them sequentially. Since an optimal crawler is intractable (Prop.~\ref{prop:np-complete}), this omniscient crawler provides an (unreachable) upper bound on crawler efficiency.
\end{inparaenum}

The \emph{Offline-Trained, Tag Path-Based Crawler} (\textsf{\small TP-OFF})~\cite{faheem2015adaptive} is a closely related focused crawler originally designed to retrieve \emph{diverse} textual content. We apply it \emph{technically unchanged}, except for redefining its reward to target retrieval.
As described in~\cite{faheem2015adaptive}, it first crawls 3k pages with BFS and groups the tag paths leading to followed links as in Sec.~\ref{subsection:states_actions}. Page benefits are computed immediately, and tag path groups are stored in a priority queue ordered by average benefit. After these 3k pages, \cite{faheem2015adaptive} \emph{only considers links matching existing tag path groups} (ordered by priority), other are ignored. 
In our context, benefit computation requires knowing if links lead to targets. To enable this baseline, \emph{we provide it with the true benefit} (number of targets behind a page's links) as if given by an oracle, for the initial 3k pages. Beyond this point, newly formed tag path groups receive a fixed benefit of~0. This baseline, given an unfair advantage in its first stage, can be seen as an \emph{ablated} version of ours, learning \emph{offline} instead of \emph{online} with our RL method.

The \emph{Focused Crawler} (\textsf{\small FOCUSED}) represents early generic focused crawling approaches that rely on a classifier to estimate the likelihood that a newly discovered hyperlink leads to a target~\cite{chakrabarti1999focused,diligenti2000focused}. The frontier is maintained as a \emph{priority queue},  favoring links predicted to be relevant. It relies on a standard \emph{logistic regression}, periodically retrained on crawled pages at no extra HTTP cost.
Its features follow standard focused crawler practice: the (approximated) depth of the source page, a 2-gram BoW representation of the URL (similarly to Sec.~\ref{subsection:url_classifier}), and a 2-gram BoW representation of the link's \emph{anchor} text~\cite{chakrabarti1999focused,diligenti2000focused}. \emph{Topic-oriented} features are intentionally excluded (further discussed in Sec.~\ref{section:related_work}). Overall, \textsf{\small FOCUSED} can be viewed as an \emph{ablated} version of our crawler, as it neither exploits tag-path structure (Sec.~\ref{subsection:url_classifier}) nor uses RL (Sec.~\ref{subsection:crawling_algorithm}).

{\textsf{\small TRES}}~\cite{kontogiannis2022tree} is a topical, RL-based focused crawler using a Bi-LSTM classifier to assess HTML page relevance. Like our approach, it employs RL to learn where interesting content resides in a website. However, {\textsf{\small TRES}} defines interestingness through a user-specified topic, whereas we target SDs regardless of subject. As it targets topic-relevant HTML pages only, it requires input keywords, positive HTML examples, and is limited to one language.
To include {\textsf{\small TRES}} as a baseline for SD retrieval, we made changes related to its focus, \emph{without altering its core logic or technical components}. Specifically, we give it three \emph{unfair advantages}: \begin{inparaenum}[(i)]
\item \textsf{\small TRES} expects a list of topic-specific keywords to initialize its classifier's training: we supply 74 hand-crafted terms likely to appear in links' anchors to targets to initialize its classifier.  
\item We provide 1k HTML pages with links to targets from the evaluation websites as positive examples (based on prior crawls), as required to pre-train its classifier, giving it partial access to the solution.  
\item \textsf{\small TRES} must classify URLs as HTML or not to manage its frontier, as it only accepts HTML pages. Although costly in practice (see Sec.~\ref{subsection:url_classifier}), we gave it an oracle doing this classification at no cost, giving it its chance while not affecting its original behavior.
\end{inparaenum}
Since its RL agent and classifier rely only on textual HTML features, {\textsf{\small TRES}} cannot directly detect targets. We therefore added two adjustments: links are added to the frontier as in~\cite{kontogiannis2022tree}, while others (which TRES would normally ignore) are visited immediately and targets counted if found; and its language filter was disabled, preventing coarse URL-based false exclusions. The modified code is available at~\cite{tres_modified_repository}.

\begin{toappendix}
  \subsection{Baselines}
    We use the following list of keywords as initial input to
  \textsf{\small TRES} to train its classifier:
    \begin{multicols}{4}
      \noindent
      pdf\\
xls\\
csv\\
tar\\
zip\\
rar\\
rdf\\
json\\
doc\\
xml\\
yaml\\
txt\\
tsv\\
ppt\\
ods\\
dta\\
7z\\
ttl\\
file\\
document\\
report\\
publication\\
dataset\\
data\\
download\\
archive\\
spreadsheet\\
table\\
list\\
resource\\
annex\\
supplement\\
attachment\\
proceedings\\
survey\\
material\\
output\\
content\\
statistics\\
article\\
paper\\
metadata\\
fact\\
download file\\
download document\\
available for download\\
access data\\
view report\\
get dataset\\
data file\\
read more\\
resource list\\
get document\\
download pulication\\
document archive\\
supporting materials\\
export data\\
download CSV\\
download PDF\\
download XLS\\
dataset download\\
attached document\\
official documents\\
browse files\\
download statistics\\
download article\\
annual report\\
white paper\\
technical documentation\\
technical report\\
raw data\\
metadata file\\
open data\\
fact sheet\\

    \end{multicols}
  
\end{toappendix}

\subsection{Practical Crawling in our Experiments}
\label{sec:exp-crawl-modalities}

To evaluate six baselines and our crawler with various
hyper-parameters, repeated crawls per website are infeasible
due to time constraints (e.g., large files or slow connections), and crawling ethics. Thus, for each website, all baselines and our crawler run in a setting where \emph{each first checks if the resource is already stored in a local database}. If so, we use it; otherwise, we fetch it via HTTP~GET and the URL, HTTP status, headers, and response body are stored.

\emph{For evaluation purposes only}, we stop crawling a website under any of the following conditions:
\begin{inparaenum}[(i)]
	\item One crawler terminates before reaching 1M pages, yielding a complete local replication: all others then rely solely on database look-ups.
	\item All crawlers reach 1M visited webpages (possibly not the
          same pages across crawlers).
    \item A 3-week time limit is reached, as crawling can be extremely slow for some websites (across 18 sites, 22.2M distinct pages were retrieved).
	\item A crawler exceeds 1 minute between two requests, excluding network latency (only affecting \textsf{\small TRES}).
\end{inparaenum}

We conduct and present our experiments as follows:
\begin{inparaenum}[(i)]
    \item We carry hyper-parameter studies and evaluate our URL classifier on fully-crawled websites with \emph{fewer than 1M pages}. Full replication enables parallel runs under multiple parameter settings and provides complete knowledge for both (unrealistic) \textsf{\small SB-ORACLE} crawler and \textsf{\small TRES} baseline (see Sec.~\ref{subsubsection:main_result}).
	\item On websites where \emph{all baselines reach 1M webpages}, crawlers are compared on the subset of the website each of them visited. Note that their target sets, and the retrieved data volume, may differ: a good crawler prioritizes \emph{more target-rich} pages.
    \item On websites where \emph{at least one baseline did not crawl 1M pages within the time limit}, crawlers are compared on the smallest crawl size achieved: e.g., if three crawlers visit 400k, 200k, and 800k pages, respectively, we compare them on their first 200k pages.
    \item On websites where \emph{at least one baseline fails to reach 1M pages} within the time limit, crawlers are compared on the smallest crawl size achieved: e.g., if crawlers visit 400k, 200k, and 800k pages, results are computed on their first 200k pages.

\end{inparaenum}

We
exclude retrieval times from our results, 
since these vary out of our control. 
We focus on the \emph{number of requests and data volumes}; crawl time
can be estimated from these, knowing the bandwidth and the ethics waiting
time. To illustrate retrieval times, on the medium-sized website
\emph{ed}, \textsf{\small SB-CLASSIFIER} requires 3h~16min to collect 5k
targets and 10h~52min to collect 10k, whereas \textsf{\small BFS} requires
5h~13min and 48h~45min respectively (1.6$\times$ and 4.5$\times$ more).

\subsection{Comparison with Baselines}\label{subsubsection:main_result}

\begin{toappendix}
  \subsection{Comparison with Baselines}

In the experiments, we apply filters on MIME types and URL
extensions to avoid downloading multimedia content. The MIME types in the
blocklist are those for multimedia content, namely \texttt{image/*},
\texttt{audio/*}, and \texttt{video/*}. The associated URL extensions are
the following:

{\small\ttfamily
\begin{multicols}{12}
\noindent
.3g2\\
.3ga\\
.3gp2\\
.3gp\\
.3gpa\\
.3gpp2\\
.3gpp\\
.aac\\
.aacp\\
.adp\\
.aff\\
.aif\\
.aiff\\
.arw\\
.asf\\
.asx\\
.avi\\
.avif\\
.avifs\\
.bmp\\
.btif\\
.cgm\\
.cmx\\
.cr2\\
.crw\\
.dcr\\
.djv\\
.djvu\\
.dng\\
.dts\\
.dtshd\\
.dwg\\
.dxf\\
.ecelp4800\\
.ecelp7470\\
.ecelp9600\\
.eol\\
.erf\\
.f4v\\
.fbs\\
.fh4\\
.fh5\\
.fh7\\
.fh\\
.fhc\\
.flac\\
.fli\\
.flv\\
.fpx\\
.fst\\
.fvt\\
.g3\\
.gif\\
.h261\\
.h263\\
.h264\\
.heic\\
.heif\\
.icns\\
.ico\\
.ief\\
.jfi\\
.jfif-tbnl\\
.jfif\\
.jif\\
.jpe\\
.jpeg\\
.jpg\\
.jpgm\\
.jpgv\\
.jpm\\
.k25\\
.kar\\
.kdc\\
.lvp\\
.m1v\\
.m2a\\
.m2v\\
.m3a\\
.m3u\\
.m4a\\
.m4b\\
.m4p\\
.m4r\\
.m4u\\
.m4v\\
.mdi\\
.mid\\
.midi\\
.mj2\\
.mjp2\\
.mka\\
.mkv\\
.mmr\\
.mov\\
.movie\\
.mp2\\
.mp2a\\
.mp3\\
.mp4\\
.mp4v\\
.mpa\\
.mpe\\
.mpeg\\
.mpg4\\
.mpg\\
.mpga\\
.mrw\\
.mxu\\
.nef\\
.npx\\
.oga\\
.ogg\\
.ogv\\
.opus\\
.orf\\
.pbm\\
.pct\\
.pcx\\
.pef\\
.pgm\\
.pic\\
.pjpg\\
.png\\
.pnm\\
.ppm\\
.psd\\
.ptx\\
.pya\\
.pyv\\
.qt\\
.ra\\
.raf\\
.ram\\
.ras\\
.raw\\
.rgb\\
.rlc\\
.rmi\\
.rmp\\
.rw2\\
.rwl\\
.snd\\
.spx\\
.sr2\\
.srf\\
.svg\\
.svgz\\
.tif\\
.tiff\\
.ts\\
.viv\\
.wav\\
.wax\\
.wbmp\\
.weba\\
.webm\\
.webp\\
.wm\\
.wma\\
.wmv\\
.wmx\\
.wvx\\
.x3f\\
.xbm\\
.xif\\
.xpm\\
.xwd\\

\end{multicols}}

Note that URL extension filters are not useful on all websites (on some,
URLs don't usually include any extension, see the discussion at the
beginning of Section~\ref{subsection:url_classifier}).
\end{toappendix}

\begin{figure*}
	\captionsetup[subfigure]{labelformat=simple, labelsep=period}
    \renewcommand{\thesubfigure}{\arabic{subfigure}}
	\begin{subfigure}[t]{\textwidth}
		\includegraphics[width=\linewidth]{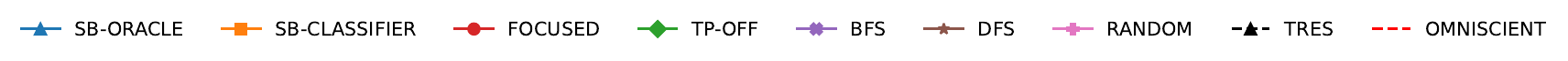}
	\end{subfigure}
	\foreach \i [count=\n from 1] in \websitesSinglePage {
		\begin{subfigure}[t]{0.47\textwidth}
			\includegraphics[width=\linewidth]{plots/oracle_vs_classifier/\i.pdf}

			\vspace{-1em}
			\caption{\getfullname{\n}}
		\end{subfigure}
		\qquad
	}

	\vspace{-.75em}
  \caption{Comparison of different crawler performance for 10
    selected websites presented in
Table~\ref{tab:websites_characteristics}; for {\small TRES}, experiments
are only shown for fully-crawled websites. {\small SB-CLASSIFIER} is the proposed approach.}
	\label{fig:10_sites_plots}
\end{figure*}

\begin{table*}
	\centering
	\caption{Percentage of requests that each crawler performs to retrieve 90\% of the targets, for websites in Table~\ref{tab:websites_characteristics}. Below double horizontal rule, percentage of saved requests and percentage of lost targets due to early-stopping mechanism (see Sec.~\ref{sec:early_stopping}).}

	\setlength{\tabcolsep}{3pt}
	\begin{tabular*}{\textwidth}{@{\extracolsep{\fill}}r*{18}{c}}
		\toprule
		\bfseries \diagbox{\scriptsize Crawler}{\scriptsize Website}         & {\emph{ab}}     & {\emph{as}}     & {\emph{be}}     & {\emph{ce}}     & {\emph{cl}}     & {\emph{cn}}     & {\emph{ed}}     & {\emph{il}}     & {\emph{in}}     & {\emph{is}}     & {\emph{jp}}     & {\emph{ju}}     & {\emph{nc}}     & {\emph{oe}}     & {\emph{ok}}     & {\emph{qa}}     & {\emph{wh}}     & {\emph{wo}}     \\
		\midrule
		\textsf{\small SB-ORACLE} & {NA}            & {NA}            & {72.6}          & {NA}            & {70.7}          & {70.3}          & {48.0}          & {NA}            & {12.8}          & {73.8}          & {NA}            & {34.1}          & {50.8}         & {55.8}          & {13.8}          & {47.3}          & {NA}            & {NA}            \\
		\textsf{\small SB-CLASS.} & {\textbf{31.2}} & {\textbf{35.1}} & {\textbf{75.7}} & {\textbf{23.5}} & {74.4} & {\textbf{70.9}} & {\textbf{51.5}} & {\textbf{14.2}} & {\textbf{11.9}} & {\textbf{76.0}} & {\textbf{37.7}} & {\textbf{35.8}} & {\textbf{51.6}} & {\textbf{59.2}} & {\textbf{15.5}} & {\textbf{57.7}} & {\textbf{19.7}} & {\textbf{18.6}} \\
		\midrule
		\textsf{\small FOCUSED}   & {68.2}          & {$+\infty$}     & {87.8}          & {36.0}          & {88.9}          & {82.7}          & {86.7}          & {$+\infty$}     & {62.8}          & {86.9}          & {42.0}         & {91.1}          & {92.8}          & {84.9}          & {51.8}          & {71.0}          & {$+\infty$}     & {$+\infty$}     \\
		\textsf{\small TP-OFF}    & {96.4}          & {50.3}          & {86.2}          & {34.7}          & {81.8}          & {88.2}          & {95.6}          & {$+\infty$}     & {99.7}          & {88.0}          & {$+\infty$}     & {74.4}          & {93.0}          & {88.7}          & {76.2}          & {88.6}          & {$+\infty$}     & {$+\infty$}     \\
		\textsf{\small BFS}       & {97.4}          & {90.8}          & {89.1}          & {73.5}          & {87.5}         & {80.0}          & {94.6}          & {33.2}          & {99.3}          & {92.7}          & {45.2}          & {80.8}          & {81.8}         & {96.5}          & {66.8}          & {70.6}          & {79.0}          & {92.0}          \\
		\textsf{\small DFS}       & {83.7}          & {$+\infty$}     & {85.2}          & {74.9}          & {\textbf{70.6}} & {84.6}          & {90.5}          & {$+\infty$}     & {99.7}          & {87.7}          & {45.6}          & {80.2}          & {93.7}          & {88.7}          & {80.5}          & {74.4}          & {$+\infty$}     & {$+\infty$}     \\
		\textsf{\small RANDOM}    & {$+\infty$}     & {98.2}          & {92.4}          & {44.5}          & {89.2}          & {85.1}          & {95.0}          & {$+\infty$}     & {99.0}          & {92.7}          & {$+\infty$}     & {83.2}          & {87.9}          & {96.8}          & {85.0}          & {77.8}          & {71.0}          & {$+\infty$}     \\
		\midrule
    \addlinespace[-0.2ex]
		\midrule
		{Saved req.} & {34.4} & {\00.0} & {\00.0} & {\00.0} & {\00.0} & {\00.0} & {27.4} & {\00.0} & {82.6} & {\02.2} & {39.0} & {18.8} & {20.4} & {\00.0} & {73.1} & {\00.0} & {\00.0} & {\00.0} \\
		{Lost targets} & {13.5} & {\00.0} & {\00.0} & {\00.0} & {\00.0} & {\00.0} & {\00.1} & {\00.0} & {\00.0} & {\00.0} & {\02.5} & {\00.4} & {\00.1} & {\00.0} & {\02.0} & {\00.0} & {\00.0} & {\00.0} \\
		\bottomrule
	\end{tabular*}
	\label{tab:all_crawlers_all_sites_nb_requests}
\end{table*}

\begin{toappendix}
  \bigskip

	\def\websitesAppendix{{ab}, {as}, {be}, {cn}, {is}, {jp}, {oe}, {qa}}

	\begin{figure*}[b]
	\captionsetup[subfigure]{labelformat=simple, labelsep=period}
    \renewcommand{\thesubfigure}{\arabic{subfigure}}
		\begin{subfigure}[t]{\textwidth}
			\includegraphics[width=\linewidth]{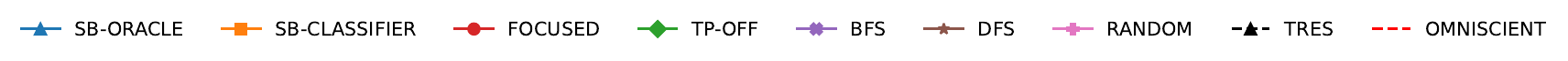}
		\end{subfigure}
		\foreach \short in \websitesAppendix {
			\begin{subfigure}[t]{0.47\textwidth}
				\includegraphics[width=\linewidth]{plots/oracle_vs_classifier/\short.pdf}

				\vspace{-1em}
				\caption{\emph{\short}}
			\end{subfigure}
			\qquad
		}

		\caption{Comparison of different crawler performance for the 8 websites not presented in Figure~\ref{fig:10_sites_plots} due to space reasons;
for {\small TRES}, experiments
are only shown for fully-crawled websites}
		\label{fig:8_last_sites_plots}
	\end{figure*}
	Figure~\ref{fig:8_last_sites_plots} completes
  Figure~\ref{fig:10_sites_plots} with the eight remaining websites plots
  that could not fit in the paper (see
  Table~\ref{tab:websites_characteristics} for characteristics of
  websites). Figures~\ref{fig:hyper_param_alpha_1} to \ref{fig:hyper_param_similarity_threshold_2} present exhaustive graphical results of the hyper-parameters studies conducted on the 11 fully-crawled websites. Especially, Figures~\ref{fig:hyper_param_alpha_1} and \ref{fig:hyper_param_alpha_2} study the exploration--exploitation coefficient $\alpha$, Figures~\ref{fig:hyper_param_n_grams_1} and \ref{fig:hyper_param_n_grams_2} study the choice of $n$ in $n$-grams used in tag paths vector representation, and  Figures~\ref{fig:hyper_param_similarity_threshold_1} and \ref{fig:hyper_param_similarity_threshold_2} study the similarity threshold $\theta$.
\end{toappendix}

\begin{table*}
	\centering
	\caption{Fraction of non-target volume each crawler retrieves before reaching 90\% of total target volume, for websites in Table~\ref{tab:websites_characteristics}}.

	\setlength{\tabcolsep}{3pt}
	\begin{tabular*}{\textwidth}{@{\extracolsep{\fill}}r*{18}{c}}
		\toprule
		\bfseries  \diagbox{\scriptsize Crawler}{\scriptsize Website}         & {\emph{ab}}    & {\emph{as}}     & {\emph{be}}     & {\emph{ce}}     & {\emph{cl}}     & {\emph{cn}}     & {\emph{ed}}     & {\emph{il}}      & {\emph{in}}     & {\emph{is}}     & {\emph{jp}}     & {\emph{ju}}     & {\emph{nc}}     & {\emph{oe}}     & {\emph{ok}}     & {\emph{qa}}     & {\emph{wh}}     & {\emph{wo}}     \\
		\midrule
		\textsf{\small SB-ORACLE} & {NA}            & {NA}            & {24.2}          & {NA}            & {56.3}          & {24.6}          & {49.2}          & {NA}             & {12.5}          & {59.7}          & {NA}            & {22.9}          & {29.5}          & {48.0}          & {33.2}          & {30.2}          & {NA}            & {NA}            \\
		\textsf{\small SB-CLASS.} & {\textbf{20.4}} & {\textbf{21.4}} & {\textbf{29.5}} & {\textbf{29.1}} & {56.0}          & {\textbf{29.0}} & {\textbf{49.2}} & {53.2}           & {\textbf{23.6}} & {\textbf{64.7}}          & {\textbf{18.6}} & {\textbf{23.1}} & {\textbf{34.5}} & {\textbf{49.5}} & {\textbf{34.9}} & {\textbf{33.2}} & {\textbf{32.7}} & {\textbf{35.8}} \\
		\midrule
		\textsf{\small FOCUSED}   & {$+\infty$}     & {$+\infty$}     & {85.2}          & {97.0}          & {76.3}          & {74.7}          & {86.4}          & {$+\infty$}      & {67.3}          & {73.8} & {66.8}          & {72.2}          & {84.9}          & {72.7}          & {49.8}          & {80.3}          & {$+\infty$}     & {$+\infty$}     \\
		\textsf{\small TP-OFF}    & {$+\infty$}     & {$+\infty$}     & {92.3}          & {64.4}          & {65.0}          & {94.7}          & {92.9}          & {$+\infty$}      & {98.8}          & {89.7}          & {$+\infty$}     & {72.3}          & {89.2}          & {89.0}          & {73.6}          & {46.9}          & {$+\infty$}     & {$+\infty$}     \\
		\textsf{\small BFS}       & {81.8}          & {75.7}          & {66.5}          & {98.5}          & {80.8}          & {50.4}          & {93.2}          & {\textbf{\03.6}} & {99.0}          & {93.8}          & {54.0}          & {68.0}          & {84.5}          & {97.5}          & {63.3}          & {87.3}          & {91.5}          & {98.3}          \\
		\textsf{\small DFS}       & {98.6}          & {$+\infty$}     & {64.2}          & {97.0}          & {\textbf{45.0}} & {82.4}          & {90.8}          & {$+\infty$}     & {98.1}          & {85.0}          & {59.5}          & {68.6}          & {96.1}          & {90.5}          & {97.0}          & {75.0}          & {$+\infty$}     & {$+\infty$}     \\
		\textsf{\small RANDOM}    & {71.6}          & {$+\infty$}     & {83.4}          & {$+\infty$}     & {89.3}          & {82.7}          & {92.9}          & {$+\infty$}      & {95.8}          & {98.3}          & {$+\infty$}     & {70.1}          & {88.2}          & {98.1}          & {86.6}          & {77.8}          & {$+\infty$}     & {$+\infty$}     \\
		\bottomrule
	\end{tabular*}
	\label{tab:all_crawlers_all_sites_volume}
\end{table*}

Figure~\ref{fig:10_sites_plots} compares our crawler's performance with the baselines on 10 diverse, representative websites; the other plots are in the extended version~\cite{github_repository}. Default settings are $n=2$ ($n$-grams), $\theta=0.75$, $\alpha=2\sqrt{2}$, $m=12$ (path vector dimension), $w=15$ (hash function), and $b=10$ (batch size). MIME type and extension blocklists are chosen to filter out multimedia content (image, audio, video), usually large in volume. Full lists are in~\cite{github_repository}. 

In addition, for each crawler and all 18 websites, Table~\ref{tab:all_crawlers_all_sites_nb_requests} (above the double rule) shows \emph{the percentage of requests needed to retrieve 90\% of targets}, and Table~\ref{tab:all_crawlers_all_sites_volume} \emph{the percentage of non-target page volume retrieved before reaching 90\% of total target volume}. Lower values indicate greater efficiency. For partially crawled sites, metrics are computed on pages visited by BFS.
 If a crawler never reaches the 90\%, we show $+\infty$.
The best-performing crawler per website is in bold, excluding \textsf{\small SB-ORACLE} (as unrealistic). 
 For each website in Figure~\ref{fig:10_sites_plots}, two graphs show crawler performance: on the left, the \emph{number of crawled targets v.s.\ number of HTTP requests} (both GET
and HEAD, in thousands); 
on the right, the \emph{volume of
	target responses (GB) v.s.\ volume of non-target ones}. In both
graphs, \emph{a higher curve is better}.

For \emph{fully-crawled websites},
we run two variants of our crawler:  \textsf{\small SB-CLASSIFIER} 
and \textsf{\small SB-ORACLE} (see Sec.~\ref{subsection:baselines}).
This assesses the impact of the URL classifier on crawl performance. 
Results are \emph{averaged over 15 runs}, with shaded areas indicating $\pm$ one STD. STD are in most cases low, indicating stable behavior. Variability arises mainly from: randomness in link selection within chosen actions; the stochastic SGD-trained classifier; and the single-state RL agent, whose design is crucial for stabilization (see~\cite{github_repository}).

For \emph{partially crawled websites}, we report a single run of \textsf{\small SB-CLASSIFIER}, lacking perfect classifier information, and since multiple crawls are unfeasible 
(Sec.~\ref{sec:exp-crawl-modalities}). All baselines but \textsf{\small RANDOM} and \textsf{\small TRES} are deterministic, thus one run suffices. \textsf{\small TRES} is only evaluated on the 10 smallest fully-crawled websites, as it requires oracle access and does not scale: beyond small websites (\emph{be}, \emph{cl}, \emph{cn}, and \emph{qa}), it exceeds the 1-minute-per-request threshold and is stopped before completing the crawl of larger websites. Crawling 100k additional pages at that point would require at the very least 10 more weeks.

The plots and Tables~\ref{tab:all_crawlers_all_sites_nb_requests} and \ref{tab:all_crawlers_all_sites_volume} show that \textbf{our crawler with the URL
classifier outperforms all baselines}, with one exception: on \emph{cl}, \textsf{\small DFS} is slightly better near the end. On \emph{il} (volume only) \textsf{\small BFS} briefly surpasses ours, but overall our crawler retrieves twice more targets.  
\textbf{For websites \emph{as}, \emph{ce}, \emph{ed}, \emph{il}, \emph{in}, \emph{ju}, \emph{nc}, \emph{wh}, and \emph{wo}, \textsf{\small SB-CLASSIFIER}} significantly reduces resource usage (time or bandwidth) to reach a fixed number of targets or maximize targets under a budget. 
On \emph{wo}, the best baseline (\textsf{\small BFS}) would require 3 months to match our targets (linear extrapolation), and \textsf{\small FOCUSED} 9 months; on \emph{wh}, the best baseline (\textsf{\small RANDOM}) would require 3 months and the worst (\textsf{\small TP-OFF}) 6 years. On \emph{il}, the top-3 baselines find no targets in the last 650k iterations, whereas ours continues until the end.

On other sites (\emph{be}, \emph{cn}, \emph{cl}, \emph{jp}, \emph{qa}), while some baselines approach \textsf{\small SB-CLASSIFIER} performance, it remains superior overall, outperforming all baselines on a wide variety of websites, with respect to the size, depth, target distribution, etc.
Small sites (\emph{be}, \emph{cl}, \emph{cn}, \emph{qa}) are traversed quickly, allowing less learning, yet our approach adapts effectively.  

\textsf{\small TP-OFF} performs poorly, especially on \emph{ed}, \emph{il}, \emph{jp}, \emph{wh}, and \emph{wo}.
Learning from just 3k pages proves insufficient on large websites. 
On \emph{cn}, \textsf{\small TP-OFF} underperforms BFS after the initial 3k pages, despite only three times more pages remaining. Increasing the training set could help but would approach the simplicity of a standard \textsf{\small BFS} crawler.
\textsf{\small FOCUSED} is outperformed by ours on all websites, showing early focused-crawlers are too basic to efficiently solve our problem.
On 10 websites out of 18, it is even beaten by \textsf{\small BFS}, \textsf{\small DFS} or \textsf{\small RANDOM}.

Despite being given three unfair advantages (hand-picked keywords, access to training pages with target links, and an oracle for URL type prediction), 
\textsf{\small TRES} fails on 9 out of 10 websites, performing reasonably only on \emph{oe} and only until it had to be stopped. More critically, \textsf{\small TRES} cannot scale to medium or large sites, even when run fully locally. Crawl iterations slow dramatically due to exhaustive feature-based evaluations during tree expansion.
These results confirm that topical crawlers are fundamentally ill-suited for direct target retrieval, even when heavily adapted and unfairly favored.

Comparing our \textsf{\small SB-CLASSIFIER} with \textsf{\small SB-ORACLE}
shows that our classifier is close to the (virtual)
	perfect oracle. Further analysis (our classifier's impact) is
  provided in the extended version~\cite{github_repository}.

\subsection{Hyper-Parameters}\label{subsubsection:metaparameters_result}

We present the impact of three key hyper-parameters ($\alpha$, $n$, and $\theta$) on crawl performance. Table~\ref{tab:parameter_study} reports the number of requests (resp. response volume for non-target pages) needed to retrieve 90\% of the targets (resp. target volume);
full plots are available in~\cite{github_repository}. When varying one parameter, others remain at default values (Sec.~\ref{subsubsection:main_result}). Additional parameters, including the projection dimension~$D$ (Sec.~\ref{subsection:group-links}), showed no significant effect.

\begin{inparaenum}
	\item \emph{Impact of $\alpha$.} $\alpha$ controls the exploration--exploitation trade-off (Sec.~\ref{subsection:group-links}). We test $\alpha \in \{0.1,2\sqrt{2},30\}$ (Table~\ref{tab:parameter_study}, top), as optimality of $2\sqrt{2}$ is not guaranteed in our non-standard MAB setting. Smaller $\alpha$ generally improves performance, with best results for $\alpha=2\sqrt{2}$. Large $\alpha$ leads to excessive \emph{exploration}, especially on low or moderate-reward websites, neglecting useful, already discovered actions.
	\item \emph{Impact of $n$ in $n$-grams for tag path representation.} For merging tag paths (Sec.~\ref{subsection:states_actions}), we test $n \in \{1,2,3\}$ ($n=1$ represents paths as sets of HTML tags). Table~\ref{tab:parameter_study} (middle) shows that $n=2$ and $3$ generally outperform $n=1$, confirming that groups of tag paths provide a better learning basis than sets of tags. We choose $n=2$, as $3$ yields similar performance but incurs higher computational cost due to exponential growth feature space growth.
	\item \emph{Impact of $\theta$.} The similarity threshold $\theta$ controls tag path grouping (Sec.~\ref{subsection:states_actions}). We test $\theta \in \{0.55,0.75,0.95\}$ (Table~\ref{tab:parameter_study}, bottom). High $\theta$ often performs poorly, especially when no similarity $>95\%$ exists, as with websites adding unique IDs in tags. For example, on \emph{ed} it caused an OOM error by creating as many actions as HTML pages. Best results occur with $\theta \in \{0.55,0.75\}$, with $\theta=0.75$ usually superior: when $0.55$ performs better, $0.75$ remains comparable, but not conversely (e.g., on \emph{ju}, $\theta=0.75$ improves both requests and volume by 40\%).

\end{inparaenum}

\begin{table*}
	\centering
	\caption{Percentage of requests that an SB crawler with oracle
		performs to retrieve 90\% of the targets (left of |).\\
		Right of |:
		percentage of the volume of non-target pages retrieved, before
  retrieving 90\% of the total target volume. \\Hyper-parameter study on
$\alpha$ (top), $n$ (in $n$-grams, center), and $\theta$ (bottom), for
the 11 fully-crawled websites.}

	\footnotesize
	\setlength{\tabcolsep}{2.5pt}

	\begin{tabular*}{\textwidth}{
	@{\extracolsep{\fill}}
	r
	*{11}{r@{\hspace*{0.3em}\textbar\hspace*{-.75em}}r}
	}
	\toprule
	\bfseries Crawler
	& \multicolumn{2}{c}{\emph{be}}
	& \multicolumn{2}{c}{\emph{cl}}
	& \multicolumn{2}{c}{\emph{cn}}
	& \multicolumn{2}{c}{\emph{ed}}
	& \multicolumn{2}{c}{\emph{in}}
	& \multicolumn{2}{c}{\emph{is}}
	& \multicolumn{2}{c}{\emph{ju}}
	& \multicolumn{2}{c}{\emph{nc}}
	& \multicolumn{2}{c}{\emph{oe}}
	& \multicolumn{2}{c}{\emph{ok}}
	& \multicolumn{2}{c}{\emph{qa}} \\
	\midrule
	$\alpha = 0.1$
	& 86.3 & 26.2
	& \textbf{75.9} & \textbf{42.3}
	& 74.3 & 35.5
	& 53.7 & 54.1
	& \textbf{\09.8} & \textbf{10.2}
	& 77.1 & 66.2
	& 37.1 & 35.0
	& 51.6 & \textbf{26.2}
	& \textbf{55.6} & \textbf{34.4}
	& 14.3 & 33.2
	& \textbf{67.7} & 32.1 \\

	$\alpha = 2\sqrt{2}$
	& 84.7 & \textbf{24.2}
	& 76.4 & 56.3
	& \textbf{71.8} & \textbf{24.6}
	& \textbf{53.0} & 49.2
	& 11.1 & 11.0
	& \textbf{74.2} & \textbf{58.9}
	& \textbf{35.0} & \textbf{22.9}
	& \textbf{51.4} & 29.5
	& 59.2 & 48.0
	& \textbf{10.3} & \textbf{19.0}
	& 68.9 & 33.9 \\

	$\alpha = 30$
	& \textbf{83.8} & 36.7
	& 79.6 & 58.9
	& 75.3 & 32.4
	& 66.2 & \textbf{41.5}
	& 11.6 & 11.8
	& 80.9 & 66.4
	& 43.3 & 28.8
	& 67.3 & 29.5
	& 68.8 & 72.9
	& 36.7 & 71.3
	& 71.8 & \textbf{30.4} \\

	\midrule
	$n = 1$
	& 84.5 & 27.1
	& 77.2 & \textbf{48.5}
	& 78.6 & 56.3
	& 57.3 & 55.1
	& \textbf{\09.9} & 10.7
	& 78.2 & 69.6
	& 35.7 & \textbf{17.6}
	& 54.8 & 33.5
	& \textbf{52.6} & \textbf{28.1}
	& 13.6 & 27.2
	& 68.9 & 34.7 \\

	$n = 2$
	& 84.7 & \textbf{24.2}
	& \textbf{76.4} & 56.3
	& 71.8 & \textbf{24.6}
	& \textbf{53.0} & \textbf{49.2}
	& 11.1 & 11.0
	& 74.2 & 58.9
	& \textbf{35.0} & 22.9
	& 51.4 & 29.5
	& 59.2 & 48.0
	& 10.3 & 19.0
	& \textbf{68.3} & \textbf{33.9} \\

	$n = 3$
	& \textbf{84.1} & 32.8
	& 78.2 & 51.2
	& \textbf{71.3} & 25.7
	& 57.0 & 53.1
	& 10.7 & \textbf{10.5}
	& \textbf{71.3} & \textbf{49.2}
	& 37.0 & 26.9
	& \textbf{51.2} & \textbf{27.0}
	& 79.6 & 79.0
	& \textbf{\06.0} & \textbf{\08.8}
	& 70.0 & 34.9 \\

	\midrule
	$\theta = 0.55$
	& \textbf{81.2} & 42.0
	& 76.8 & \textbf{50.5}
	& 76.6 & 41.9
	& 56.5 & 53.1
	& \textbf{\08.2} & \textbf{\09.4}
	& 78.7 & 65.5
	& 80.6 & 65.4
	& 56.1 & 35.5
	& \textbf{52.4} & \textbf{30.9}
	& 12.5 & 25.7
	& \textbf{67.8} & 26.0 \\

	$\theta = 0.75$
	& 84.7 & \textbf{24.2}
	& \textbf{76.4} & 56.3
	& \textbf{71.8} & \textbf{24.6}
	& \textbf{53.0} & \textbf{49.2}
	& 11.1 & 11.0
	& 74.2 & 58.9
	& \textbf{35.0} & \textbf{22.9}
	& \textbf{51.4} & \textbf{29.5}
	& 59.2 & 48.0
	& \textbf{10.3} & \textbf{18.7}
	& 68.9 & 33.9 \\

	$\theta = 0.95$
	& 82.4 & 47.7
	& 84.3 & 72.1
	& 73.1 & 44.7
	& OOM & OOM
	& \09.8 & 11.0
	& \textbf{71.0} & \textbf{54.9}
	& 73.3 & 66.5
	& 57.3 & 33.2
	& 90.2 & 87.2
	& 12.4 & 19.0
	& 68.3 & \textbf{25.9} \\
	\bottomrule
	\end{tabular*}
	\label{tab:parameter_study}
	\end{table*}

All crawls with \textsf{\small SB-CLASSIFIER} so far use a logistic
regression classifier (\textsc{LR}) with URL features
(\textsc{URL\_ONLY}). To study model and feature impacts, we also
evaluate a linear SVM (\textsc{SVM}), multinomial Naive Bayes
(\textsc{NB}), and a passive--aggressive
classifier~\cite{crammer2006online} (\textsc{PA}), restricting to
lightweight online models and excluding deep approaches whose cost would
shift the bottleneck from network latency to local CPU/GPU time. Each
model is used with two feature sets: \textsc{URL\_ONLY} and
\textsc{URL\_CONT} (URL plus anchor text, DOM path, and surrounding
text). All features use character-level 2-gram BoW encoding, yielding
eight variants. Experiments are run on fully-crawled websites, with 15
runs per configuration and per site. Table~\ref{tab:expes_url_classifier}
reports, for each website and variant, the \emph{intra-website} crawl
metric of Table~\ref{tab:all_crawlers_all_sites_nb_requests} (number of
requests to reach 90\% of targets); variants requiring at least 2\% fewer
requests than \textsc{URL\_ONLY--LR} are in bold. The last column
(\textbf{``MR''}) reports \emph{inter-website} misclassification rates on
HTML pages and targets from averaged confusion matrices. On 10 of the 11
websites, no variant improves over \textsc{URL\_ONLY--LR} by more than
2\%; the sole exception is \emph{qa}, where two variants achieve a $\leq
2.3\%$ gain. This lack of consistent gains, also seen in \textbf{``MR''}, shows that neither richer features nor more complex models yield meaningful gains, justifying the choice of \textsc{URL\_ONLY--LR}.

\begin{table*}
	\centering
	\caption{Intra-website crawl metric (90\% targets reached) and inter-website misclassification rates for URL classifier variants.}
	\begin{tabular}{lcccccccccccc}
		\toprule
		\textbf{Variant} & \emph{be} & \emph{cl} & \emph{cn} & \emph{ed} & \emph{in} & \emph{is} & \emph{ju} & \emph{nc} & \emph{oe} & \emph{ok} & \emph{qa} & \bfseries MR \\
		\midrule
		\small URL\_ONLY-LR & 82.1 & 75.1 & 71.3 & 53.2 & 11.7 & 76.1 & 36.5 & 52.6 & 60.7 & 15.9 & 62.3 & 2.62 \\
		\small URL\_ONLY-SVM & 82.7 & 75.7 & 71.8 & 63.6 & 11.3 & 76.0 & 37.4 & 52.2 & 63.5 & 16.7 & 61.5 & 2.99 \\
		\small URL\_ONLY-NB & 82.9 & 75.2 & 72.1 & 53.7 & 11.4 & 76.3 & 35.8 & 52.7 & 59.7 & 18.0 & 63.1 & 2.92 \\
		\small URL\_ONLY-PA & 82.3 & 74.4 & 71.7 & 53.3 & 11.1 & 75.8 & 36.7 & 51.6 & 60.5 & 15.9 & 60.9 & 2.56 \\
		\small URL\_CONT-LR & 82.2 & 74.4 & 71.9 & 54.3 & 11.3 & 76.4 & 37.8 & 52.9 & 64.7 & 16.8 & \textbf{60.0} & 5.93 \\
		\small URL\_CONT-SVM & 82.6 & 75.0 & 71.8 & 52.8 & 11.6 & 76.4 & 38.8 & 53.1 & 61.1 & 18.7 & \textbf{60.1} & 6.36 \\
		\small URL\_CONT-NB & 84.1 & 74.7 & 71.9 & 53.6 & 11.4 & 75.7 & 35.5 & 52.3 & 59.9 & 19.1 & 60.4 & 7.15 \\
		\small URL\_CONT-PA & 82.5 & 75.1 & 71.9 & 53.6 & 11.6 & 76.2 & 38.4 & 52.1 & 62.6 & 16.1 & 60.6 & 4.12 \\
		\bottomrule
	\end{tabular}
	\label{tab:expes_url_classifier}
\end{table*}

\begin{toappendix}

	\def\websitesHyperParamOne{{be}, {cl}, {cn}, {ed}, {in}, {is}}
	\def\websitesHyperParamTwo{{ju}, {nc}, {oe}, {ok}, {qa}}

	\newcounter{plotcount}
	\begin{figure}
	\captionsetup[subfigure]{labelformat=simple, labelsep=period}
\renewcommand{\thesubfigure}{\arabic{subfigure}}
		\begin{subfigure}[t]{\textwidth}
			\includegraphics[width=\linewidth]{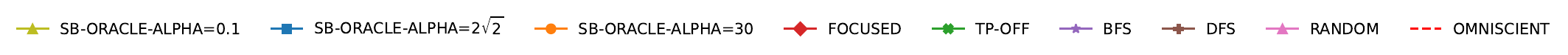}
		\end{subfigure}

		\setcounter{plotcount}{0}
		\foreach \short in \websitesHyperParamOne {
			\stepcounter{plotcount}
			\ifnum\value{plotcount}<9
			\begin{subfigure}[t]{0.47\textwidth}
				\includegraphics[width=\linewidth]{plots/alpha/\short.pdf}
				\vspace{-2em}
				\caption{\emph{\short}}
			\end{subfigure}
			\ifodd\value{plotcount}
			\hfill
			\else
			\par\vspace{0.5em}
			\fi
			\else

			\begin{center}
				\begin{subfigure}[t]{0.47\textwidth}
					\includegraphics[width=\linewidth]{plots/alpha/\short.pdf}
					\vspace{-2em}
					\caption{\emph{\short}}
				\end{subfigure}
			\end{center}
			\fi
		}

		\caption{Crawler performance for hyper-parameter study on
			exploration--exploitation coefficient $\alpha$, for websites \emph{be}, \emph{cl}, \emph{cn}, \emph{ed}, \emph{in}, and \emph{is}}
		\label{fig:hyper_param_alpha_1}
	\end{figure}

	\begin{figure}
	\captionsetup[subfigure]{labelformat=simple, labelsep=period}
\renewcommand{\thesubfigure}{\arabic{subfigure}}
		\begin{subfigure}[t]{\textwidth}
			\includegraphics[width=\linewidth]{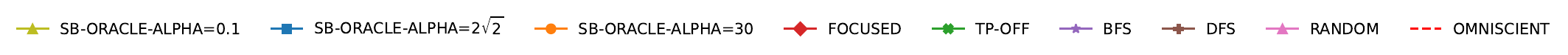}
		\end{subfigure}

		\setcounter{plotcount}{0}
		\foreach \short in \websitesHyperParamTwo {
			\stepcounter{plotcount}
			\ifnum\value{plotcount}<9
			\begin{subfigure}[t]{0.47\textwidth}
				\includegraphics[width=\linewidth]{plots/alpha/\short.pdf}
				\vspace{-2em}
				\caption{\emph{\short}}
			\end{subfigure}
			\ifodd\value{plotcount}
			\hfill
			\else
			\par\vspace{0.5em}
			\fi
			\else

			\begin{center}
				\begin{subfigure}[t]{0.47\textwidth}
					\includegraphics[width=\linewidth]{plots/alpha/\short.pdf}
					\vspace{-2em}
					\caption{\emph{\short}}
				\end{subfigure}
			\end{center}
			\fi
		}

		\caption{Crawler performance for hyper-parameter study on
			exploration--exploitation coefficient $\alpha$, for websites \emph{ju}, \emph{nc}, \emph{oe}, \emph{ok}, and \emph{qa}}
		\label{fig:hyper_param_alpha_2}
	\end{figure}

	\begin{figure}
	\captionsetup[subfigure]{labelformat=simple, labelsep=period}
\renewcommand{\thesubfigure}{\arabic{subfigure}}
		\begin{subfigure}[t]{\textwidth}
			\includegraphics[width=\linewidth]{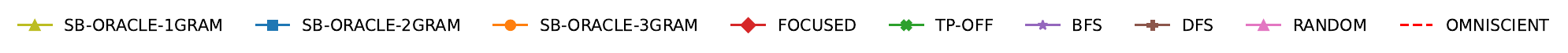}
		\end{subfigure}

		\setcounter{plotcount}{0}
		\foreach \short in \websitesHyperParamOne {
			\stepcounter{plotcount}
			\ifnum\value{plotcount}<9
			\begin{subfigure}[t]{0.47\textwidth}
				\includegraphics[width=\linewidth]{plots/n_grams/\short.pdf}
				\vspace{-2em}
				\caption{\emph{\short}}
			\end{subfigure}
			\ifodd\value{plotcount}
			\hfill
			\else
			\par\vspace{0.5em}
			\fi
			\else

			\begin{center}
				\begin{subfigure}[t]{0.47\textwidth}
					\includegraphics[width=\linewidth]{plots/n_grams/\short.pdf}
					\vspace{-2em}
					\caption{\emph{\short}}
				\end{subfigure}
			\end{center}
			\fi
		}

		\caption{Crawler performance for impact study of the choice of $n$ in $n$-grams used in tag path vector representation, for websites \emph{be}, \emph{cl}, \emph{cn}, \emph{ed}, \emph{in}, and \emph{is}}
		\label{fig:hyper_param_n_grams_1}
	\end{figure}

	\begin{figure}
	\captionsetup[subfigure]{labelformat=simple, labelsep=period}
\renewcommand{\thesubfigure}{\arabic{subfigure}}
		\begin{subfigure}[t]{\textwidth}
			\includegraphics[width=\linewidth]{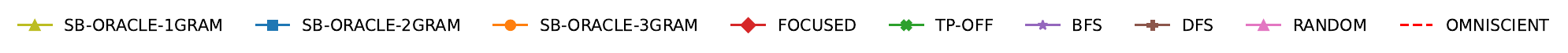}
		\end{subfigure}

		\setcounter{plotcount}{0}
		\foreach \short in \websitesHyperParamTwo {
			\stepcounter{plotcount}
			\ifnum\value{plotcount}<9
			\begin{subfigure}[t]{0.47\textwidth}
				\includegraphics[width=\linewidth]{plots/n_grams/\short.pdf}
				\vspace{-2em}
				\caption{\emph{\short}}
			\end{subfigure}
			\ifodd\value{plotcount}
			\hfill
			\else
			\par\vspace{0.5em}
			\fi
			\else

			\begin{center}
				\begin{subfigure}[t]{0.47\textwidth}
					\includegraphics[width=\linewidth]{plots/n_grams/\short.pdf}
					\vspace{-2em}
					\caption{\emph{\short}}
				\end{subfigure}
			\end{center}
			\fi
		}

		\caption{Crawler performance for impact study of the choice of $n$ in $n$-grams used in tag path vector representation, for websites \emph{ju}, \emph{nc}, \emph{oe}, \emph{ok}, and \emph{qa}}
		\label{fig:hyper_param_n_grams_2}
	\end{figure}

	\begin{figure}
	\captionsetup[subfigure]{labelformat=simple, labelsep=period}
\renewcommand{\thesubfigure}{\arabic{subfigure}}
		\begin{subfigure}[t]{\textwidth}
			\includegraphics[width=\linewidth]{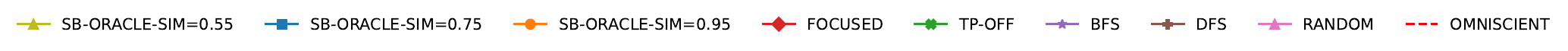}
		\end{subfigure}

		\setcounter{plotcount}{0}
		\foreach \short in \websitesHyperParamOne {
			\stepcounter{plotcount}
			\ifnum\value{plotcount}<9
			\begin{subfigure}[t]{0.47\textwidth}
				\includegraphics[width=\linewidth]{plots/similarity_threshold/\short.pdf}
				\vspace{-2em}
				\caption{\emph{\short}}
			\end{subfigure}
			\ifodd\value{plotcount}
			\hfill
			\else
			\par\vspace{0.5em}
			\fi
			\else

			\begin{center}
				\begin{subfigure}[t]{0.47\textwidth}
					\includegraphics[width=\linewidth]{plots/similarity_threshold/\short.pdf}
					\vspace{-2em}
					\caption{\emph{\short}}
				\end{subfigure}
			\end{center}
			\fi
		}

		\caption{Crawler performance for impact study on similarity threshold $\theta$, \emph{cl}, \emph{cn}, \emph{ed}, \emph{in}, and \emph{is}}
		\label{fig:hyper_param_similarity_threshold_1}
	\end{figure}

	\begin{figure}
		\captionsetup[subfigure]{labelformat=simple, labelsep=period}
\renewcommand{\thesubfigure}{\arabic{subfigure}}
		\begin{subfigure}[t]{\textwidth}
			\includegraphics[width=\linewidth]{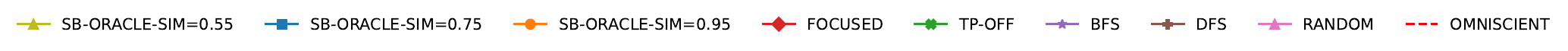}
		\end{subfigure}

		\setcounter{plotcount}{0}
		\foreach \short in \websitesHyperParamTwo {
			\stepcounter{plotcount}
			\ifnum\value{plotcount}<9
			\begin{subfigure}[t]{0.47\textwidth}
				\includegraphics[width=\linewidth]{plots/similarity_threshold/\short.pdf}
				\vspace{-2em}
				\caption{\emph{\short}}
			\end{subfigure}
			\ifodd\value{plotcount}
			\hfill
			\else
			\par\vspace{0.5em}
			\fi
			\else

			\begin{center}
				\begin{subfigure}[t]{0.47\textwidth}
					\includegraphics[width=\linewidth]{plots/similarity_threshold/\short.pdf}
					\vspace{-2em}
					\caption{\emph{\short}}
				\end{subfigure}
			\end{center}
			\fi
		}

		\caption{Crawler performance for impact study on similarity threshold $\theta$, for websites \emph{ju}, \emph{nc}, \emph{oe}, \emph{ok}, and \emph{qa}}
		\label{fig:hyper_param_similarity_threshold_2}
	\end{figure}
	\clearpage

\def\websitesFullyCrawled{{be}, {cl}, {cn}, {ed}, {in}, {is}, {ju}, {nc}, {oe}, {ok}, {qa}}

Figure~\ref{fig:url_classifier_models_features_study} shows the crawl performance of \textsf{\small SB-CLASSIFIER} across different URL classifier configurations, combining various learning models (LR: logistic regression; SVM: linear support vector machine; NB: multinomial Naive Bayes; PA: passive-aggressive classifier) and feature sets (URL\_ONLY: character-level 2-gram bag-of-words of the URL alone; URL\_CONT: URL\_ONLY augmented with character-level 2-gram bag-of-words of the anchor text, DOM path, and surrounding text), evaluated on the 11 fully-crawled websites.

\begin{figure}
\captionsetup[subfigure]{labelformat=simple, labelsep=period}
\renewcommand{\thesubfigure}{\arabic{subfigure}}
		\begin{subfigure}[t]{\textwidth}
			\includegraphics[width=\linewidth]{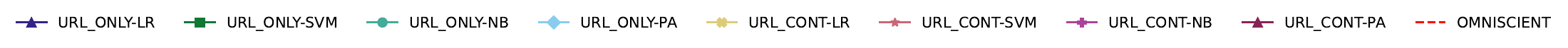}
		\end{subfigure}

		\setcounter{plotcount}{0}
		\foreach \short in \websitesFullyCrawled {
			\stepcounter{plotcount}
			\begin{subfigure}[t]{0.24\textwidth}
				\includegraphics[width=\linewidth]{plots/url_classifier/\short.pdf}
				\vspace{-2em}
				\caption{\emph{\short}}
			\end{subfigure}
			\ifnum\value{plotcount}=4
			\par\vspace{0.5em}
			\setcounter{plotcount}{0}
			\else
			\hfill
			\fi
		}

	\caption{Crawler performance for impact study on URL classifier models and feature sets, for all 11 fully-crawled websites}
	\label{fig:url_classifier_models_features_study}
\end{figure}
\clearpage

Tables~\ref{tab:cm_pcts_url_only_lr}~to~\ref{tab:cm_pcts_url_cont_pa} report the confusion matrices for each URL classifier variant, averaged across the 11 fully-crawled websites and over 15 runs. Detailed confusion matrices for each variant on individual sites are available in the repository~\cite{github_repository}.

\begin{table}[h]
	\centering
	\caption{Confusion matrix of the URL classifier URL\_ONLY-LR, inter-site average over the 11 fully-crawled websites (each averaged over 15 runs)}
	\vspace*{-.5em}
	\begin{tabular}{lrrr}
		\toprule
		\bfseries \diagbox{True}{Predicted} & \bfseries HTML (\%) &
		\bfseries Target (\%) & \bfseries Neither (\%) \\
		\midrule
		\bfseries HTML    & 58.73 & \01.69 & 0.00 \\
		\bfseries Target  & \00.77  & 32.73 & 0.00 \\
		\bfseries Neither & \04.50  & \01.58 & 0.00 \\
		\bottomrule
	\end{tabular}
	\label{tab:cm_pcts_url_only_lr}
\end{table}

\begin{table}[h]
	\centering
	\vspace{-1em}
	\caption{Confusion matrix of the URL classifier URL\_ONLY-SVM, inter-site average over the 11 fully-crawled websites (each averaged over 15 runs)}
	\vspace*{-.5em}
	\begin{tabular}{lrrr}
		\toprule
		\bfseries \diagbox{True}{Predicted} & \bfseries HTML (\%) &
		\bfseries Target (\%) & \bfseries Neither (\%) \\
		\midrule
		\bfseries HTML    & 58.59 & \01.84 & 0.00 \\
		\bfseries Target  & \00.98  & 32.53 & 0.00 \\
		\bfseries Neither & \04.48  & \01.60 & 0.00 \\
		\bottomrule
	\end{tabular}
	\label{tab:cm_pcts_url_only_svm}
\end{table}

\begin{table}[h]
	\centering
	\vspace{-1em}
	\caption{Confusion matrix of the URL classifier URL\_ONLY-NB, inter-site average over the 11 fully-crawled websites (each averaged over 15 runs)}
	\vspace*{-.5em}
	\begin{tabular}{lrrr}
		\toprule
		\bfseries \diagbox{True}{Predicted} & \bfseries HTML (\%) &
		\bfseries Target (\%) & \bfseries Neither (\%) \\
		\midrule
		\bfseries HTML    & 58.44 & \01.97 & 0.00 \\
		\bfseries Target  & \00.77  & 32.73 & 0.00 \\
		\bfseries Neither & \04.47  & \01.61 & 0.00 \\
		\bottomrule
	\end{tabular}
	\label{tab:cm_pcts_url_only_nb}
\end{table}

\begin{table}[h]
	\centering
	\vspace{-1em}
	\caption{Confusion matrix of the URL classifier URL\_ONLY-PA, inter-site average over the 11 fully-crawled websites (each averaged over 15 runs)}
	\vspace*{-.5em}
	\begin{tabular}{lrrr}
		\toprule
		\bfseries \diagbox{True}{Predicted} & \bfseries HTML (\%) &
		\bfseries Target (\%) & \bfseries Neither (\%) \\
		\midrule
		\bfseries HTML    & 58.60 & \01.82 & 0.00 \\
		\bfseries Target  & \00.58  & 32.92 & 0.00 \\
		\bfseries Neither & \04.50  & \01.57 & 0.00 \\
		\bottomrule
	\end{tabular}
	\label{tab:cm_pcts_url_only_pa}
\end{table}

\begin{table}[h]
	\centering
	\vspace{-1em}
	\caption{Confusion matrix of the URL classifier URL\_CONT-LR, inter-site average over the 11 fully-crawled websites (each averaged over 15 runs)}
	\vspace*{-.5em}
	\begin{tabular}{lrrr}
		\toprule
		\bfseries \diagbox{True}{Predicted} & \bfseries HTML (\%) &
		\bfseries Target (\%) & \bfseries Neither (\%) \\
		\midrule
		\bfseries HTML    & 57.62 & \02.79 & 0.00 \\
		\bfseries Target  & \02.78  & 30.73 & 0.00 \\
		\bfseries Neither & \04.58  & \01.49 & 0.00 \\
		\bottomrule
	\end{tabular}
	\label{tab:cm_pcts_url_cont_lr}
\end{table}

\begin{table}[h]
	\centering
	\vspace{-1em}
	\caption{Confusion matrix of the URL classifier URL\_CONT-SVM, inter-site average over the 11 fully-crawled websites (each averaged over 15 runs)}
	\vspace*{-.5em}
	\begin{tabular}{lrrr}
		\toprule
		\bfseries \diagbox{True}{Predicted} & \bfseries HTML (\%) &
		\bfseries Target (\%) & \bfseries Neither (\%) \\
		\midrule
		\bfseries HTML    & 57.47 & \02.94 & 0.00 \\
		\bfseries Target  & \03.03  & 30.48 & 0.00 \\
		\bfseries Neither & \04.61  & \01.46 & 0.00 \\
		\bottomrule
	\end{tabular}
	\label{tab:cm_pcts_url_cont_svm}
\end{table}

\begin{table}[h]
	\centering
	\vspace{-1em}
	\caption{Confusion matrix of the URL classifier URL\_CONT-NB, inter-site average over the 11 fully-crawled websites (each averaged over 15 runs)}
	\vspace*{-.5em}
	\begin{tabular}{lrrr}
		\toprule
		\bfseries \diagbox{True}{Predicted} & \bfseries HTML (\%) &
		\bfseries Target (\%) & \bfseries Neither (\%) \\
		\midrule
		\bfseries HTML    & 58.34 & \02.08 & 0.00 \\
		\bfseries Target  & \04.64 & 28.87 & 0.00 \\
		\bfseries Neither & \04.58 & \01.49 & 0.00 \\
		\bottomrule
	\end{tabular}
	\label{tab:cm_pcts_url_cont_nb}
\end{table}

\begin{table}[h]
	\centering
	\vspace{-1em}
	\caption{Confusion matrix of the URL classifier URL\_CONT-PA, inter-site average over the 11 fully-crawled websites (each averaged over 15 runs)}
	\vspace*{-.5em}
	\begin{tabular}{lrrr}
		\toprule
		\bfseries \diagbox{True}{Predicted} & \bfseries HTML (\%) &
		\bfseries Target (\%) & \bfseries Neither (\%) \\
		\midrule
		bfseries HTML    & 57.96 & \02.45 & 0.00 \\
		\bfseries Target  & \01.42 & 32.09 & 0.00 \\
		\bfseries Neither & \04.54 & \01.53 & 0.00 \\
		\bottomrule
	\end{tabular}
	\label{tab:cm_pcts_url_cont_pa}
\end{table}

\clearpage

\end{toappendix}

\subsection{Effectiveness of SB Learning} \label{sec:exp-sb-algo}

\begin{figure}
	\centering
	\includegraphics[width=\linewidth]{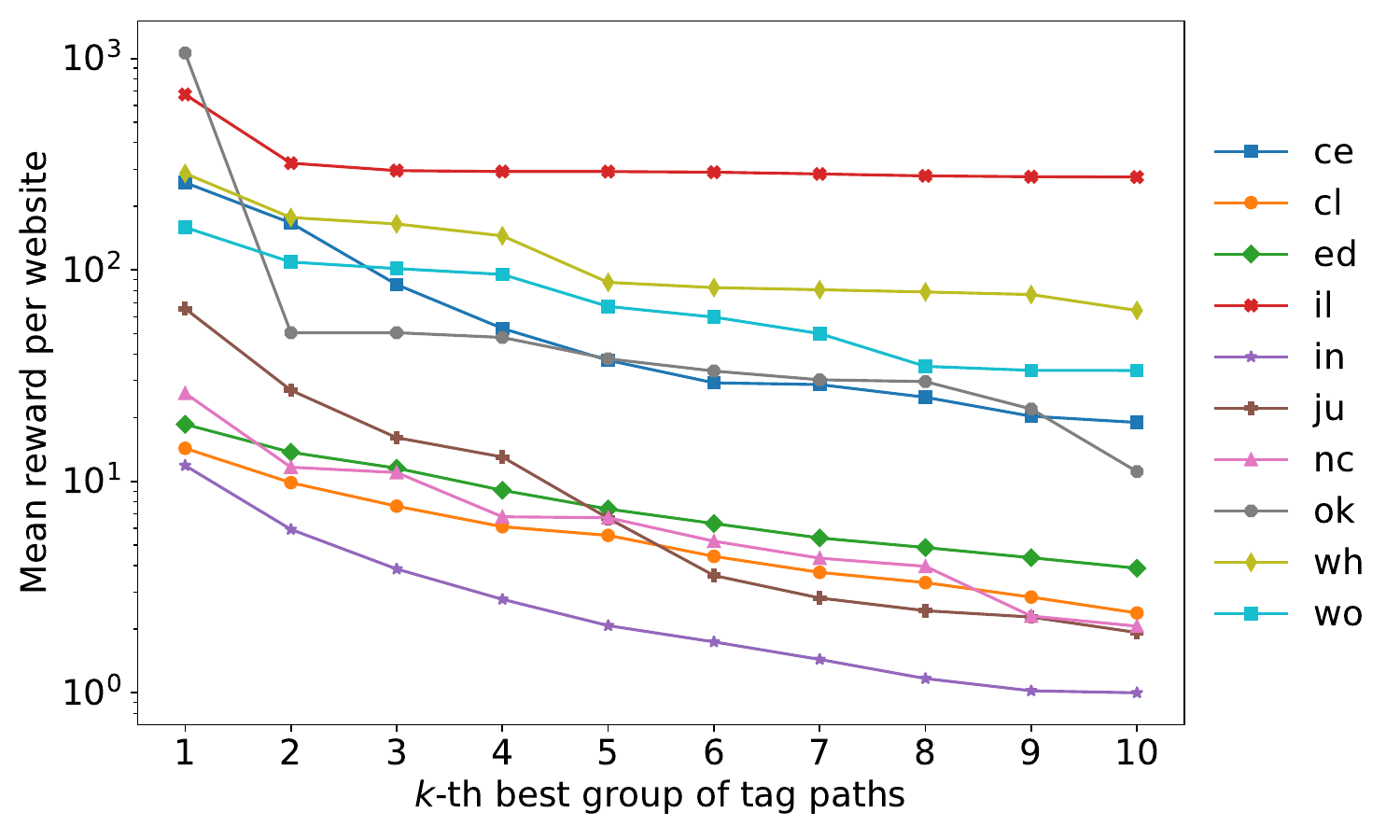}
	\caption{Mean rewards of the top-10 groups of tag paths for the ten
  selected websites from Figure~\ref{fig:10_sites_plots} (log $y$-scale)}
	\label{fig:top_k_dom_groups}
\end{figure}

\begin{table*}
	\centering
	\caption{Mean and STD of non-zero rewards of the
  learning agent on each website}
	\begin{tabular}{ccccccccccccccccccc}
		\toprule
		               & \emph{ab} & \emph{as} & \emph{be} & \emph{ce} & \emph{cl} & \emph{cn} & \emph{ed} & \emph{il} & \emph{in} & \emph{is} & \emph{jp} & \emph{ju} & \emph{nc} & \emph{oe} & \emph{ok} & \emph{qa} & \emph{wh} & \emph{wo} \\
		\midrule
		\bfseries Mean & \01.7     & 1.5       & \04.5     & \030.2    & 12.4      & 4.2       & 2.5       & \03.1     & 1.6       & \03.5     & \03.5     & \05.4     & \02.0     & \02.5     & \05.5     & 15.4      & \03.0     & \02.1     \\
		\bfseries Std  & 16.8      & 5.35      & 20.9      & 290.3     & \02.8     & 8.9       & 7.1       & 53.9      & 4.2       & 11.1      & 17.4      & 10.5      & \08.7     & \09.3     & 13.9      & 18.8      & 22.0      & 43.5      \\
		\bottomrule
	\end{tabular}
	\label{tab:websites_rewards}
\end{table*}

We first study the reward associated by our algorithm to tag paths.
Figure~\ref{fig:top_k_dom_groups} shows the mean reward of the top 10
path groups (logarithmic $y$-axis), for the ten websites in Figure~\ref{fig:10_sites_plots}.
Top groups have high rewards: across websites, the best group averages 258, followed by 89, 74, 67, and 41 for the 10th, showing that \textbf{our SB agent effectively identifies tag path groups leading to HTML pages with target links}. Although the first group often dominates, several groups still yield substantial reward, so restricting to the top one is insufficient. Comparing Figure~\ref{fig:top_k_dom_groups} with the
mean rewards for each website (on groups with non-zero rewards)
in Table~\ref{tab:websites_rewards}, the top 10 groups
generally far exceed the overall mean
(e.g., for \emph{wo}, the the 10th group scores 33.5 v.s.\
 2.1 for the mean). This is also clear
from the high STD values from Table~\ref{tab:websites_rewards} (over 20 times the mean for \emph{wo})
implying significant differences in the rewards of different
	groups: rewards are not normally distributed across groups but more
closely resemble a power law.
Table~\ref{tab:websites_rewards} also shows large cross-website reward disparities, leading to the impossibility of setting an optimal exploration--exploitation trade-off coefficient~$\alpha$. This supports our pragmatic choice of $\alpha=2\sqrt{2}$, effective in practice. These disparities further highlight strong heterogeneity in target concentration across websites, complementing Sec.~\ref{subsection:dataset_characteristics}.

Examining top-group tag paths often reveals clues about target-rich areas. In \emph{nc}, a typical path includes \texttt{div.container.w-iap/}
\noindent\texttt{div.iap-content}, linked to the \emph{International Activities Program}, a program focused on collecting international education statistics. In \emph{wo}, paths include \texttt{collections-sief}, related to the \emph{Strategic Impact Evaluation Fund}, a trust fund producing data and reports. Some groups directly include target-retrieval keywords, e.g., \texttt{fr-link--download} in \emph{ju} or \texttt{status-publish} in \emph{ok}. Other websites' top paths are not  human-interpretable yet still achieve high rewards (examples in~\cite{github_repository}). This shows that our agent detects useful patterns even without semantic meaning, adapting to diverse structures without prior knowledge and supporting an online, per-website learning approach.
It further demonstrates how our approach is language-independent: the RL agent learns structural patterns in HTML pages regardless of semantic meaning, enabling efficient target retrieval even on non-English or multilingual websites.

Assessing the concrete precision of our approach to retrieving SDs
requires determining the number of SDs collected in the targets.
Detecting statistics tables in heterogeneous formats is costly: even
state-of-the-art methods require roughly one second per PDF
page~\cite{soric2025benchmarkingtableextractionheterogeneous};
spreadsheets pose similar challenges~\cite{vitagliano2021detecting}.
Performing this online or at scale remains an open problem. As an empirical peak into the SD retrieval precision, we manually analyze a random sample of $7\times 40=280$ targets from 7 diverse websites in Table~\ref{tab:stats_tables}, counting the percentage of targets containing at least one statistics table (\textbf{``SD Yield''}) and their mean number per target (\textbf{``Mean \# SDs / Target''}). Results show that even on non-statistical websites (\emph{ed}, \emph{in}, \emph{oe}, \emph{wh}), an important fraction of targets contains SDs; although many targets contain none, SDs concentrate on a subset, yielding in most cases more than two SDs per collected target.

\begin{table}
	\centering
	\caption{SDs retrieval across sample targets}
	\begin{tabular}{lccccccc}
		\toprule
		& \emph{be} & \emph{ed} & \emph{is} & \emph{in} & \emph{nc} & \emph{oe} & \emph{wh} \\
		\midrule
		\bfseries SD Yield (\%) & 82 & 35 & 93 & 40 & 83 & 60 & 40 \\
		\bfseries Mean \# SDs / Target & 9.1 & 2.8 & 2.9 & 2.1 & 2.1 & 4.9 & 1.4 \\
		\bottomrule
	\end{tabular}
	\label{tab:stats_tables}
\end{table}

\begin{toappendix}
  \subsection{Effectiveness of SB Learning}

  A typical tag path containing reference to IAP in the website \emph{nc}
  is:
  \begin{quote}
    \path{/html/body/div.nces/div.container.w-iap/div.iap-content/div.row/div.col-md-6/h4/a}
  \end{quote}
  and to SIEF in the website \emph{wo}:
  \begin{quote}
  \path{/html/body/div.wp-page-body.container.default-wrapper page-collections.collections-sief/div.body-content-wrap.theme-nada-2/div.about-collection/div.repository-container/div.body/div/p/a}
  \end{quote}

  Examples of tag paths including keywords related to target retrieval are
  \begin{quote}
  \path{/html/body.path-node.page-node-type-ressource/div.dialog-off-canvas-main-canvas/div.layout-container/main#main-content/div.layout-content/div.region.region-content/div.block.block-system.block-system-main-block#block-open-theme-contenudelapageprincipale/article.mj-resource.mj-break-word.node--promoted/div.fr-container.mj-ressource__content/div.fr-grid-row.fr-grid-row--gutters.mj-content__corps/div.fr-col-lg-8.fr-col-offset-lg-2.fr-col-12.content-container__paragraph/section.fr-downloads-group.fr-downloads-group--multiple-links/ul/li/a.fr-link fr-link--download}
  \end{quote}
for \emph{ju}, and
  \begin{quote}
  \path{/html/body.home.page-template.page-template-onecolumn-page.page-template-onecolumn-page-php.page.page-id-7/main.content/div.container/div.row/div.main col-md-12/article.post.post-7.page.type-page.status-publish.hentry#post-7/div.entry-content/div#stcpDiv/div/strong/a}
  \end{quote}
for \emph{ok}.

Many other websites however have typical target-rich tag paths that cannot be interpreted by humans: for instance on \emph{jp} we have
  \begin{quote}
  \path{/html/body.top/div#wrapper/div#groval_navi/ul#groval_menu/li.menu-item-has-children/ul.sub-menu/li/a}
  \end{quote}
 on \emph{il} we have
  \begin{quote}
  \path{/html/body/div.container.s-lib-side-borders.pad-top-med#s-lg-side-nav-content/div.row/div.col-md-9/div.s-lg-tab-content/div.tab-pane.active#s-lg-guide-main/div.row.s-lg-row/div.col-md-12#s-lg-col-1/div.s-lg-col-boxes/div.s-lg-box-wrapper-17054143#s-lg-box-wrapper-17054143/div.s-lib-box-container#s-lg-box-14451639-container/div.s-lib-box.s-lib-box-std#s-lg-box-14451639/div#s-lg-box-collapse-14451639/div.s-lib-box-content/div.clearfix#s-lg-content-31285074/ul/li/a}
  \end{quote}
 and on \emph{ed} we have
  \begin{quote}
  \path{/html/body.cf-rtm/div.accueil-portail#page/div.container#main-ermes-container/main#main/div#readspeaker-container/div#portal/div.row/div.col-md-12.cms-inner-layout#layout-2/div.row/div.col-md-6.cms-inner-zone#zone-4/div#frame-224/div.frame.frame-portalcarouselwebframefactory.hidden-print/div.frame-standard.panel.panel-front.webframe-ermes-carousel/div.panel-body/div.ermes-frame-html#carousel-4AD04CE72B556A3CCF5DFCF7FB70964E/div.rsItem/div.modele_13.model-html/table/tbody/tr/td/a.vide}
  \end{quote}

\end{toappendix}

\begin{toappendix}
\subsection{URL Classification Quality}\label{subsubsection:url_classifier_results}

\begin{table}[h]
	\centering
  \vspace{-1em}
	\caption{Confusion matrix of the URL classifier, on the 11
		fully-crawled websites (average over 15 runs)}
	\vspace*{-.5em}
	\begin{tabular}{lrrr}
		\toprule
    \bfseries \diagbox{True}{Predicted} & \bfseries HTML (\%)    &
		\bfseries Target (\%)    & \bfseries Neither (\%)              \\
		\midrule
		\bfseries HTML           & 58.04                  & \01.37 & 0.00 \\
		\bfseries Target         & \00.75                 & 32.19  & 0.00 \\
		\bfseries Neither        & \05.34                 & \02.41 & 0.00 \\
		\bottomrule
	\end{tabular}
	\label{tab:cm_pcts_11_sites}
\end{table}

The confusion matrix of our URL classifier, averaged over 15 runs, for all fully-crawled websites, is presented
in Table~\ref{tab:cm_pcts_11_sites}.

We observe that classification errors are extremely marginal on ``HTML''
	and ``Target'' URLs. Never classifying as ``Neither'' leads to some
classification errors (discussed in Sec.~\ref{subsection:url_classifier}),
ultimately responsible for the difference between the number of requests
made by
\textsf{\small SB-CLASSIFIER} and \textsf{\small SB-ORACLE} on the 11 fully-crawled websites. However, since the classifier oracle is unfeasible in
practice, the performance of
\textsf{\small SB-CLASSIFIER} is satisfactory.
\end{toappendix}

\subsection{Early-Stopping}\label{sec:early_stopping}

To prevent the crawler from continuing on websites with few or no remaining targets, we add an early-stopping mechanism. Every $\nu$ iterations, we compute a slope $\sigma = \tfrac{y_t - y_{t-\nu}}{\nu}$, representing the growth rate in the number $y_t$ of targets at crawling step~$t$. We then maintain an exponential moving average $\mu = \gamma \cdot \sigma + (1-\gamma)\cdot \mu$, with $\gamma$ the decay rate. If $\mu$ stays below a threshold $\epsilon$ for $\kappa$ consecutive slopes (i.e., $\kappa \cdot \nu$ iterations), crawling stops. Parameters ($\nu=1000$, $\epsilon=0.2$, $\gamma=0.05$, $\kappa=15$) provide a good balance between avoiding premature termination and efficiently stopping on exhausted sites.

The lower part of Table~\ref{tab:all_crawlers_all_sites_nb_requests} (below the double rule) shows early-stopping results: the percentage of requests avoided and of targets missed. High values are better for the former, low for the latter. We observe three behaviors:
\begin{inparaenum}[(i)]
  \item Medium to large websites where target discovery slows and stopping occurs effectively (\emph{ab}, \emph{ed}, \emph{in}, \emph{jp}, \emph{ju}, \emph{nc}, \emph{ok}). Smaller sites take longer (in \%) to stop due to the required $\kappa \cdot \nu$ iterations;
  \item Large websites with continuous target discovery where stopping conditions are never met before manual termination (\emph{as}, \emph{ce}, \emph{il}, \emph{oe}, \emph{wh}, \emph{wo});
	\item Small sites (\emph{be}, \emph{cl}, \emph{cn}, \emph{qa}) where the crawl ends before early stopping can trigger within $\kappa \cdot \nu = 15$k iterations; this is not an issue, as these sites are quickly fully crawled.
\end{inparaenum}

\begin{toappendix}
  \subsection{Early-Stopping}
  \begin{figure}[h]
	\begin{subfigure}[t]{0.49\linewidth}
		\includegraphics[width=\linewidth]{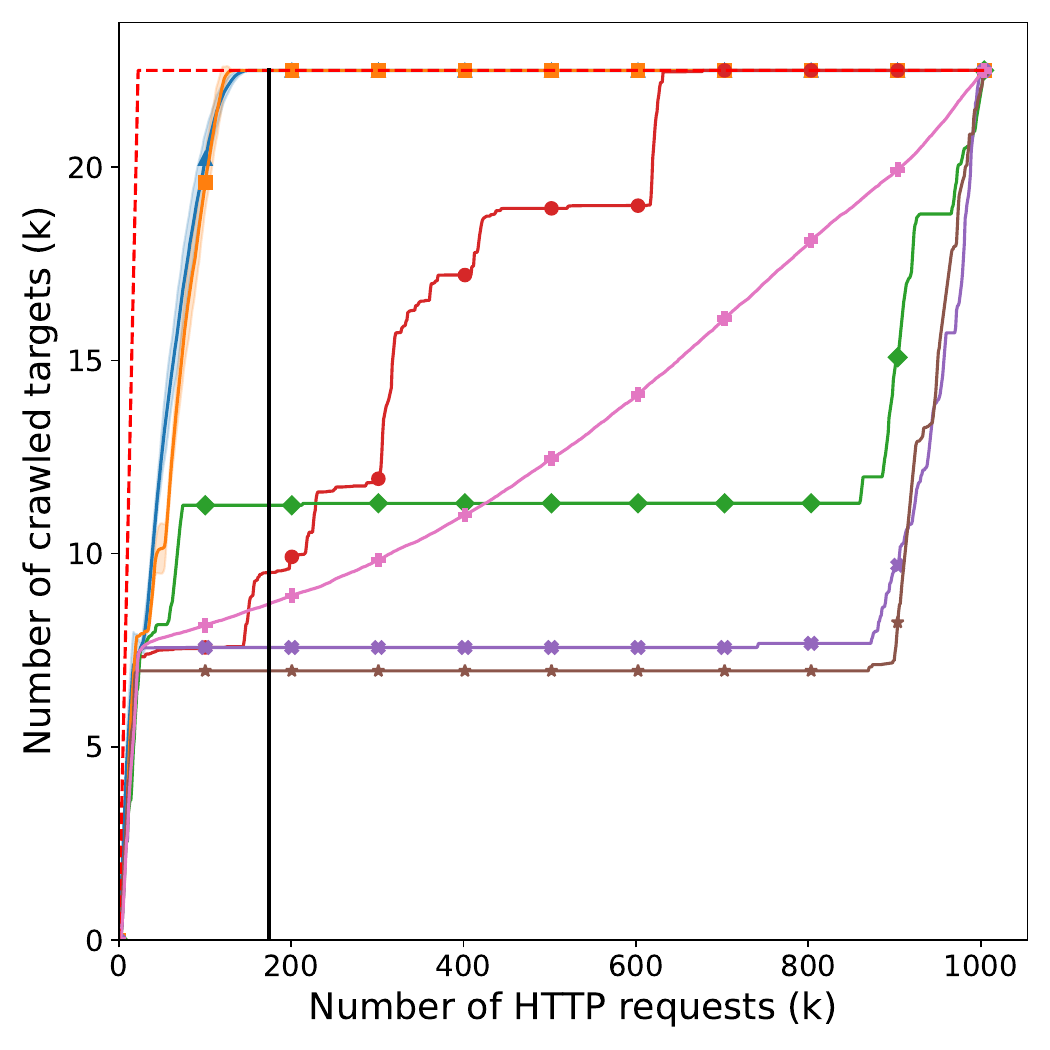}

		\vspace{-1em}
		\caption{\emph{in}}
		\label{fig:in_early_stopping}
	\end{subfigure}
	\begin{subfigure}[t]{0.49\linewidth}
		\includegraphics[width=\linewidth]{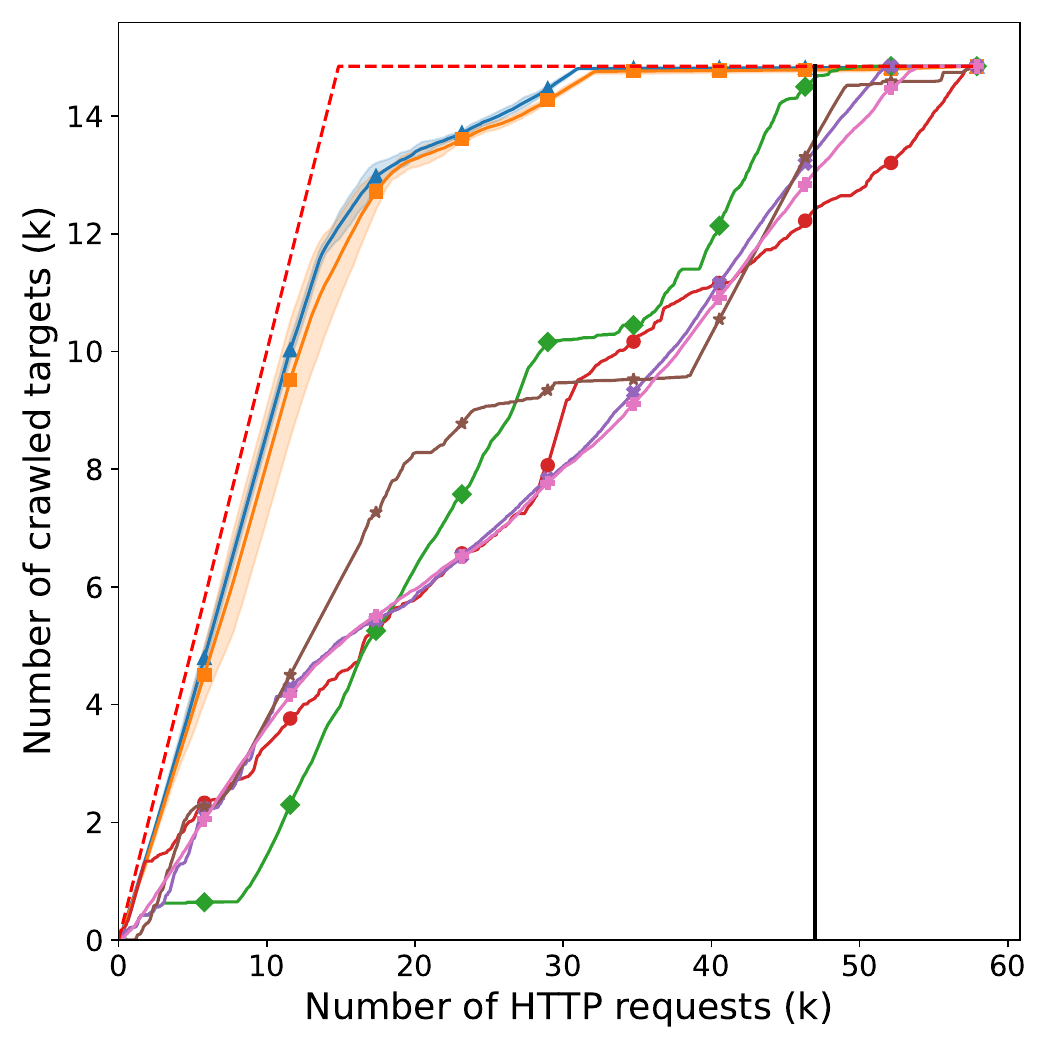}

		\vspace{-1em}
		\caption{\emph{ju}}
		\label{fig:ju_early_stopping}
	\end{subfigure}
	\vspace{-.75em}
	\caption{Visualization of early-stopping time (black rule) for websites \emph{in} and \emph{ju}}
	\label{fig:early_stopping_in_ju}
\end{figure}

	Figure~\ref{fig:early_stopping_in_ju} presents a visualization of
  the early-stopping mechanism applied to two websites: \emph{in} and
\emph{ju} (right). Both fall under the first behavior described in
Sec.~\ref{sec:early_stopping}, illustrating that smaller websites take
longer (in \%) to stop after the first signs of target convergence. For
\emph{in}, stopping occurs shortly after the point where a human would
likely have cut, but for \emph{ju}, the delay is more important. Despite
early signs of convergence, stopping is postponed due to the fixed
$\kappa \cdot \nu =15\,000$ threshold, which represents a much larger fraction of the site for \emph{ju} (60k pages) than for \emph{in} (1M pages).

\bigskip
\end{toappendix}

\section{Related Work}\label{section:related_work}
\paragraph*{Web crawlers} We discuss several aspects of Web
crawlers, see the extended version~\cite{github_repository} for other relevant work and a summary table outlining the characteristics 
 of existing focused crawlers. \begin{asparaenum} 

\item \emph{Focused crawlers} prioritize some pages during the crawl, often based on predefined \emph{topics}. Traditional focused crawlers \cite{chakrabarti1999focused,diligenti2000focused} mainly train a classifier to predict the likelihood that a hyperlink leads to a relevant page; we adapted this approach to our target retrieval in \textsf{\small FOCUSED}. Modern focused crawlers typically employ RL to adapt the crawl: we review them here. We distinguish two kinds of focused crawlers by their \emph{focus}: most are \emph{topical} crawlers \cite{gouriten2014scalable,han2018focused,zhang2021dsdd,kontogiannis2022tree}, which focus on a predefined \emph{topic} via keywords or sample documents. \emph{Non-topical} focused crawlers are rarer: besides this work (which focuses on \emph{targets} of a given file type), examples include \textsf{\small Anthelion} \cite{meusel2014focused}, which focuses on semantic annotations in Web pages, and \textsf{\small ACEBot} \cite{faheem2015adaptive}, which accumulates as much diverse textual content as possible. Anthelion's focus is more specific than ours and ACEBot's focus differs but we adapted it for target retrieval in \textsf{\small TP-OFF}. 
Topical crawlers are ill-suited for target retrieval as shown in Sec.~\ref{subsubsection:main_result}, since they target HTML content and require language-specific elements defining the topic. 
Our SD application requires crawling sites with content about different domains (economics, education, health, etc.) and in multiple languages (Sec.~\ref{subsection:dataset_characteristics}). Another key difference is that assumptions of topical locality \cite{10.1145/345508.345597} (similar-topic pages are close) and crawl direction \cite{han2018focused} (page scores increase near topic-relevant pages) do not typically hold in our case.
  Existing systems differ in the way they use learning: many choose to have
  a simple single-state representation
  \cite{gouriten2014scalable,meusel2014focused,zhang2021dsdd} in the form
  of classical MABs, while some introduce a multiple-state MDP
  representation as in \cite{han2018focused} or
  \cite{kontogiannis2022tree} which directly extends on
  \cite{han2018focused}. In order to support an MDP representation,
  \cite{han2018focused,kontogiannis2022tree} have to define states that
  cannot just be the current state of the crawl but a summary thereof (as
  RL policy optimization requires visiting many times each state). For
  this, they use features characterizing the topic relevance of the last crawled
  page and its neighborhood. This does not have a direct
  equivalent in the setting of target retrieval (again, because of
  non-locality). 
  We differ from prior work by choosing a SB representation, a mathematically well-founded way to both use a single state and model that not all actions are available at every step. It also allows faster convergence of action estimation than multiple-state approaches. Unlike our method, \cite{faheem2015adaptive} does not use RL but relies on offline training on part of the website (see Sec.~\ref{subsection:baselines}).
  In most cases, including ours, actions correspond to selecting the next URL to retrieve. Two works differ: \cite{meusel2014focused} applies RL only to select \emph{sites}, to crawl the next URL from, with URL choice handled separately, while \cite{zhang2021dsdd} additionally allows actions that query a SE.
  Finally, except for \cite{faheem2015adaptive}, our work differs from all cited approaches in targeting a focused crawl of a single given website: the most common approach crawl across sites without strict scope constraints. This distinction is crucial for our application, which requires retrieving datasets exclusively from trusted sources.

\begin{toappendix}
  \begin{table*}
    \caption{Characteristics of focused crawler works
    (LI = Language-Independent; Code = Code publicly available)}
    \label{tab:related-work}
    \vspace{-.5em}
    \begin{tabular}{rlllllcc}
      \toprule
      \textbf{System}&\textbf{Focus}&\textbf{Learning
      Model}&\textbf{States}&\textbf{Actions}&\textbf{Scope}&\textbf{LI}&\textbf{Code}\\
      \midrule
      Gouriten et al. \cite{gouriten2014scalable}&Topic&Bandit&—&URL
      selection&Across sites&\xmark&\xmark\\
      \textsf{\small{Anthelion}}
      \cite{meusel2014focused}&Semantic annot.&Bandit&---&Site
      selection&Across sites&\cmark&\cmark\\
      \textsf{\small ACEBot} \cite{faheem2015adaptive}&Text
      content&Offline&—&URL selection&Site-by-site&\cmark&\xmark\\
      Han et al. \cite{han2018focused}&Topic&MDP&Relevance features&URL
      selection&Across sites&\xmark&\xmark\\
      \textsf{\small DSDD} \cite{zhang2021dsdd}&Topic&Bandit&---&URL selection
      or SE use&Across sites&\xmark&\xmark\\
      \textsf{\small TRES} \cite{kontogiannis2022tree}&Topic&MDP&Relevance
      features&URL selection&Across sites&\xmark&\cmark\\
    \textbf{\textsf{\small FOCUSED}}&Targets&Online&---&URL
      selection&Site-by-site&\xmark&\cmark\\
    \textbf{\textsf{\small SB-CLASSIFIER}}&Targets&Sleeping Bandit&Available tag paths&URL
      selection&Site-by-site&\cmark&\cmark\\
      \bottomrule
    \end{tabular}
  \end{table*}

    \paragraph*{Web crawlers}
    Table~\ref{tab:related-work} presents the main characteristics of focused crawler works. We then discuss other relevant literature on Web crawling apart from focused and incremental crawling.
    \begin{asparaenum}
	\item \emph{Distributed} or \emph{parallel} Web crawlers use multiple
	crawlers concurrently to speed up page retrieval. Key challenges lie in resource (especially, network)
	management. \cite{cho2002parallel} introduced general architectures
	for parallel crawling, while \cite{boldi2004ubicrawler} proposed
	\textsf{\small UbiCrawler}, a decentralized and distributed system using consistent hashing for domain partitioning. Other works \cite{chau2007parallel, bosnjak2012twitterecho} focus on social network crawling. These  works aim to efficiently crawl a given set of pages, whereas we focus on minimizing that set. The two could be combined to improve crawling speed and reduce network load.
      \item \emph{Web crawlers targeting specific website types} (such as forums, blogs, or CMS) are built to leverage specific and common structural patterns within these sites. For instance, \cite{guo2006board, cai2008irobot} are focusing on forums. While effective, these crawlers are less general, relying on assumptions about site structure. In contrast, our approach adapts to each website dynamically, 
        and needs no assumptions (see Sec.~\ref{section:problem_statement}). 
	\item \emph{Hidden-} or \emph{deep-}Web crawlers explore content not reachable via standard hyperlinks, often requiring user interactions such as filling out forms. A comparative study of state-of-the-art deep-Web crawler appears in~\cite{hernandez2019deep}. In contrast, our work focuses on acquiring specific targets from official websites. In our context, form-based interfaces serve mainly as filters, which are not useful for large-scale retrieval. Moreover, on many institutional websites (see Sec.~\ref{subsection:dataset_characteristics}), such data is accessible through standard navigation, not portals. Extending our approach to deep-Web crawling remains a natural direction for future work (see Sec.~\ref{section:conclusion}).
  \end{asparaenum}
\end{toappendix}

\item \emph{Incremental} or \emph{revisit policy-based} crawling views a website as an evolving set of pages and selectively revisits them to maximize updated content while minimizing unnecessary requests. \cite{cho2000evolution} presents core challenges, and \cite{sigurdhsson2005incremental} introduces an incremental version of \textsf{\small Heritrix}~\cite{mohr2004introduction}, used by the \href{https://archive.org/}{Internet Archive}. More recent works propose learning-based revisit policies: \cite{dang2023look, avrachenkov2022online, kolobov2019optimal} predict emergence of new outlinks from page features. Others use RL: \cite{schulam2023improving} shows Thompson Sampling (TS) outperforms MAB alternatives and the \textsf{\small Pham Crawler}~\cite{pham2018learning}, and \cite{kolobov2019staying} introduces a learning algorithm to compute optimal revisit policies. While our method is a single-shot crawl, we plan to build upon it to implement incremental strategies.

\end{asparaenum}

\paragraph*{Tag paths} Tag paths, e.g., \emph{root-to-link} paths
in the DOM of each HTML page, have been exploited for Web crawling.
\cite{crescenzi2003fine, crescenzi2005clustering} take advantage of the website structure to cluster webpages, with applications to \emph{Web scrapers} aiming to automatically extract and structure data from HTML pages. \cite{faheem2015adaptive} directly uses paths in the DOM of each HTML page in order to do focused Web crawling.

\begin{toappendix}
\paragraph*{MAB algorithms} While UCB~\cite{auer2002finite} and its
variants are the state of the art for solving MAB problems, simpler approaches like $\epsilon$-greedy~\cite{sutton2018reinforcement} randomly select actions with probability $\epsilon$ and otherwise choose the highest scoring action. We also considered {Bayesian} Bandits, especially TS, a probabilistic method based on posterior distribution estimation. We excluded those approaches to ensure our crawler's {\em stability}, i.e., ability to give the same output if run several times independently of how the site can change from one crawl to
another. Also, Bayesian approaches require prior distributions, unavailable to us due to lack of prior knowledge about the websites. We also did not adopt \emph{linear} or \emph{kernel}-based bandit methods, despite their generalization capabilities. These approaches require stable, qualitative feature representations, extracted from webpages, which are difficult to define in our context. Indeed, while we hypothesize that links appearing in similar tag paths lead to similar content, this assumption does not guarantee the existence of consistent features extracted from the \emph{content} (not structure) of the pages. We would also need to choose \emph{universal} features, i.e., computable for any page of any website we want our crawler to visit. Moreover, most feature-based methods are language-dependent, which conflicts with our aim of being language-independent. Given these limitations, and the good performance we already observed, we focused on AUER~\cite{kleinberg2010regret}, the sleeping variant of UCB.
\end{toappendix}

\section{Conclusion}\label{section:conclusion}

We addressed the problem of scalable web data acquisition by developing an efficient crawling that aims to maximize the retrieval of targets (interesting pages or files) while minimizing computational resource usage in terms of time and bandwidth, and respecting crawling ethics. 
Our solution, based on RL, outperforms standard baselines on an heterogeneous set of websites.

In our future work, we would like to explore more complex reward functions than the simple number of targets reachable. 
Also, integrating deep-Web crawling techniques could enhance our crawler's ability to access data behind forms or within portals more efficiently, thereby further improving our web data acquisition at scale.
Finally, while our crawler focuses on the initial acquisition of
targets, it does not handle updates or newly published resources: an
important limitation for statistical data journalism, where timely
information is essential. We leave extending our crawler with
\emph{incremental revisits} for future work, combining the knowledge acquired by our RL-agent with existing re-crawling strategies.

\begin{acks}
  We thank Antoine Krempf (Radio~France) for suggesting the SD retrieval problem.
  This work was funded in part by the French government under management of
Agence Nationale de la  Recher-che (ANR) as part of the ``France~2030'' program,
reference ANR-23-IACL-0008 (PR[AI]RIE-PSAI) and ANR-23-IACL-005 (Hi! PARIS Cluster 2030).
\end{acks}

\section*{Artifacts}

The repository~\cite{github_repository} provides a complete experiment reproducibility kit. The \texttt{README.md} file describes the repository contents and explains how it is organized. The kit mainly allows running the crawlers \textsf{\small SB-ORACLE}, \textsf{\small SB-CLASSIFIER}, \textsf{\small FOCUSED}, \textsf{\small TP-OFF} (sometimes referred to as \verb|dom_off|), \textsf{\small BFS}, \textsf{\small DFS}, and \textsf{\small RANDOM}. Three execution modes are available:
\begin{itemize}
  \item \emph{Local crawling}: used when a website has already been fully replicated in a local database (see Sec.~\ref{sec:exp-crawl-modalities}).
  \item \emph{Online-to-local crawling}: creates a local copy of a website using a naive crawler.
  \item \emph{Semi-online crawling}: first checks the local database and fetches the resource only if it is not present (applied to non-fully crawled websites, also described in Sec.~\ref{sec:exp-crawl-modalities}).
\end{itemize}

The folder \texttt{url\_classifier\_confusion\_matrices} contains all confusions matrices used to compute the ``MR'' metric in Sec.~\ref{subsubsection:metaparameters_result} for the URL classifier models.
The repository also contains code to reproduce the plots, information about the websites used in our experiments, and an extended version of this paper. The extended version includes: the proof of Prop.~\ref{prop:np-complete}; additional details about the websites' characteristics; the MIME type list defining the targets used in the experiments; the initial keyword set for \textsf{\small TRES}; and a complete blocklist of URL extensions and MIME types. It also provides the plots of the 8 websites that could not be included in Figure~\ref{fig:10_sites_plots}; exhaustive plots from the hyper-parameter studies; extra examples of typical tag paths for 6 websites; plots and confusion matrices for the URL classifier models evaluation; additional results on URL classification quality; a visualization of the early-stopping mechanism for two websites; a table summarizing the main characteristics of related focused crawlers; and an extended discussion of other crawler types, as well as alternatives to AUER for MAB algorithms.

A second repository~\cite{tres_modified_repository} contains a fork of the \textsf{\small TRES}~\cite{kontogiannis2022tree} crawler code, adapted for the target retrieval task described in Sec.~\ref{subsection:baselines}. 

\bibliographystyle{ACM-Reference-Format}
\bibliography{main}

\newcommand{\etalchar}[1]{$^{#1}$}
\begin{thebibliography}{BOM{\etalchar{+}}12}

\bibitem[ACBF02]{auer2002finite}
Peter Auer, Nicol{\`o} Cesa-Bianchi, and Paul Fischer.
\newblock \href{
  https://link.springer.com/content/pdf/10.1023/A:1013689704352.pdf}{
  Finite-time Analysis of the Multiarmed Bandit Problem}.
\newblock {\em Machine Learning}, 47(2):235--256, 2002.

\bibitem[BCSV04]{boldi2004ubicrawler}
Paolo Boldi, Bruno Codenotti, Massimo Santini, and Sebastiano Vigna.
\newblock Ubicrawler: a scalable fully distributed web crawler.
\newblock {\em Softw. Pract. Exp.}, 34(8):711--726, 2004.

\bibitem[BOM{\etalchar{+}}12]{bosnjak2012twitterecho}
Matko Bosnjak, Eduardo Oliveira, Jos{\'{e}} Martins, Eduarda~Mendes Rodrigues,
  and Lu{\'{\i}}s Sarmento.
\newblock Twitterecho: a distributed focused crawler to support open research
  with twitter data.
\newblock In Alain Mille, Fabien Gandon, Jacques Misselis, Michael Rabinovich,
  and Steffen Staab, editors, {\em Proceedings of the 21st World Wide Web
  Conference, {WWW} 2012, Lyon, France, April 16-20, 2012 (Companion Volume)},
  pages 1233--1240. {ACM}, 2012.

\bibitem[CG02]{cho2002parallel}
Junghoo Cho and Hector Garcia{-}Molina.
\newblock Parallel crawlers.
\newblock In David Lassner, David~De Roure, and Arun Iyengar, editors, {\em
  Proceedings of the Eleventh International World Wide Web Conference, {WWW}
  2002, May 7-11, 2002, Honolulu, Hawaii, {USA}}, pages 124--135. {ACM}, 2002.

\bibitem[CPWF07]{chau2007parallel}
Duen~Horng Chau, Shashank Pandit, Samuel Wang, and Christos Faloutsos.
\newblock Parallel crawling for online social networks.
\newblock In Carey~L. Williamson, Mary~Ellen Zurko, Peter~F. Patel{-}Schneider,
  and Prashant~J. Shenoy, editors, {\em Proceedings of the 16th International
  Conference on World Wide Web, {WWW} 2007, Banff, Alberta, Canada, May 8-12,
  2007}, pages 1283--1284. {ACM}, 2007.

\bibitem[CYL{\etalchar{+}}08]{cai2008irobot}
Rui Cai, Jiang{-}Ming Yang, Wei Lai, Yida Wang, and Lei Zhang.
\newblock irobot: an intelligent crawler for web forums.
\newblock In Jinpeng Huai, Robin Chen, Hsiao{-}Wuen Hon, Yunhao Liu, Wei{-}Ying
  Ma, Andrew Tomkins, and Xiaodong Zhang, editors, {\em Proceedings of the 17th
  International Conference on World Wide Web, {WWW} 2008, Beijing, China, April
  21-25, 2008}, pages 447--456. {ACM}, 2008.

\bibitem[FS15]{faheem2015adaptive}
Muhammad Faheem and Pierre Senellart.
\newblock Adaptive web crawling through structure-based link classification.
\newblock In Robert~B. Allen, Jane Hunter, and Marcia~Lei Zeng, editors, {\em
  Digital Libraries: Providing Quality Information - 17th International
  Conference on Asia-Pacific Digital Libraries, {ICADL} 2015, Seoul, Korea,
  December 9-12, 2015, Proceedings}, volume 9469 of {\em Lecture Notes in
  Computer Science}, pages 39--51. Springer, 2015.

\bibitem[GJ79]{johnson1979computers}
M.~R. Garey and David~S. Johnson.
\newblock {\em Computers and Intractability: {A} Guide to the Theory of
  NP-Completeness}.
\newblock W. H. Freeman, 1979.

\bibitem[GLZZ06]{guo2006board}
Yan Guo, Kui Li, Kai Zhang, and Gang Zhang.
\newblock Board forum crawling: {A} web crawling method for web forum.
\newblock In {\em 2006 {IEEE} / {WIC} / {ACM} International Conference on Web
  Intelligence {(WI} 2006), 18-22 December 2006, Hong Kong, China}, pages
  745--748. {IEEE} Computer Society, 2006.

\bibitem[GMS14]{gouriten2014scalable}
Georges Gouriten, Silviu Maniu, and Pierre Senellart.
\newblock Scalable, generic, and adaptive systems for focused crawling.
\newblock In Leo Ferres, Gustavo Rossi, Virg{\'{\i}}lio A.~F. Almeida, and
  Eelco Herder, editors, {\em 25th {ACM} Conference on Hypertext and Social
  Media, {HT} '14, Santiago, Chile, September 1-4, 2014}, pages 35--45. {ACM},
  2014.

\bibitem[GMS25]{github_repository}
Antoine Gauquier, Ioana Manolescu, and Pierre Senellart.
\newblock {Efficient Crawling for Scalable Web Data Acquisition (experiment
  reproducibility kit)}.
\newblock
  \url{https://github.com/AntoineGauquier/efficient_crawler_for_scalable_web_data_acquisition/},
  2025.

\bibitem[HRR19]{hernandez2019deep}
Inma Hern{\'{a}}ndez, Carlos~R. Rivero, and David Ruiz.
\newblock Deep web crawling: a survey.
\newblock {\em World Wide Web}, 22(4):1577--1610, 2019.

\bibitem[HWS18]{han2018focused}
Miyoung Han, Pierre{-}Henri Wuillemin, and Pierre Senellart.
\newblock Focused crawling through reinforcement learning.
\newblock In Tommi Mikkonen, Ralf Klamma, and Juan Hern{\'{a}}ndez, editors,
  {\em Web Engineering - 18th International Conference, {ICWE} 2018,
  C{\'{a}}ceres, Spain, June 5-8, 2018, Proceedings}, volume 10845 of {\em
  Lecture Notes in Computer Science}, pages 261--278. Springer, 2018.

\bibitem[KKP{\etalchar{+}}21]{kontogiannis2022tree}
Andreas Kontogiannis, Dimitrios Kelesis, Vasilis Pollatos, Georgios Paliouras,
  and George Giannakopoulos.
\newblock Tree-based focused web crawling with reinforcement learning.
\newblock {\em ArXiv preprint}, abs/2112.07620, 2021.

\bibitem[KNS08]{kleinberg2010regret}
Robert~D. Kleinberg, Alexandru Niculescu{-}Mizil, and Yogeshwer Sharma.
\newblock Regret bounds for sleeping experts and bandits.
\newblock In Rocco~A. Servedio and Tong Zhang, editors, {\em 21st Annual
  Conference on Learning Theory - {COLT} 2008, Helsinki, Finland, July 9-12,
  2008}, pages 425--436. Omnipress, 2008.

\bibitem[MMB14]{meusel2014focused}
Robert Meusel, Peter Mika, and Roi Blanco.
\newblock Focused crawling for structured data.
\newblock In Jianzhong Li, Xiaoyang~Sean Wang, Minos~N. Garofalakis, Ian
  Soboroff, Torsten Suel, and Min Wang, editors, {\em Proceedings of the 23rd
  {ACM} International Conference on Conference on Information and Knowledge
  Management, {CIKM} 2014, Shanghai, China, November 3-7, 2014}, pages
  1039--1048. {ACM}, 2014.

\bibitem[SB98]{sutton2018reinforcement}
Richard~S. Sutton and Andrew~G. Barto.
\newblock {\em Reinforcement learning - an introduction}.
\newblock Adaptive computation and machine learning. {MIT} Press, 1998.

\bibitem[WW16]{watel2016greedy}
Dimitri Watel and Marc{-}Antoine Weisser.
\newblock A practical greedy approximation for the directed steiner tree
  problem.
\newblock {\em J. Comb. Optim.}, 32(4):1327--1370, 2016.

\bibitem[ZSF21]{zhang2021dsdd}
Haoxiang Zhang, A{\'{e}}cio S.~R. Santos, and Juliana Freire.
\newblock {DSDD:} domain-specific dataset discovery on the web.
\newblock In Gianluca Demartini, Guido Zuccon, J.~Shane Culpepper, Zi~Huang,
  and Hanghang Tong, editors, {\em {CIKM} '21: The 30th {ACM} International
  Conference on Information and Knowledge Management, Virtual Event,
  Queensland, Australia, November 1 - 5, 2021}, pages 2527--2536. {ACM}, 2021.

\end{thebibliography}



\begin{thebibliography}{59}


\ifx \showCODEN    \undefined \def \showCODEN     #1{\unskip}     \fi
\ifx \showDOI      \undefined \def \showDOI       #1{#1}\fi
\ifx \showISBNx    \undefined \def \showISBNx     #1{\unskip}     \fi
\ifx \showISBNxiii \undefined \def \showISBNxiii  #1{\unskip}     \fi
\ifx \showISSN     \undefined \def \showISSN      #1{\unskip}     \fi
\ifx \showLCCN     \undefined \def \showLCCN      #1{\unskip}     \fi
\ifx \shownote     \undefined \def \shownote      #1{#1}          \fi
\ifx \showarticletitle \undefined \def \showarticletitle #1{#1}   \fi
\ifx \showURL      \undefined \def \showURL       {\relax}        \fi
\providecommand\bibfield[2]{#2}
\providecommand\bibinfo[2]{#2}
\providecommand\natexlab[1]{#1}
\providecommand\showeprint[2][]{arXiv:#2}

\bibitem[\protect\citeauthoryear{Alshukri, Coenen, and Zito}{Alshukri
  et~al\mbox{.}}{2011}]%
        {alshukri2010website}
\bibfield{author}{\bibinfo{person}{Ayesh Alshukri}, \bibinfo{person}{Frans
  Coenen}, {and} \bibinfo{person}{Michele Zito}.}
  \bibinfo{year}{2011}\natexlab{}.
\newblock \showarticletitle{Incremental Web-Site Boundary Detection Using
  Random Walks}. In \bibinfo{booktitle}{\emph{Machine Learning and Data Mining
  in Pattern Recognition - 7th International Conference, {MLDM} 2011, New York,
  NY, USA, August 30 - September 3, 2011. Proceedings}}
  \emph{(\bibinfo{series}{Lecture Notes in Computer Science})},
  \bibfield{editor}{\bibinfo{person}{Petra Perner}} (Ed.),
  Vol.~\bibinfo{volume}{6871}. \bibinfo{publisher}{Springer},
  \bibinfo{pages}{414--427}.
\newblock
\urldef\tempurl%
\url{https://doi.org/10.1007/978-3-642-23199-5\_31}
\showDOI{\tempurl}


\bibitem[\protect\citeauthoryear{Arslan, Hassan, Li, and Tremayne}{Arslan
  et~al\mbox{.}}{2020}]%
        {claimbuster}
\bibfield{author}{\bibinfo{person}{Fatma Arslan}, \bibinfo{person}{Naeemul
  Hassan}, \bibinfo{person}{Chengkai Li}, {and} \bibinfo{person}{Mark
  Tremayne}.} \bibinfo{year}{2020}\natexlab{}.
\newblock \showarticletitle{A Benchmark Dataset of Check-Worthy Factual
  Claims}. In \bibinfo{booktitle}{\emph{Proceedings of the Fourteenth
  International {AAAI} Conference on Web and Social Media, {ICWSM} 2020, Held
  Virtually, Original Venue: Atlanta, Georgia, USA, June 8-11, 2020}},
  \bibfield{editor}{\bibinfo{person}{Munmun~De Choudhury},
  \bibinfo{person}{Rumi Chunara}, \bibinfo{person}{Aron Culotta}, {and}
  \bibinfo{person}{Brooke~Foucault Welles}} (Eds.). \bibinfo{publisher}{{AAAI}
  Press}, \bibinfo{pages}{821--829}.
\newblock
\urldef\tempurl%
\url{https://ojs.aaai.org/index.php/ICWSM/article/view/7346}
\showURL{%
\tempurl}


\bibitem[\protect\citeauthoryear{Auer, Cesa-Bianchi, and Fischer}{Auer
  et~al\mbox{.}}{2002}]%
        {auer2002finite}
\bibfield{author}{\bibinfo{person}{Peter Auer}, \bibinfo{person}{Nicol{\`o}
  Cesa-Bianchi}, {and} \bibinfo{person}{Paul Fischer}.}
  \bibinfo{year}{2002}\natexlab{}.
\newblock \showarticletitle{\href{
  https://link.springer.com/content/pdf/10.1023/A:1013689704352.pdf}{
  Finite-time Analysis of the Multiarmed Bandit Problem}}.
\newblock \bibinfo{journal}{\emph{Machine Learning}} \bibinfo{volume}{47},
  \bibinfo{number}{2} (\bibinfo{year}{2002}), \bibinfo{pages}{235--256}.
\newblock


\bibitem[\protect\citeauthoryear{Auer, Bizer, Kobilarov, Lehmann, Cyganiak, and
  Ives}{Auer et~al\mbox{.}}{2007}]%
        {DBLP:conf/semweb/AuerBKLCI07}
\bibfield{author}{\bibinfo{person}{S{\"{o}}ren Auer},
  \bibinfo{person}{Christian Bizer}, \bibinfo{person}{Georgi Kobilarov},
  \bibinfo{person}{Jens Lehmann}, \bibinfo{person}{Richard Cyganiak}, {and}
  \bibinfo{person}{Zachary~G. Ives}.} \bibinfo{year}{2007}\natexlab{}.
\newblock \showarticletitle{DBpedia: {A} Nucleus for a Web of Open Data}. In
  \bibinfo{booktitle}{\emph{The Semantic Web, 6th International Semantic Web
  Conference, 2nd Asian Semantic Web Conference, {ISWC} 2007 + {ASWC} 2007,
  Busan, Korea, November 11-15, 2007}} \emph{(\bibinfo{series}{Lecture Notes in
  Computer Science})}, \bibfield{editor}{\bibinfo{person}{Karl Aberer},
  \bibinfo{person}{Key{-}Sun Choi}, \bibinfo{person}{Natasha~Fridman Noy},
  \bibinfo{person}{Dean Allemang}, \bibinfo{person}{Kyung{-}Il Lee},
  \bibinfo{person}{Lyndon J.~B. Nixon}, \bibinfo{person}{Jennifer Golbeck},
  \bibinfo{person}{Peter Mika}, \bibinfo{person}{Diana Maynard},
  \bibinfo{person}{Riichiro Mizoguchi}, \bibinfo{person}{Guus Schreiber}, {and}
  \bibinfo{person}{Philippe Cudr{\'{e}}{-}Mauroux}} (Eds.),
  Vol.~\bibinfo{volume}{4825}. \bibinfo{publisher}{Springer},
  \bibinfo{pages}{722--735}.
\newblock
\urldef\tempurl%
\url{https://doi.org/10.1007/978-3-540-76298-0\_52}
\showDOI{\tempurl}


\bibitem[\protect\citeauthoryear{Avrachenkov, Patil, and Thoppe}{Avrachenkov
  et~al\mbox{.}}{2022}]%
        {avrachenkov2022online}
\bibfield{author}{\bibinfo{person}{Konstantin Avrachenkov},
  \bibinfo{person}{Kishor Patil}, {and} \bibinfo{person}{Gugan Thoppe}.}
  \bibinfo{year}{2022}\natexlab{}.
\newblock \showarticletitle{Online algorithms for estimating change rates of
  web pages}.
\newblock \bibinfo{journal}{\emph{Perform. Evaluation}}  \bibinfo{volume}{153}
  (\bibinfo{year}{2022}), \bibinfo{pages}{102261}.
\newblock
\urldef\tempurl%
\url{https://doi.org/10.1016/J.PEVA.2021.102261}
\showDOI{\tempurl}


\bibitem[\protect\citeauthoryear{Balalau, Ebel, Galhardas, Galizzi, and
  Manolescu}{Balalau et~al\mbox{.}}{2024}]%
        {balalau2024star}
\bibfield{author}{\bibinfo{person}{Oana Balalau}, \bibinfo{person}{Simon Ebel},
  \bibinfo{person}{Helena Galhardas}, \bibinfo{person}{Th{\'{e}}o Galizzi},
  {and} \bibinfo{person}{Ioana Manolescu}.} \bibinfo{year}{2024}\natexlab{}.
\newblock \showarticletitle{STaR: Space and Time-aware Statistic Query
  Answering}. In \bibinfo{booktitle}{\emph{Proceedings of the 33rd {ACM}
  International Conference on Information and Knowledge Management, {CIKM}
  2024, Boise, ID, USA, October 21-25, 2024}},
  \bibfield{editor}{\bibinfo{person}{Edoardo Serra} {and}
  \bibinfo{person}{Francesca Spezzano}} (Eds.). \bibinfo{publisher}{{ACM}},
  \bibinfo{pages}{5190--5194}.
\newblock
\urldef\tempurl%
\url{https://doi.org/10.1145/3627673.3679209}
\showDOI{\tempurl}


\bibitem[\protect\citeauthoryear{Balalau, Ebel, Galizzi, Manolescu, Massonnat,
  Deiana, Gautreau, Krempf, Pontillon, Roux, and Yakin}{Balalau
  et~al\mbox{.}}{2022}]%
        {DBLP:conf/cikm/BalalauEGMMDGKP22}
\bibfield{author}{\bibinfo{person}{Oana Balalau}, \bibinfo{person}{Simon Ebel},
  \bibinfo{person}{Th{\'{e}}o Galizzi}, \bibinfo{person}{Ioana Manolescu},
  \bibinfo{person}{Quentin Massonnat}, \bibinfo{person}{Antoine Deiana},
  \bibinfo{person}{Emilie Gautreau}, \bibinfo{person}{Antoine Krempf},
  \bibinfo{person}{Thomas Pontillon}, \bibinfo{person}{G{\'{e}}rald Roux},
  {and} \bibinfo{person}{Joanna Yakin}.} \bibinfo{year}{2022}\natexlab{}.
\newblock \showarticletitle{Statistical Claim Checking: StatCheck in Action}.
  In \bibinfo{booktitle}{\emph{Proceedings of the 31st {ACM} International
  Conference on Information {\&} Knowledge Management, Atlanta, GA, USA,
  October 17-21, 2022}}, \bibfield{editor}{\bibinfo{person}{Mohammad~Al Hasan}
  {and} \bibinfo{person}{Li~Xiong}} (Eds.). \bibinfo{publisher}{{ACM}},
  \bibinfo{pages}{4798--4802}.
\newblock
\urldef\tempurl%
\url{https://doi.org/10.1145/3511808.3557198}
\showDOI{\tempurl}


\bibitem[\protect\citeauthoryear{Bottou}{Bottou}{2010}]%
        {bottou2010large}
\bibfield{author}{\bibinfo{person}{L{\'{e}}on Bottou}.}
  \bibinfo{year}{2010}\natexlab{}.
\newblock \showarticletitle{Large-Scale Machine Learning with Stochastic
  Gradient Descent}. In \bibinfo{booktitle}{\emph{19th International Conference
  on Computational Statistics, {COMPSTAT} 2010, Paris, France, August 22-27,
  2010 - Keynote, Invited and Contributed Papers}},
  \bibfield{editor}{\bibinfo{person}{Yves Lechevallier} {and}
  \bibinfo{person}{Gilbert Saporta}} (Eds.).
  \bibinfo{publisher}{Physica-Verlag}, \bibinfo{pages}{177--186}.
\newblock
\urldef\tempurl%
\url{https://doi.org/10.1007/978-3-7908-2604-3\_16}
\showDOI{\tempurl}


\bibitem[\protect\citeauthoryear{Cao, Manolescu, and Tannier}{Cao
  et~al\mbox{.}}{2018}]%
        {DBLP:conf/webdb/CaoMT18}
\bibfield{author}{\bibinfo{person}{Tien~Duc Cao}, \bibinfo{person}{Ioana
  Manolescu}, {and} \bibinfo{person}{Xavier Tannier}.}
  \bibinfo{year}{2018}\natexlab{}.
\newblock \showarticletitle{Searching for Truth in a Database of Statistics}.
  In \bibinfo{booktitle}{\emph{Proceedings of the 21st International Workshop
  on the Web and Databases, Houston, TX, USA, June 10, 2018}}.
  \bibinfo{publisher}{{ACM}}, \bibinfo{pages}{4:1--4:6}.
\newblock
\urldef\tempurl%
\url{https://doi.org/10.1145/3201463.3201467}
\showDOI{\tempurl}


\bibitem[\protect\citeauthoryear{Chakrabarti, van~den Berg, and
  Dom}{Chakrabarti et~al\mbox{.}}{1999}]%
        {chakrabarti1999focused}
\bibfield{author}{\bibinfo{person}{Soumen Chakrabarti}, \bibinfo{person}{Martin
  van~den Berg}, {and} \bibinfo{person}{Byron Dom}.}
  \bibinfo{year}{1999}\natexlab{}.
\newblock \showarticletitle{Focused Crawling: {A} New Approach to
  Topic-Specific Web Resource Discovery}.
\newblock \bibinfo{journal}{\emph{Comput. Networks}} \bibinfo{volume}{31},
  \bibinfo{number}{11-16} (\bibinfo{year}{1999}), \bibinfo{pages}{1623--1640}.
\newblock
\urldef\tempurl%
\url{https://doi.org/10.1016/S1389-1286(99)00052-3}
\showDOI{\tempurl}


\bibitem[\protect\citeauthoryear{Cho and Garcia{-}Molina}{Cho and
  Garcia{-}Molina}{2000}]%
        {cho2000evolution}
\bibfield{author}{\bibinfo{person}{Junghoo Cho} {and} \bibinfo{person}{Hector
  Garcia{-}Molina}.} \bibinfo{year}{2000}\natexlab{}.
\newblock \showarticletitle{The Evolution of the Web and Implications for an
  Incremental Crawler}. In \bibinfo{booktitle}{\emph{{VLDB} 2000, Proceedings
  of 26th International Conference on Very Large Data Bases, September 10-14,
  2000, Cairo, Egypt}}, \bibfield{editor}{\bibinfo{person}{Amr~El Abbadi},
  \bibinfo{person}{Michael~L. Brodie}, \bibinfo{person}{Sharma Chakravarthy},
  \bibinfo{person}{Umeshwar Dayal}, \bibinfo{person}{Nabil Kamel},
  \bibinfo{person}{Gunter Schlageter}, {and} \bibinfo{person}{Kyu{-}Young
  Whang}} (Eds.). \bibinfo{publisher}{Morgan Kaufmann},
  \bibinfo{pages}{200--209}.
\newblock
\urldef\tempurl%
\url{http://www.vldb.org/conf/2000/P200.pdf}
\showURL{%
\tempurl}


\bibitem[\protect\citeauthoryear{Christensen, Leventidis, Lissandrini, Rocco,
  Miller, and Hose}{Christensen et~al\mbox{.}}{2025}]%
        {christensen_fantastic_2025}
\bibfield{author}{\bibinfo{person}{Martin~Pek{\'{a}}r Christensen},
  \bibinfo{person}{Aristotelis Leventidis}, \bibinfo{person}{Matteo
  Lissandrini}, \bibinfo{person}{Laura~Di Rocco},
  \bibinfo{person}{Ren{\'{e}}e~J. Miller}, {and} \bibinfo{person}{Katja Hose}.}
  \bibinfo{year}{2025}\natexlab{}.
\newblock \showarticletitle{Fantastic Tables and Where to Find Them: Table
  Search in Semantic Data Lakes}. In \bibinfo{booktitle}{\emph{Proceedings 28th
  International Conference on Extending Database Technology, {EDBT} 2025,
  Barcelona, Spain, March 25-28, 2025}},
  \bibfield{editor}{\bibinfo{person}{Alkis Simitsis}, \bibinfo{person}{Bettina
  Kemme}, \bibinfo{person}{Anna Queralt}, \bibinfo{person}{Oscar Romero}, {and}
  \bibinfo{person}{Petar Jovanovic}} (Eds.).
  \bibinfo{publisher}{OpenProceedings.org}, \bibinfo{pages}{397--410}.
\newblock
\urldef\tempurl%
\url{https://doi.org/10.48786/EDBT.2025.32}
\showDOI{\tempurl}


\bibitem[\protect\citeauthoryear{Christmann, Roy, and Weikum}{Christmann
  et~al\mbox{.}}{2024}]%
        {Christmann2023CompMixAB}
\bibfield{author}{\bibinfo{person}{Philipp Christmann},
  \bibinfo{person}{Rishiraj~Saha Roy}, {and} \bibinfo{person}{Gerhard Weikum}.}
  \bibinfo{year}{2024}\natexlab{}.
\newblock \showarticletitle{CompMix: {A} Benchmark for Heterogeneous Question
  Answering}. In \bibinfo{booktitle}{\emph{Companion Proceedings of the {ACM}
  on Web Conference 2024, {WWW} 2024, Singapore, Singapore, May 13-17, 2024}},
  \bibfield{editor}{\bibinfo{person}{Tat{-}Seng Chua},
  \bibinfo{person}{Chong{-}Wah Ngo}, \bibinfo{person}{Roy~Ka{-}Wei Lee},
  \bibinfo{person}{Ravi Kumar}, {and} \bibinfo{person}{Hady~W. Lauw}} (Eds.).
  \bibinfo{publisher}{{ACM}}, \bibinfo{pages}{1091--1094}.
\newblock
\urldef\tempurl%
\url{https://doi.org/10.1145/3589335.3651444}
\showDOI{\tempurl}


\bibitem[\protect\citeauthoryear{Crescenzi, Merialdo, and Missier}{Crescenzi
  et~al\mbox{.}}{2003}]%
        {crescenzi2003fine}
\bibfield{author}{\bibinfo{person}{Valter Crescenzi}, \bibinfo{person}{Paolo
  Merialdo}, {and} \bibinfo{person}{Paolo Missier}.}
  \bibinfo{year}{2003}\natexlab{}.
\newblock \showarticletitle{Fine-grain web site structure discovery}. In
  \bibinfo{booktitle}{\emph{Fifth {ACM} {CIKM} International Workshop on Web
  Information and Data Management {(WIDM} 2003), New Orleans, Louisiana, USA,
  November 7-8, 2003}}, \bibfield{editor}{\bibinfo{person}{Roger H.~L. Chiang},
  \bibinfo{person}{Alberto H.~F. Laender}, {and} \bibinfo{person}{Ee{-}Peng
  Lim}} (Eds.). \bibinfo{publisher}{{ACM}}, \bibinfo{pages}{15--22}.
\newblock
\urldef\tempurl%
\url{https://doi.org/10.1145/956699.956703}
\showDOI{\tempurl}


\bibitem[\protect\citeauthoryear{Crescenzi, Merialdo, and Missier}{Crescenzi
  et~al\mbox{.}}{2005}]%
        {crescenzi2005clustering}
\bibfield{author}{\bibinfo{person}{Valter Crescenzi}, \bibinfo{person}{Paolo
  Merialdo}, {and} \bibinfo{person}{Paolo Missier}.}
  \bibinfo{year}{2005}\natexlab{}.
\newblock \showarticletitle{Clustering Web pages based on their structure}.
\newblock \bibinfo{journal}{\emph{Data Knowl. Eng.}} \bibinfo{volume}{54},
  \bibinfo{number}{3} (\bibinfo{year}{2005}), \bibinfo{pages}{279--299}.
\newblock
\urldef\tempurl%
\url{https://doi.org/10.1016/J.DATAK.2004.11.004}
\showDOI{\tempurl}


\bibitem[\protect\citeauthoryear{Dang, Bucur, Atil, Pitel, Ruis, Kadkhodaei,
  and Litvak}{Dang et~al\mbox{.}}{2023}]%
        {dang2023look}
\bibfield{author}{\bibinfo{person}{Thi Kim~Nhung Dang}, \bibinfo{person}{Doina
  Bucur}, \bibinfo{person}{Berk Atil}, \bibinfo{person}{Guillaume Pitel},
  \bibinfo{person}{Frank Ruis}, \bibinfo{person}{Hamid~Reza Kadkhodaei}, {and}
  \bibinfo{person}{Nelly Litvak}.} \bibinfo{year}{2023}\natexlab{}.
\newblock \showarticletitle{Look back, look around: {A} systematic analysis of
  effective predictors for new outlinks in focused Web crawling}.
\newblock \bibinfo{journal}{\emph{Knowl. Based Syst.}}  \bibinfo{volume}{260}
  (\bibinfo{year}{2023}), \bibinfo{pages}{110126}.
\newblock
\urldef\tempurl%
\url{https://doi.org/10.1016/J.KNOSYS.2022.110126}
\showDOI{\tempurl}


\bibitem[\protect\citeauthoryear{Davison}{Davison}{2000}]%
        {10.1145/345508.345597}
\bibfield{author}{\bibinfo{person}{Brian~D. Davison}.}
  \bibinfo{year}{2000}\natexlab{}.
\newblock \showarticletitle{Topical locality in the Web}. In
  \bibinfo{booktitle}{\emph{{SIGIR} 2000: Proceedings of the 23rd Annual
  International {ACM} {SIGIR} Conference on Research and Development in
  Information Retrieval, July 24-28, 2000, Athens, Greece}},
  \bibfield{editor}{\bibinfo{person}{Emmanuel~J. Yannakoudakis},
  \bibinfo{person}{Nicholas~J. Belkin}, \bibinfo{person}{Peter Ingwersen},
  {and} \bibinfo{person}{Mun{-}Kew Leong}} (Eds.). \bibinfo{publisher}{{ACM}},
  \bibinfo{pages}{272--279}.
\newblock
\urldef\tempurl%
\url{https://doi.org/10.1145/345508.345597}
\showDOI{\tempurl}


\bibitem[\protect\citeauthoryear{Deng, Chai, Cao, Yuan, Chen, Yu, Sun, Wang,
  Li, Cao, Jin, Zhang, Jiang, Zhang, Wang, Yuan, Wang, and Tang}{Deng
  et~al\mbox{.}}{2024}]%
        {deng_lakebench_2024}
\bibfield{author}{\bibinfo{person}{Yuhao Deng}, \bibinfo{person}{Chengliang
  Chai}, \bibinfo{person}{Lei Cao}, \bibinfo{person}{Qin Yuan},
  \bibinfo{person}{Siyuan Chen}, \bibinfo{person}{Yanrui Yu},
  \bibinfo{person}{Zhaoze Sun}, \bibinfo{person}{Junyi Wang},
  \bibinfo{person}{Jiajun Li}, \bibinfo{person}{Ziqi Cao},
  \bibinfo{person}{Kaisen Jin}, \bibinfo{person}{Chi Zhang},
  \bibinfo{person}{Yuqing Jiang}, \bibinfo{person}{Yuanfang Zhang},
  \bibinfo{person}{Yuping Wang}, \bibinfo{person}{Ye Yuan},
  \bibinfo{person}{Guoren Wang}, {and} \bibinfo{person}{Nan Tang}.}
  \bibinfo{year}{2024}\natexlab{}.
\newblock \showarticletitle{LakeBench: {A} Benchmark for Discovering Joinable
  and Unionable Tables in Data Lakes}.
\newblock \bibinfo{journal}{\emph{Proc. {VLDB} Endow.}} \bibinfo{volume}{17},
  \bibinfo{number}{8} (\bibinfo{year}{2024}), \bibinfo{pages}{1925--1938}.
\newblock
\urldef\tempurl%
\url{https://doi.org/10.14778/3659437.3659448}
\showDOI{\tempurl}


\bibitem[\protect\citeauthoryear{Diligenti, Coetzee, Lawrence, Giles, Gori,
  et~al\mbox{.}}{Diligenti et~al\mbox{.}}{2000}]%
        {diligenti2000focused}
\bibfield{author}{\bibinfo{person}{Michelangelo Diligenti},
  \bibinfo{person}{Frans Coetzee}, \bibinfo{person}{Steve Lawrence},
  \bibinfo{person}{C~Lee Giles}, \bibinfo{person}{Marco Gori}, {et~al\mbox{.}}}
  \bibinfo{year}{2000}\natexlab{}.
\newblock \showarticletitle{{Focused Crawling Using Context Graphs}}. In
  \bibinfo{booktitle}{\emph{VLDB}}. \bibinfo{pages}{527--534}.
\newblock


\bibitem[\protect\citeauthoryear{Faheem and Senellart}{Faheem and
  Senellart}{2015}]%
        {faheem2015adaptive}
\bibfield{author}{\bibinfo{person}{Muhammad Faheem} {and}
  \bibinfo{person}{Pierre Senellart}.} \bibinfo{year}{2015}\natexlab{}.
\newblock \showarticletitle{Adaptive Web Crawling Through Structure-Based Link
  Classification}. In \bibinfo{booktitle}{\emph{Digital Libraries: Providing
  Quality Information - 17th International Conference on Asia-Pacific Digital
  Libraries, {ICADL} 2015, Seoul, Korea, December 9-12, 2015, Proceedings}}
  \emph{(\bibinfo{series}{Lecture Notes in Computer Science})},
  \bibfield{editor}{\bibinfo{person}{Robert~B. Allen}, \bibinfo{person}{Jane
  Hunter}, {and} \bibinfo{person}{Marcia~Lei Zeng}} (Eds.),
  Vol.~\bibinfo{volume}{9469}. \bibinfo{publisher}{Springer},
  \bibinfo{pages}{39--51}.
\newblock
\urldef\tempurl%
\url{https://doi.org/10.1007/978-3-319-27974-9\_5}
\showDOI{\tempurl}


\bibitem[\protect\citeauthoryear{Fan, Wang, Li, Zhang, and Miller}{Fan
  et~al\mbox{.}}{2023}]%
        {fan_semantics-aware_2023}
\bibfield{author}{\bibinfo{person}{Grace Fan}, \bibinfo{person}{Jin Wang},
  \bibinfo{person}{Yuliang Li}, \bibinfo{person}{Dan Zhang}, {and}
  \bibinfo{person}{Ren{\'{e}}e~J. Miller}.} \bibinfo{year}{2023}\natexlab{}.
\newblock \showarticletitle{Semantics-aware Dataset Discovery from Data Lakes
  with Contextualized Column-based Representation Learning}.
\newblock \bibinfo{journal}{\emph{Proc. {VLDB} Endow.}} \bibinfo{volume}{16},
  \bibinfo{number}{7} (\bibinfo{year}{2023}), \bibinfo{pages}{1726--1739}.
\newblock
\urldef\tempurl%
\url{https://doi.org/10.14778/3587136.3587146}
\showDOI{\tempurl}


\bibitem[\protect\citeauthoryear{Furche, Grasso, Kravchenko, and
  Schallhart}{Furche et~al\mbox{.}}{2012}]%
        {furche2012turn}
\bibfield{author}{\bibinfo{person}{Tim Furche}, \bibinfo{person}{Giovanni
  Grasso}, \bibinfo{person}{Andrey Kravchenko}, {and}
  \bibinfo{person}{Christian Schallhart}.} \bibinfo{year}{2012}\natexlab{}.
\newblock \showarticletitle{Turn the Page: Automated Traversal of Paginated
  Websites}. In \bibinfo{booktitle}{\emph{Web Engineering - 12th International
  Conference, {ICWE} 2012, Berlin, Germany, July 23-27, 2012. Proceedings}}
  \emph{(\bibinfo{series}{Lecture Notes in Computer Science})},
  \bibfield{editor}{\bibinfo{person}{Marco Brambilla},
  \bibinfo{person}{Takehiro Tokuda}, {and} \bibinfo{person}{Robert Tolksdorf}}
  (Eds.), Vol.~\bibinfo{volume}{7387}. \bibinfo{publisher}{Springer},
  \bibinfo{pages}{332--346}.
\newblock
\urldef\tempurl%
\url{https://doi.org/10.1007/978-3-642-31753-8\_27}
\showDOI{\tempurl}


\bibitem[\protect\citeauthoryear{Garey and Johnson}{Garey and Johnson}{1979}]%
        {johnson1979computers2}
\bibfield{author}{\bibinfo{person}{M.~R. Garey} {and} \bibinfo{person}{David~S.
  Johnson}.} \bibinfo{year}{1979}\natexlab{}.
\newblock \bibinfo{booktitle}{\emph{Computers and Intractability: {A} Guide to
  the Theory of NP-Completeness}}.
\newblock \bibinfo{publisher}{W. H. Freeman}.
\newblock
\showISBNx{0-7167-1044-7}


\bibitem[\protect\citeauthoryear{Gauquier, Manolescu, and Senellart}{Gauquier
  et~al\mbox{.}}{2025a}]%
        {github_repository}
\bibfield{author}{\bibinfo{person}{Antoine Gauquier}, \bibinfo{person}{Ioana
  Manolescu}, {and} \bibinfo{person}{Pierre Senellart}.}
  \bibinfo{year}{2025}\natexlab{a}.
\newblock \bibinfo{title}{{Efficient Crawling for Scalable Web Data Acquisition
  (experiment reproducibility kit)}}.
\newblock
  \bibinfo{howpublished}{\url{https://github.com/AntoineGauquier/efficient_crawler_for_scalable_web_data_acquisition/}}.
\newblock


\bibitem[\protect\citeauthoryear{Gauquier, Manolescu, and Senellart}{Gauquier
  et~al\mbox{.}}{2025b}]%
        {tres_modified_repository}
\bibfield{author}{\bibinfo{person}{Antoine Gauquier}, \bibinfo{person}{Ioana
  Manolescu}, {and} \bibinfo{person}{Pierre Senellart}.}
  \bibinfo{year}{2025}\natexlab{b}.
\newblock \bibinfo{title}{{TRES System Adapted to the Target Retrieval Task}}.
\newblock
  \bibinfo{howpublished}{\url{hhttps://github.com/AntoineGauquier/tres_modified_for_target_retrieval_non_anonymous}}.
\newblock


\bibitem[\protect\citeauthoryear{Gauquier, Manolescu, and Senellart}{Gauquier
  et~al\mbox{.}}{2026}]%
        {gauquier2026efficient2}
\bibfield{author}{\bibinfo{person}{Antoine Gauquier}, \bibinfo{person}{Ioana
  Manolescu}, {and} \bibinfo{person}{Pierre Senellart}.}
  \bibinfo{year}{2026}\natexlab{}.
\newblock \showarticletitle{{Efficient Crawling for Scalable Web Data
  Acquisition}}. In \bibinfo{booktitle}{\emph{Proceedings of the 29th
  International Conference on Extending Database Technology, {EDBT} 2026}}.
\newblock


\bibitem[\protect\citeauthoryear{Gouriten, Maniu, and Senellart}{Gouriten
  et~al\mbox{.}}{2014}]%
        {gouriten2014scalable}
\bibfield{author}{\bibinfo{person}{Georges Gouriten}, \bibinfo{person}{Silviu
  Maniu}, {and} \bibinfo{person}{Pierre Senellart}.}
  \bibinfo{year}{2014}\natexlab{}.
\newblock \showarticletitle{Scalable, generic, and adaptive systems for focused
  crawling}. In \bibinfo{booktitle}{\emph{25th {ACM} Conference on Hypertext
  and Social Media, {HT} '14, Santiago, Chile, September 1-4, 2014}},
  \bibfield{editor}{\bibinfo{person}{Leo Ferres}, \bibinfo{person}{Gustavo
  Rossi}, \bibinfo{person}{Virg{\'{\i}}lio A.~F. Almeida}, {and}
  \bibinfo{person}{Eelco Herder}} (Eds.). \bibinfo{publisher}{{ACM}},
  \bibinfo{pages}{35--45}.
\newblock
\urldef\tempurl%
\url{https://doi.org/10.1145/2631775.2631795}
\showDOI{\tempurl}


\bibitem[\protect\citeauthoryear{Guan, Annaswamy, and Tseng}{Guan
  et~al\mbox{.}}{2020}]%
        {howard1960dynamic}
\bibfield{author}{\bibinfo{person}{Yue Guan}, \bibinfo{person}{Anuradha~M.
  Annaswamy}, {and} \bibinfo{person}{H.~Eric Tseng}.}
  \bibinfo{year}{2020}\natexlab{}.
\newblock \showarticletitle{Towards Dynamic Pricing for Shared Mobility on
  Demand using Markov Decision Processes and Dynamic Programming}. In
  \bibinfo{booktitle}{\emph{23rd {IEEE} International Conference on Intelligent
  Transportation Systems, {ITSC} 2020, Rhodes, Greece, September 20-23, 2020}}.
  \bibinfo{publisher}{{IEEE}}, \bibinfo{pages}{1--7}.
\newblock
\urldef\tempurl%
\url{https://doi.org/10.1109/ITSC45102.2020.9294685}
\showDOI{\tempurl}


\bibitem[\protect\citeauthoryear{Han, Wuillemin, and Senellart}{Han
  et~al\mbox{.}}{2018}]%
        {han2018focused}
\bibfield{author}{\bibinfo{person}{Miyoung Han},
  \bibinfo{person}{Pierre{-}Henri Wuillemin}, {and} \bibinfo{person}{Pierre
  Senellart}.} \bibinfo{year}{2018}\natexlab{}.
\newblock \showarticletitle{Focused Crawling Through Reinforcement Learning}.
  In \bibinfo{booktitle}{\emph{Web Engineering - 18th International Conference,
  {ICWE} 2018, C{\'{a}}ceres, Spain, June 5-8, 2018, Proceedings}}
  \emph{(\bibinfo{series}{Lecture Notes in Computer Science})},
  \bibfield{editor}{\bibinfo{person}{Tommi Mikkonen}, \bibinfo{person}{Ralf
  Klamma}, {and} \bibinfo{person}{Juan Hern{\'{a}}ndez}} (Eds.),
  Vol.~\bibinfo{volume}{10845}. \bibinfo{publisher}{Springer},
  \bibinfo{pages}{261--278}.
\newblock
\urldef\tempurl%
\url{https://doi.org/10.1007/978-3-319-91662-0\_20}
\showDOI{\tempurl}


\bibitem[\protect\citeauthoryear{Herzig, M{\"u}ller, Krichene, and
  Eisenschlos}{Herzig et~al\mbox{.}}{2021}]%
        {herzig-etal-2021-open}
\bibfield{author}{\bibinfo{person}{Jonathan Herzig}, \bibinfo{person}{Thomas
  M{\"u}ller}, \bibinfo{person}{Syrine Krichene}, {and} \bibinfo{person}{Julian
  Eisenschlos}.} \bibinfo{year}{2021}\natexlab{}.
\newblock \showarticletitle{Open Domain Question Answering over Tables via
  Dense Retrieval}. In \bibinfo{booktitle}{\emph{Proceedings of the 2021
  Conference of the North American Chapter of the Association for Computational
  Linguistics: Human Language Technologies}},
  \bibfield{editor}{\bibinfo{person}{Kristina Toutanova}, \bibinfo{person}{Anna
  Rumshisky}, \bibinfo{person}{Luke Zettlemoyer}, \bibinfo{person}{Dilek
  Hakkani-Tur}, \bibinfo{person}{Iz~Beltagy}, \bibinfo{person}{Steven Bethard},
  \bibinfo{person}{Ryan Cotterell}, \bibinfo{person}{Tanmoy Chakraborty}, {and}
  \bibinfo{person}{Yichao Zhou}} (Eds.). \bibinfo{publisher}{Association for
  Computational Linguistics}, \bibinfo{address}{Online},
  \bibinfo{pages}{512--519}.
\newblock
\urldef\tempurl%
\url{https://doi.org/10.18653/v1/2021.naacl-main.43}
\showDOI{\tempurl}


\bibitem[\protect\citeauthoryear{Hulsebos, Lin, Shankar, and
  Parameswaran}{Hulsebos et~al\mbox{.}}{2024}]%
        {Hulsebos2024ItTL}
\bibfield{author}{\bibinfo{person}{Madelon Hulsebos}, \bibinfo{person}{Wenjing
  Lin}, \bibinfo{person}{Shreya Shankar}, {and} \bibinfo{person}{Aditya~G.
  Parameswaran}.} \bibinfo{year}{2024}\natexlab{}.
\newblock \showarticletitle{It Took Longer than {I} was Expecting: Why is
  Dataset Search Still so Hard?}. In \bibinfo{booktitle}{\emph{Proceedings of
  the 2024 Workshop on Human-In-the-Loop Data Analytics, {HILDA} 24, Santiago,
  Chile, 14 June 2024}}, \bibfield{editor}{\bibinfo{person}{Jean{-}Daniel
  Fekete}, \bibinfo{person}{Behrooz Omidvar{-}Tehrani}, \bibinfo{person}{Kexin
  Rong}, {and} \bibinfo{person}{Roee Shraga}} (Eds.).
  \bibinfo{publisher}{{ACM}}, \bibinfo{pages}{1--4}.
\newblock
\urldef\tempurl%
\url{https://doi.org/10.1145/3665939.3665959}
\showDOI{\tempurl}


\bibitem[\protect\citeauthoryear{James, Witten, Hastie, Tibshirani,
  et~al\mbox{.}}{James et~al\mbox{.}}{2013}]%
        {james2013introduction}
\bibfield{author}{\bibinfo{person}{Gareth James}, \bibinfo{person}{Daniela
  Witten}, \bibinfo{person}{Trevor Hastie}, \bibinfo{person}{Robert
  Tibshirani}, {et~al\mbox{.}}} \bibinfo{year}{2013}\natexlab{}.
\newblock \bibinfo{booktitle}{\emph{{An introduction to statistical
  learning}}}. Vol.~\bibinfo{volume}{112}.
\newblock Chapter 4.3.
\newblock


\bibitem[\protect\citeauthoryear{Karagiannis, Saeed, Papotti, and
  Trummer}{Karagiannis et~al\mbox{.}}{2020}]%
        {DBLP:journals/pvldb/Karagiannis0PT20}
\bibfield{author}{\bibinfo{person}{Georgios Karagiannis},
  \bibinfo{person}{Mohammed Saeed}, \bibinfo{person}{Paolo Papotti}, {and}
  \bibinfo{person}{Immanuel Trummer}.} \bibinfo{year}{2020}\natexlab{}.
\newblock \showarticletitle{Scrutinizer: {A} Mixed-Initiative Approach to
  Large-Scale, Data-Driven Claim Verification}.
\newblock \bibinfo{journal}{\emph{Proc. {VLDB} Endow.}} \bibinfo{volume}{13},
  \bibinfo{number}{11} (\bibinfo{year}{2020}), \bibinfo{pages}{2508--2521}.
\newblock
\urldef\tempurl%
\url{http://www.vldb.org/pvldb/vol13/p2508-karagiannis.pdf}
\showURL{%
\tempurl}


\bibitem[\protect\citeauthoryear{Kleinberg, Niculescu{-}Mizil, and
  Sharma}{Kleinberg et~al\mbox{.}}{2008}]%
        {kleinberg2010regret}
\bibfield{author}{\bibinfo{person}{Robert~D. Kleinberg},
  \bibinfo{person}{Alexandru Niculescu{-}Mizil}, {and}
  \bibinfo{person}{Yogeshwer Sharma}.} \bibinfo{year}{2008}\natexlab{}.
\newblock \showarticletitle{Regret Bounds for Sleeping Experts and Bandits}. In
  \bibinfo{booktitle}{\emph{21st Annual Conference on Learning Theory - {COLT}
  2008, Helsinki, Finland, July 9-12, 2008}},
  \bibfield{editor}{\bibinfo{person}{Rocco~A. Servedio} {and}
  \bibinfo{person}{Tong Zhang}} (Eds.). \bibinfo{publisher}{Omnipress},
  \bibinfo{pages}{425--436}.
\newblock
\urldef\tempurl%
\url{http://colt2008.cs.helsinki.fi/papers/114-Kleinberg.pdf}
\showURL{%
\tempurl}


\bibitem[\protect\citeauthoryear{Kolobov, Peres, Lu, and Horvitz}{Kolobov
  et~al\mbox{.}}{2019a}]%
        {kolobov2019staying}
\bibfield{author}{\bibinfo{person}{Andrey Kolobov}, \bibinfo{person}{Yuval
  Peres}, \bibinfo{person}{Cheng Lu}, {and} \bibinfo{person}{Eric~Joel
  Horvitz}.} \bibinfo{year}{2019}\natexlab{a}.
\newblock \showarticletitle{Staying up to Date with Online Content Changes
  Using Reinforcement Learning for Scheduling}. In
  \bibinfo{booktitle}{\emph{Advances in Neural Information Processing Systems
  32: Annual Conference on Neural Information Processing Systems 2019, NeurIPS
  2019, December 8-14, 2019, Vancouver, BC, Canada}},
  \bibfield{editor}{\bibinfo{person}{Hanna~M. Wallach}, \bibinfo{person}{Hugo
  Larochelle}, \bibinfo{person}{Alina Beygelzimer}, \bibinfo{person}{Florence
  d'Alch{\'{e}}{-}Buc}, \bibinfo{person}{Emily~B. Fox}, {and}
  \bibinfo{person}{Roman Garnett}} (Eds.). \bibinfo{pages}{579--589}.
\newblock
\urldef\tempurl%
\url{https://proceedings.neurips.cc/paper/2019/hash/ad13a2a07ca4b7642959dc0c4c740ab6-Abstract.html}
\showURL{%
\tempurl}


\bibitem[\protect\citeauthoryear{Kolobov, Peres, Lubetzky, and Horvitz}{Kolobov
  et~al\mbox{.}}{2019b}]%
        {kolobov2019optimal}
\bibfield{author}{\bibinfo{person}{Andrey Kolobov}, \bibinfo{person}{Yuval
  Peres}, \bibinfo{person}{Eyal Lubetzky}, {and} \bibinfo{person}{Eric
  Horvitz}.} \bibinfo{year}{2019}\natexlab{b}.
\newblock \showarticletitle{Optimal Freshness Crawl Under Politeness
  Constraints}. In \bibinfo{booktitle}{\emph{Proceedings of the 42nd
  International {ACM} {SIGIR} Conference on Research and Development in
  Information Retrieval, {SIGIR} 2019, Paris, France, July 21-25, 2019}},
  \bibfield{editor}{\bibinfo{person}{Benjamin Piwowarski}, \bibinfo{person}{Max
  Chevalier}, \bibinfo{person}{{\'{E}}ric Gaussier}, \bibinfo{person}{Yoelle
  Maarek}, \bibinfo{person}{Jian{-}Yun Nie}, {and} \bibinfo{person}{Falk
  Scholer}} (Eds.). \bibinfo{publisher}{{ACM}}, \bibinfo{pages}{495--504}.
\newblock
\urldef\tempurl%
\url{https://doi.org/10.1145/3331184.3331241}
\showDOI{\tempurl}


\bibitem[\protect\citeauthoryear{Kontogiannis, Kelesis, Pollatos, Paliouras,
  and Giannakopoulos}{Kontogiannis et~al\mbox{.}}{2021}]%
        {kontogiannis2022tree}
\bibfield{author}{\bibinfo{person}{Andreas Kontogiannis},
  \bibinfo{person}{Dimitrios Kelesis}, \bibinfo{person}{Vasilis Pollatos},
  \bibinfo{person}{Georgios Paliouras}, {and} \bibinfo{person}{George
  Giannakopoulos}.} \bibinfo{year}{2021}\natexlab{}.
\newblock \showarticletitle{Tree-based Focused Web Crawling with Reinforcement
  Learning}.
\newblock \bibinfo{journal}{\emph{ArXiv preprint}}
  \bibinfo{volume}{abs/2112.07620} (\bibinfo{year}{2021}).
\newblock
\urldef\tempurl%
\url{https://arxiv.org/abs/2112.07620}
\showURL{%
\tempurl}


\bibitem[\protect\citeauthoryear{Lockard, Shiralkar, Dong, and
  Hajishirzi}{Lockard et~al\mbox{.}}{2020}]%
        {DBLP:conf/wsdm/LockardSDH20}
\bibfield{author}{\bibinfo{person}{Colin Lockard}, \bibinfo{person}{Prashant
  Shiralkar}, \bibinfo{person}{Xin~Luna Dong}, {and} \bibinfo{person}{Hannaneh
  Hajishirzi}.} \bibinfo{year}{2020}\natexlab{}.
\newblock \showarticletitle{Web-scale Knowledge Collection}. In
  \bibinfo{booktitle}{\emph{{WSDM} '20: The Thirteenth {ACM} International
  Conference on Web Search and Data Mining, Houston, TX, USA, February 3-7,
  2020}}, \bibfield{editor}{\bibinfo{person}{James Caverlee},
  \bibinfo{person}{Xia~(Ben) Hu}, \bibinfo{person}{Mounia Lalmas}, {and}
  \bibinfo{person}{Wei Wang}} (Eds.). \bibinfo{publisher}{{ACM}},
  \bibinfo{pages}{888--889}.
\newblock
\urldef\tempurl%
\url{https://doi.org/10.1145/3336191.3371878}
\showDOI{\tempurl}


\bibitem[\protect\citeauthoryear{Malkov and Yashunin}{Malkov and
  Yashunin}{2016}]%
        {malkov2018efficient}
\bibfield{author}{\bibinfo{person}{Yury~A. Malkov} {and}
  \bibinfo{person}{Dmitry~A. Yashunin}.} \bibinfo{year}{2016}\natexlab{}.
\newblock \showarticletitle{Efficient and robust approximate nearest neighbor
  search using Hierarchical Navigable Small World graphs}.
\newblock \bibinfo{journal}{\emph{ArXiv preprint}}
  \bibinfo{volume}{abs/1603.09320} (\bibinfo{year}{2016}).
\newblock
\urldef\tempurl%
\url{https://arxiv.org/abs/1603.09320}
\showURL{%
\tempurl}


\bibitem[\protect\citeauthoryear{Meusel, Mika, and Blanco}{Meusel
  et~al\mbox{.}}{2014}]%
        {meusel2014focused}
\bibfield{author}{\bibinfo{person}{Robert Meusel}, \bibinfo{person}{Peter
  Mika}, {and} \bibinfo{person}{Roi Blanco}.} \bibinfo{year}{2014}\natexlab{}.
\newblock \showarticletitle{Focused Crawling for Structured Data}. In
  \bibinfo{booktitle}{\emph{Proceedings of the 23rd {ACM} International
  Conference on Conference on Information and Knowledge Management, {CIKM}
  2014, Shanghai, China, November 3-7, 2014}},
  \bibfield{editor}{\bibinfo{person}{Jianzhong Li},
  \bibinfo{person}{Xiaoyang~Sean Wang}, \bibinfo{person}{Minos~N. Garofalakis},
  \bibinfo{person}{Ian Soboroff}, \bibinfo{person}{Torsten Suel}, {and}
  \bibinfo{person}{Min Wang}} (Eds.). \bibinfo{publisher}{{ACM}},
  \bibinfo{pages}{1039--1048}.
\newblock
\urldef\tempurl%
\url{https://doi.org/10.1145/2661829.2661902}
\showDOI{\tempurl}


\bibitem[\protect\citeauthoryear{Miao, Tatemura, Hsiung, Sawires, and
  Moser}{Miao et~al\mbox{.}}{2009}]%
        {miao2009extracting}
\bibfield{author}{\bibinfo{person}{Gengxin Miao}, \bibinfo{person}{Jun'ichi
  Tatemura}, \bibinfo{person}{Wang{-}Pin Hsiung}, \bibinfo{person}{Arsany
  Sawires}, {and} \bibinfo{person}{Louise~E. Moser}.}
  \bibinfo{year}{2009}\natexlab{}.
\newblock \showarticletitle{Extracting data records from the web using tag path
  clustering}. In \bibinfo{booktitle}{\emph{Proceedings of the 18th
  International Conference on World Wide Web, {WWW} 2009, Madrid, Spain, April
  20-24, 2009}}, \bibfield{editor}{\bibinfo{person}{Juan Quemada},
  \bibinfo{person}{Gonzalo Le{\'{o}}n}, \bibinfo{person}{Yo{\"{e}}lle~S.
  Maarek}, {and} \bibinfo{person}{Wolfgang Nejdl}} (Eds.).
  \bibinfo{publisher}{{ACM}}, \bibinfo{pages}{981--990}.
\newblock
\urldef\tempurl%
\url{https://doi.org/10.1145/1526709.1526841}
\showDOI{\tempurl}


\bibitem[\protect\citeauthoryear{Mohr, Stack, Rnitovic, Avery, and
  Kimpton}{Mohr et~al\mbox{.}}{2004}]%
        {mohr2004introduction}
\bibfield{author}{\bibinfo{person}{Gordon Mohr}, \bibinfo{person}{Michael
  Stack}, \bibinfo{person}{Igor Rnitovic}, \bibinfo{person}{Dan Avery}, {and}
  \bibinfo{person}{Michele Kimpton}.} \bibinfo{year}{2004}\natexlab{}.
\newblock \showarticletitle{{Introduction to Heritrix}}. In
  \bibinfo{booktitle}{\emph{4th International Web Archiving Workshop}}.
  \bibinfo{pages}{109--115}.
\newblock


\bibitem[\protect\citeauthoryear{Orth}{Orth}{2022}]%
        {wrongstats}
\bibfield{author}{\bibinfo{person}{Taylor Orth}.}
  \bibinfo{year}{2022}\natexlab{}.
\newblock \bibinfo{title}{From millionaires to {M}uslims, small subgroups of
  the population seem much larger to many {A}mericans}.
\newblock
  \bibinfo{howpublished}{\url{https://today.yougov.com/politics/articles/41556-americans-misestimate-small-subgroups-population}}.
\newblock


\bibitem[\protect\citeauthoryear{Pham, Santos, and Freire}{Pham
  et~al\mbox{.}}{2018}]%
        {pham2018learning}
\bibfield{author}{\bibinfo{person}{Kien Pham}, \bibinfo{person}{A{\'{e}}cio
  S.~R. Santos}, {and} \bibinfo{person}{Juliana Freire}.}
  \bibinfo{year}{2018}\natexlab{}.
\newblock \showarticletitle{Learning to Discover Domain-Specific Web Content}.
  In \bibinfo{booktitle}{\emph{Proceedings of the Eleventh {ACM} International
  Conference on Web Search and Data Mining, {WSDM} 2018, Marina Del Rey, CA,
  USA, February 5-9, 2018}}, \bibfield{editor}{\bibinfo{person}{Yi~Chang},
  \bibinfo{person}{Chengxiang Zhai}, \bibinfo{person}{Yan Liu}, {and}
  \bibinfo{person}{Yoelle Maarek}} (Eds.). \bibinfo{publisher}{{ACM}},
  \bibinfo{pages}{432--440}.
\newblock
\urldef\tempurl%
\url{https://doi.org/10.1145/3159652.3159724}
\showDOI{\tempurl}


\bibitem[\protect\citeauthoryear{Saeed and Papotti}{Saeed and Papotti}{2021}]%
        {DBLP:journals/debu/0002P21}
\bibfield{author}{\bibinfo{person}{Mohammed Saeed} {and} \bibinfo{person}{Paolo
  Papotti}.} \bibinfo{year}{2021}\natexlab{}.
\newblock \showarticletitle{Fact-Checking Statistical Claims with Tables}.
\newblock \bibinfo{journal}{\emph{{IEEE} Data Eng. Bull.}}
  \bibinfo{volume}{44}, \bibinfo{number}{3} (\bibinfo{year}{2021}),
  \bibinfo{pages}{27--38}.
\newblock
\urldef\tempurl%
\url{http://sites.computer.org/debull/A21sept/p27.pdf}
\showURL{%
\tempurl}


\bibitem[\protect\citeauthoryear{Schulam and Muslea}{Schulam and
  Muslea}{2023}]%
        {schulam2023improving}
\bibfield{author}{\bibinfo{person}{Peter Schulam} {and} \bibinfo{person}{Ion
  Muslea}.} \bibinfo{year}{2023}\natexlab{}.
\newblock \showarticletitle{Improving the Exploration/Exploitation Trade-Off in
  Web Content Discovery}. In \bibinfo{booktitle}{\emph{Companion Proceedings of
  the {ACM} Web Conference 2023, {WWW} 2023, Austin, TX, USA, 30 April 2023 - 4
  May 2023}}, \bibfield{editor}{\bibinfo{person}{Ying Ding},
  \bibinfo{person}{Jie Tang}, \bibinfo{person}{Juan~F. Sequeda},
  \bibinfo{person}{Lora Aroyo}, \bibinfo{person}{Carlos Castillo}, {and}
  \bibinfo{person}{Geert{-}Jan Houben}} (Eds.). \bibinfo{publisher}{{ACM}},
  \bibinfo{pages}{1183--1189}.
\newblock
\urldef\tempurl%
\url{https://doi.org/10.1145/3543873.3587574}
\showDOI{\tempurl}


\bibitem[\protect\citeauthoryear{{SDMX Community}}{{SDMX Community}}{2024}]%
        {sdmx}
\bibfield{author}{\bibinfo{person}{{SDMX Community}}.}
  \bibinfo{year}{2024}\natexlab{}.
\newblock \bibinfo{title}{SDMX Technical Specifications}.
\newblock \bibinfo{howpublished}{https://sdmx.org/?page\_id=5008}.
\newblock


\bibitem[\protect\citeauthoryear{Senellart}{Senellart}{2005}]%
        {senellart2005identifying}
\bibfield{author}{\bibinfo{person}{Pierre Senellart}.}
  \bibinfo{year}{2005}\natexlab{}.
\newblock \showarticletitle{Identifying Websites with Flow Simulation}. In
  \bibinfo{booktitle}{\emph{Web Engineering, 5th International Conference,
  {ICWE} 2005, Sydney, Australia, July 27-29, 2005, Proceedings}}
  \emph{(\bibinfo{series}{Lecture Notes in Computer Science})},
  \bibfield{editor}{\bibinfo{person}{David~B. Lowe} {and}
  \bibinfo{person}{Martin Gaedke}} (Eds.), Vol.~\bibinfo{volume}{3579}.
  \bibinfo{publisher}{Springer}, \bibinfo{pages}{124--129}.
\newblock
\urldef\tempurl%
\url{https://doi.org/10.1007/11531371\_18}
\showDOI{\tempurl}


\bibitem[\protect\citeauthoryear{Shalev{-}Shwartz, Crammer, Dekel, and
  Singer}{Shalev{-}Shwartz et~al\mbox{.}}{2003}]%
        {crammer2006online}
\bibfield{author}{\bibinfo{person}{Shai Shalev{-}Shwartz},
  \bibinfo{person}{Koby Crammer}, \bibinfo{person}{Ofer Dekel}, {and}
  \bibinfo{person}{Yoram Singer}.} \bibinfo{year}{2003}\natexlab{}.
\newblock \showarticletitle{Online Passive-Aggressive Algorithms}. In
  \bibinfo{booktitle}{\emph{Advances in Neural Information Processing Systems
  16 [Neural Information Processing Systems, {NIPS} 2003, December 8-13, 2003,
  Vancouver and Whistler, British Columbia, Canada]}},
  \bibfield{editor}{\bibinfo{person}{Sebastian Thrun},
  \bibinfo{person}{Lawrence~K. Saul}, {and} \bibinfo{person}{Bernhard
  Sch{\"{o}}lkopf}} (Eds.). \bibinfo{publisher}{{MIT} Press},
  \bibinfo{pages}{1229--1236}.
\newblock
\urldef\tempurl%
\url{https://proceedings.neurips.cc/paper/2003/hash/4ebd440d99504722d80de606ea8507da-Abstract.html}
\showURL{%
\tempurl}


\bibitem[\protect\citeauthoryear{Sigurðsson}{Sigurðsson}{2005}]%
        {sigurdhsson2005incremental}
\bibfield{author}{\bibinfo{person}{Kristinn Sigurðsson}.}
  \bibinfo{year}{2005}\natexlab{}.
\newblock \showarticletitle{Incremental crawling with {Heritrix}}. In
  \bibinfo{booktitle}{\emph{Proceedings of the 5th International Web Archiving
  Workshop}}. \bibinfo{address}{Vienna, Austria}.
\newblock
\urldef\tempurl%
\url{http://iwaw.europarchive.org/05/papers/iwaw05-sigurdsson.pdf}
\showURL{%
\tempurl}


\bibitem[\protect\citeauthoryear{Soric, Gracianne, Manolescu, and
  Senellart}{Soric et~al\mbox{.}}{2025}]%
        {soric2025benchmarkingtableextractionheterogeneous}
\bibfield{author}{\bibinfo{person}{Marijan Soric}, \bibinfo{person}{Cécile
  Gracianne}, \bibinfo{person}{Ioana Manolescu}, {and} \bibinfo{person}{Pierre
  Senellart}.} \bibinfo{year}{2025}\natexlab{}.
\newblock \bibinfo{title}{{Benchmarking Table Extraction from Heterogeneous
  Scientific Extraction Documents}}.
\newblock
\newblock
\urldef\tempurl%
\url{https://arxiv.org/abs/2511.16134}
\showURL{%
\tempurl}


\bibitem[\protect\citeauthoryear{Sutton and Barto}{Sutton and Barto}{1998}]%
        {sutton2018reinforcement}
\bibfield{author}{\bibinfo{person}{Richard~S. Sutton} {and}
  \bibinfo{person}{Andrew~G. Barto}.} \bibinfo{year}{1998}\natexlab{}.
\newblock \bibinfo{booktitle}{\emph{Reinforcement learning - an introduction}}.
\newblock \bibinfo{publisher}{{MIT} Press}.
\newblock
\showISBNx{978-0-262-19398-6}
\urldef\tempurl%
\url{http://www.incompleteideas.net/book/first/the-book.html}
\showURL{%
\tempurl}


\bibitem[\protect\citeauthoryear{Tanon, Weikum, and Suchanek}{Tanon
  et~al\mbox{.}}{2020}]%
        {DBLP:conf/esws/TanonWS20}
\bibfield{author}{\bibinfo{person}{Thomas~Pellissier Tanon},
  \bibinfo{person}{Gerhard Weikum}, {and} \bibinfo{person}{Fabian~M.
  Suchanek}.} \bibinfo{year}{2020}\natexlab{}.
\newblock \showarticletitle{{YAGO} 4: {A} Reason-able Knowledge Base}. In
  \bibinfo{booktitle}{\emph{The Semantic Web - 17th International Conference,
  {ESWC} 2020, Heraklion, Crete, Greece, May 31-June 4, 2020, Proceedings}}
  \emph{(\bibinfo{series}{Lecture Notes in Computer Science})},
  \bibfield{editor}{\bibinfo{person}{Andreas Harth}, \bibinfo{person}{Sabrina
  Kirrane}, \bibinfo{person}{Axel{-}Cyrille~Ngonga Ngomo},
  \bibinfo{person}{Heiko Paulheim}, \bibinfo{person}{Anisa Rula},
  \bibinfo{person}{Anna~Lisa Gentile}, \bibinfo{person}{Peter Haase}, {and}
  \bibinfo{person}{Michael Cochez}} (Eds.), Vol.~\bibinfo{volume}{12123}.
  \bibinfo{publisher}{Springer}, \bibinfo{pages}{583--596}.
\newblock
\urldef\tempurl%
\url{https://doi.org/10.1007/978-3-030-49461-2\_34}
\showDOI{\tempurl}


\bibitem[\protect\citeauthoryear{Vitagliano, Jiang, and Naumann}{Vitagliano
  et~al\mbox{.}}{2021}]%
        {vitagliano2021detecting}
\bibfield{author}{\bibinfo{person}{Gerardo Vitagliano}, \bibinfo{person}{Lan
  Jiang}, {and} \bibinfo{person}{Felix Naumann}.}
  \bibinfo{year}{2021}\natexlab{}.
\newblock \showarticletitle{Detecting layout templates in complex multiregion
  files}.
\newblock \bibinfo{journal}{\emph{Proc. VLDB Endow.}} \bibinfo{volume}{15},
  \bibinfo{number}{3} (\bibinfo{year}{2021}), \bibinfo{pages}{646–658}.
\newblock


\bibitem[\protect\citeauthoryear{Vlachos and Riedel}{Vlachos and
  Riedel}{2015}]%
        {vlachos-riedel-2015-identification}
\bibfield{author}{\bibinfo{person}{Andreas Vlachos} {and}
  \bibinfo{person}{Sebastian Riedel}.} \bibinfo{year}{2015}\natexlab{}.
\newblock \showarticletitle{Identification and Verification of Simple Claims
  about Statistical Properties}. In \bibinfo{booktitle}{\emph{Proceedings of
  the 2015 Conference on Empirical Methods in Natural Language Processing}},
  \bibfield{editor}{\bibinfo{person}{Llu{\'\i}s M{\`a}rquez},
  \bibinfo{person}{Chris Callison-Burch}, {and} \bibinfo{person}{Jian Su}}
  (Eds.). \bibinfo{publisher}{Association for Computational Linguistics},
  \bibinfo{address}{Lisbon, Portugal}, \bibinfo{pages}{2596--2601}.
\newblock
\urldef\tempurl%
\url{https://doi.org/10.18653/v1/D15-1312}
\showDOI{\tempurl}


\bibitem[\protect\citeauthoryear{Wang and Fernandez}{Wang and
  Fernandez}{2023}]%
        {DBLP:journals/pacmmod/WangF23}
\bibfield{author}{\bibinfo{person}{Qiming Wang} {and}
  \bibinfo{person}{Raul~Castro Fernandez}.} \bibinfo{year}{2023}\natexlab{}.
\newblock \showarticletitle{Solo: Data Discovery Using Natural Language
  Questions Via {A} Self-Supervised Approach}.
\newblock \bibinfo{journal}{\emph{Proc. {ACM} Manag. Data}}
  \bibinfo{volume}{1}, \bibinfo{number}{4} (\bibinfo{year}{2023}),
  \bibinfo{pages}{262:1--262:27}.
\newblock
\urldef\tempurl%
\url{https://doi.org/10.1145/3626756}
\showDOI{\tempurl}


\bibitem[\protect\citeauthoryear{Weikum, Dong, Razniewski, and Suchanek}{Weikum
  et~al\mbox{.}}{2021}]%
        {DBLP:journals/ftdb/WeikumDRS21}
\bibfield{author}{\bibinfo{person}{Gerhard Weikum}, \bibinfo{person}{Xin~Luna
  Dong}, \bibinfo{person}{Simon Razniewski}, {and} \bibinfo{person}{Fabian~M.
  Suchanek}.} \bibinfo{year}{2021}\natexlab{}.
\newblock \showarticletitle{Machine Knowledge: Creation and Curation of
  Comprehensive Knowledge Bases}.
\newblock \bibinfo{journal}{\emph{Found. Trends Databases}}
  \bibinfo{volume}{10}, \bibinfo{number}{2-4} (\bibinfo{year}{2021}),
  \bibinfo{pages}{108--490}.
\newblock
\urldef\tempurl%
\url{https://doi.org/10.1561/1900000064}
\showDOI{\tempurl}


\bibitem[\protect\citeauthoryear{{WHATWG}}{{WHATWG}}{2024}]%
        {DOM}
\bibfield{author}{\bibinfo{person}{{WHATWG}}.} \bibinfo{year}{2024}\natexlab{}.
\newblock \bibinfo{title}{{DOM}: Living Standard}.
\newblock \bibinfo{howpublished}{https://dom.spec.whatwg.org/}.
\newblock


\bibitem[\protect\citeauthoryear{Zhang, Santos, and Freire}{Zhang
  et~al\mbox{.}}{2021}]%
        {zhang2021dsdd}
\bibfield{author}{\bibinfo{person}{Haoxiang Zhang},
  \bibinfo{person}{A{\'{e}}cio S.~R. Santos}, {and} \bibinfo{person}{Juliana
  Freire}.} \bibinfo{year}{2021}\natexlab{}.
\newblock \showarticletitle{{DSDD:} Domain-Specific Dataset Discovery on the
  Web}. In \bibinfo{booktitle}{\emph{{CIKM} '21: The 30th {ACM} International
  Conference on Information and Knowledge Management, Virtual Event,
  Queensland, Australia, November 1 - 5, 2021}},
  \bibfield{editor}{\bibinfo{person}{Gianluca Demartini},
  \bibinfo{person}{Guido Zuccon}, \bibinfo{person}{J.~Shane Culpepper},
  \bibinfo{person}{Zi~Huang}, {and} \bibinfo{person}{Hanghang Tong}} (Eds.).
  \bibinfo{publisher}{{ACM}}, \bibinfo{pages}{2527--2536}.
\newblock
\urldef\tempurl%
\url{https://doi.org/10.1145/3459637.3482427}
\showDOI{\tempurl}


\end{thebibliography}

\end{document}